\documentclass[10pt, final, journal, letterpaper, oneside, twocolumn]{IEEEtran}
\usepackage{graphicx}
\usepackage{amsmath}
\usepackage{amssymb}
\usepackage{mathrsfs}
\usepackage{stfloats}
\usepackage{dsfont}
\usepackage{setspace}
\usepackage{array}
\newtheorem{theorem}{Theorem}
\newtheorem{corollary}{Corollary}
\newtheorem{proposition}{Proposition}

\newtheorem{remark}{Remark}
\begin{document}
\title{Average Rate of Downlink Heterogeneous Cellular Networks over Generalized Fading Channels --\\ A Stochastic Geometry Approach}
\author{Marco~Di~Renzo,~\IEEEmembership{Member,~IEEE},
        Alessandro~Guidotti,~\IEEEmembership{Student~Member,~IEEE}, and
        Giovanni~E.~Corazza,~\IEEEmembership{Senior~Member,~IEEE}
\thanks{Manuscript received November 18, 2012; revised March 18, 2013. The associate editor coordinating the review of this paper and approving it for publication was M. Tao. }
\thanks{M. Di Renzo is with the Laboratoire des Signaux et Syst\`emes, Unit\'e Mixte de Recherche 8506, Centre National de la Recherche Scientifique--\'Ecole
Sup\'erieure d'\'Electricit\'e--Universit\'e Paris--Sud XI, 91192 Gif--sur--Yvette Cedex, France, (e--mail: marco.direnzo@lss.supelec.fr). }
\thanks{A. Guidotti and G. E. Corazza are with the Department of Electronics, Computer Science, and Systems, the University of Bologna, Viale Risorgimento 2, 40136 Bologna, Italy, (e--mail: \{a.guidotti, giovanni.corazza\}@unibo.it). }
\thanks{This paper was presented in part at the IEEE Int. Commun. Conf. (ICC), Ottawa, Canada, June 2012.}
\thanks{This work was supported in part by the European Commission under the auspices of the FP7--PEOPLE MITN--CROSSFIRE project (grant 317126) and the FP7--ICT NoE--NEWCOM\# project (grant 318306).}
\thanks{Digital Object Identifier XX.XXXX/TCOMM.XXX.XXX.XXX.} }
\markboth{IEEE Transactions on Communications} {M. Di Renzo, A. Guidotti, and G. E. Corazza: Average Rate of Downlink Heterogeneous Cellular Networks over Generalized Fading Channels -- A Stochastic Geometry Approach}
\maketitle
\begin{abstract}
In this paper, we introduce an analytical framework to compute the average rate of downlink heterogeneous cellular networks. The framework leverages recent application of stochastic geometry to other--cell interference modeling and analysis. The heterogeneous cellular network is modeled as the superposition of many tiers of Base Stations (BSs) having different transmit power, density, path--loss exponent, fading parameters and distribution, and unequal biasing for flexible tier association. A long--term averaged maximum biased--received--power tier association is considered. The positions of the BSs in each tier are modeled as points of an independent Poisson Point Process (PPP). Under these assumptions, we introduce a new analytical methodology to evaluate the average rate, which avoids the computation of the Coverage Probability (Pcov) and needs only the Moment Generating Function (MGF) of the aggregate interference at the probe mobile terminal. The distinguishable characteristic of our analytical methodology consists in providing a tractable and numerically efficient framework that is applicable to general fading distributions, including composite fading channels with small-- and mid--scale fluctuations. In addition, our method can efficiently handle correlated Log--Normal shadowing with little increase of the computational complexity. The proposed MGF--based approach needs the computation of either a single or a two--fold numerical integral, thus reducing the complexity of Pcov--based frameworks, which require, for general fading distributions, the computation of a four--fold integral.
\end{abstract}
\section{Introduction} \label{Introduction}
\PARstart{T}{he} analytical performance modeling of cellular networks is a long--standing open issue \cite{Padovani1991}. The mathematical intractability mainly arises from the difficulty of accurately modeling other--cell interference by taking into account the spatial positions of the Base Stations (BSs) and the stochastic character of the wireless channel \cite{AndrewsNov2010}. For this reason, accurate performance analysis is usually conducted via costly, time--consuming, and often proprietary system--level simulators \cite{Baccelli1997}. This approach, however, seldom provides insightful information on system design and on the dependency of the system parameters to optimize. This situation is even exacerbated in future cellular deployments, which are becoming more heterogeneous with the introduction of new infrastructure elements, \emph{e.g.}, femto/pico BSs, fixed/mobile relays, cognitive radios, and distributed antennas \cite{Chandrasekhar2008}--\cite{Ghosh2012} and \cite{AndrewsMar2013} for a survey.
\subsection{Abstraction Models for Analysis and Design of Cellular Networks}
To circumvent this problem, communications theorists usually resort to ``abstractions'' for tractable other--cell interference modeling and for performance analysis. These abstractions usually encompass simplified spatial models for the locations of the BSs. In particular, three abstraction models are commonly adopted: i) the Wyner model \cite{Wyner1994}; ii) the single--cell interfering model \cite{ChaeJSAC}; and iii) the regular hexagonal or square grid model \cite{Macdonald1979}. These abstraction models, however, are often either over--simplistic or inaccurate \cite{AndrewsWyner}. Furthermore, in some cases, as for the regular hexagonal/square grid model, they still require either intensive numerical simulations or multi--fold numerical integrations. Finally, these abstraction models usually provide information for \emph{specific} BSs deployments, and typically fail to provide useful information for more random, unplanned, and/or clustered BSs deployments, which are typical of emerging heterogeneous cellular networks with, \emph{e.g.}, overlaid femtocells and picocells \cite{Ghosh2012}, \cite{AndrewsMar2013}. Motivated by these considerations, a new abstraction model is currently emerging and gaining popularity, according to which the positions of the BSs are modeled as points of a Poisson Point Process (PPP) and powerful tools from applied probability, such as stochastic geometry, are leveraged to develop tractable integrals and closed--form mathematical frameworks for important performance metrics (\emph{e.g.}, coverage and average rate) \cite{HaenggiBook2009}--\cite{BaccelliBook2009}.
\subsection{Stochastic Geometry based Modeling of Heterogeneous Cellular Networks}
The stochastic geometry based abstraction model for the analysis of cellular networks dates back to (at least) 1997 \cite{Baccelli1997}, \cite{Fleming1997}. Subsequently, a similar shotgun--based, \emph{i.e.}, PPP--based, abstraction model was proposed in \cite{Brown2000}, and it was shown that, compared with the traditional hexagonal grid model, the shotgun approach provides upper performance bounds. More recently, the PPP model has been used for the analysis of spatial and opportunistic Aloha protocols \cite{BaccelliSep2009}, and for the characterization of the Signal--to--Interference--Ratio (SIR) of (single--tier) cellular networks \cite{BrownGLOBE2009}. In spite of these initial and pioneering attempts of applying the PPP model and stochastic geometry to the analysis of cellular networks, only recently the random--based abstraction model for the positions of the BSs has received the attention it deserved. In particular, its emergence and widespread adoption for cellular networks analysis and design is mostly due to \cite{AndrewsNov2011}, where a comprehensive framework to compute coverage and average rate of single--tier deployments is provided. In \cite{AndrewsNov2011}, it is shown that the PPP model is as accurate as regular grid models, but it has the main advantage of being more analytically tractable. A comprehensive study based on real BSs deployments obtained from the open source project OpenCellID \cite{OpenCellID_PPP} has revealed that the PPP model can indeed be used for accurate coverage analysis in major cities worldwide. Recent results about the validation of the PPP model for real BSs deployments are available in \cite{Haenggi_Fitting}, where data collected from Ofcom, \emph{i.e.}, the independent regulator and competition authority in the UK, is used. Fueled by these encouraging results, many researchers are currently using the PPP--based abstraction model to study single-- and multi--tier cellular networks, \emph{e.g.}, \cite{DecreusefondSep2010}--\cite{AndrewsFeb2013} and references therein. The PPP--based approach is also widely adopted for network interference modeling, \emph{e.g.}, \cite{Sousa1992}--\cite{Chopra2012}.
\subsection{Analytical Computation of the Average Rate: State--of--the--Art and Paper Contribution} \label{Intro_SOTA}
In this paper, we capitalize on the emerging PPP--based abstraction model for multi--tier cellular networks, and propose a new mathematical methodology to compute the downlink average rate over general fading channels. Recent papers have developed frameworks to compute the average rate for single--tier downlink \cite{AndrewsNov2011}, \cite{GinibreJune2012}, multi--tier downlink \cite{DhillonAllerton2011}, \cite{AndrewsOct2012}, downlink multi--cell coordination \cite{AkoumSPAWC2012}, \cite{AkoumTSP2012}, and single--tier uplink cellular networks \cite{NovlanMar2012}. All these papers use the same two--step methodological approach to compute the average rate, which was originally introduced in \cite{BaccelliSep2009} and exploits the Plancherel--Parseval theorem: i) first, the Coverage Probability (Pcov) is computed; and ii) then, the average rate is obtained by integrating Pcov over the positive real axis \cite[Eq. (2.12)]{BaccelliSep2009}. Throughout this paper, this methodology is denoted by \emph{Pcov--based approach}. Although this technique avoids system--level simulations, it requires, for general fading channels, the computation of a four--fold integral \cite[Appendix C]{AndrewsNov2011}. For this reason, many authors often limit the analysis to Rayleigh fading channels and/or to interference--limited networks, where simplified frameworks can be obtained. Further details about the computational complexity of the Pcov--based approach are available in Section \ref{Pcov_vs_MGF}.

To overcome this limitation, we propose a new analytical framework which, at the same time, reduces the number of integrals to be computed, and, similar to the Pcov--based approach, is flexible enough for application to arbitrary fading distributions (including correlated composite channel models). The framework leverages the application of recent results on the computation of the ergodic capacity in the presence of interference and noise \cite{Hamdi2010}. It avoids the computation of Pcov, and needs only the Moment Generating Function (MGF) of the aggregate interference at the probe mobile terminal. Throughout this paper, this framework is denoted by \emph{MGF--based approach}. We show that it is applicable to multi--tier cellular networks with long--term averaged maximum biased--received--power tier association, and that either a single or a two--fold numerical integral need to be computed for arbitrary fading channels.
\subsection{Paper Organization}
The remainder of this paper is organized as follows. In Section \ref{SystemModel}, the system model is described. In Section \ref{SingleTier}, the MGF--based approach is introduced for single--tier cellular networks. In Section \ref{MultiTier}, the proposed methodology is applied to multi--tier cellular networks with flexible (biased) tier association. In Section \ref{NumericalResults}, extensive Monte Carlo simulations are shown to substantiate the proposed mathematical framework for various fading channel models and cellular networks deployments. Finally, Section \ref{Conclusion} concludes this paper.
\section{System Model and Problem Statement} \label{SystemModel}
We consider a downlink heterogeneous cellular networks model similar to \cite{DhillonJSAC2012}, \cite{AndrewsOct2012}, and \cite{BrownGLOBE2012}. However, the following differences hold. In \cite{DhillonJSAC2012} and \cite{BrownGLOBE2012}, the tier association policy is based on the instantaneous Signal--to--Interference--plus--Noise--Ratio (SINR). On the other hand, similar to \cite{AndrewsOct2012}, we consider a biased long--term averaged tier association policy, as described in Section \ref{TierAssociationPolicy}. Compared with \cite{DhillonJSAC2012}, \cite{AndrewsOct2012}, and \cite{BrownGLOBE2012} the analytical methodology to compute the average rate is not based on the Pcov--based approach but on the MGF--based approach.

\emph{Notation}: ${\mathbb{E}}\left\{  \cdot  \right\}$ denotes the expectation operator. $\mathcal{M}_X\left( s \right) = {\mathbb{E}}\left\{ {\exp \left( { - sX} \right)} \right\}$ is the MGF of random variable $X$. $f_X \left(  \cdot  \right)$ denotes the Probability Density Function (PDF) of random variable $X$. $\Gamma \left( x \right) = \int\nolimits_0^{ + \infty } {\exp \left( { - t} \right)t^{x - 1} dt}$ is the gamma function. ${\rm{erfc}}\left( x \right) = \left( {{2 \mathord{\left/ {\vphantom {2 {\sqrt \pi  }}} \right. \kern-\nulldelimiterspace} {\sqrt \pi  }}} \right)\int\nolimits_x^{ + \infty } {\exp \left\{ { - \xi ^2 } \right\}d\xi }$ is the complementary Gauss error function. $S_{a,b} \left(  \cdot  \right)$ is the Lommel function defined in \cite[Sec. 7.5.5]{BatemanVol2}. $G_{p,q}^{m,n} \left( {\left(  \cdot  \right)\left| {\begin{array}{*{20}c} {\left( {a_p } \right)}  \\ {\left( {b_q } \right)}  \\ \end{array}} \right.} \right)$ is the Meijer G--function defined in \cite[Sec. 2.24]{Prudnikov_Vol3}. $\Delta \left( {n,x} \right) = \left[ {{x \mathord{\left/ {\vphantom {x n}} \right. \kern-\nulldelimiterspace} n},{{\left( {x + 1} \right)} \mathord{\left/ {\vphantom {{\left( {x + 1} \right)} n}} \right. \kern-\nulldelimiterspace} n}, \ldots ,,{{\left( {x + n - 1} \right)} \mathord{\left/ {\vphantom {{\left( {x + n - 1} \right)} n}} \right.\kern-\nulldelimiterspace} n}} \right]$, with $n$ being a positive integer and $x$ a real number. The $i$--th entry of $\Delta \left( {n,x} \right)$ is denoted by $\Delta_i \left( {n,x} \right)$. ${}_2F_1 \left( { \cdot , \cdot , \cdot , \cdot } \right)$ is the Gauss hypergeometric function defined in \cite[Ch. 15]{AbramowitzStegun}. ${}_1F_1 \left( {\cdot , \cdot , \cdot } \right)$ is the confluent hypergeometric function defined in \cite[Ch. 13]{AbramowitzStegun}. $\left( { \cdot !} \right)$ is the factorial operator. $\Gamma \left( {z,x} \right) = \int\nolimits_x^{ + \infty } {t^{z - 1} \exp \left\{ { - t} \right\}dt}$ is the upper--incomplete gamma function. $\gamma \left( {z,x} \right) = \int\nolimits_0^x {t^{z - 1} \exp \left\{ { - t} \right\}dt}$ is the lower--incomplete gamma function. $\delta \left(  \cdot  \right)$ is the Dirac delta function. $I_\nu  \left(  \cdot  \right)$ is the modified Bessel function of the first kind and order $\nu$ defined in \cite[Sec. 9.6]{AbramowitzStegun}. $\mathcal{H}\left( \cdot \right)$ is the Heaviside function, \emph{i.e.}, $\mathcal{H}\left( x \right) = 1$ if $x \ge 0$ and $\mathcal{H}\left( x \right) = 0$ if $x < 0$. ${}_pF_q \left( { \cdot , \cdot , \cdot } \right)$ is the generalized hypergeometric function defined in \cite[Ch. IV]{BatemanVol1}. $j = \sqrt { - 1}$ is the imaginary unit. ${\rm{card}}\left\{  \cdot  \right\}$ denotes the cardinality of a set.
\subsection{Heterogeneous Cellular Networks Model} \label{HetNetsModel}
Let us consider the PPP--based abstraction model for the positions of the BSs in a bi--dimensional plane. Then, a heterogeneous cellular deployment can be modeled as a $T$--tier network where each tier models the BSs of a particular class. Each class of BSs is distinguished by its spatial density ($\lambda_t$ for $t=1, 2, \ldots, T$), transmit power ($P_t$ for $t=1, 2, \ldots, T$), path--loss exponent ($\alpha_t > 2$ for $t=1, 2, \ldots, T$), biasing factor ($B_t > 0$ for $t=1, 2, \ldots, T$), and fading parameters and distribution. The BSs of each PPP are assumed to have the same transmit power, the same path--loss exponent, the same biasing factor, and their fading channels are independent and identically distributed (i.i.d.). The extension to correlated and identically distributed (c.i.d.) fading is discussed in Section \ref{Correlated_LogN}. However, for mathematical generality, we assume that the fading distribution of the serving (tagged) BS is different from the fading distribution of the intra--tier interfering BSs. The BSs of each tier are assumed to be spatially distributed according to a homogeneous PPP ($\Phi_t$ for $t=1, 2, \ldots, T$). The $T$ PPPs are assumed to be spatially independent. Our analysis applies to a typical mobile terminal, as permissible in any homogeneous PPP according to the Slivnyak--Mecke's theorem \cite[vol. 1, Theorem 1.4.5]{BaccelliBook2009}. Without loss of generality, the typical Mobile Terminal (${\rm{MT}}_0$) is assumed to be located at the origin of the bi--dimensional plane. The $b$--th BS of the $t$--th tier is denoted by ${\rm{BS}}_{t,b}$. The distance from ${\rm{BS}}_{t,b}$ to ${\rm{MT}}_0$ is denoted by $d_{t,b}$. The standard path--loss function $l \left( d_{t,b} \right) = d_{t,b}^{ - \alpha _t }$ is considered. The power channel gain of the ${\rm{BS}}_{t,b}$--to--${\rm{MT}}_0$ link is denoted by $g_{t,b}  = \left| {h_{t,b} } \right|^2$, where ${h_{t,b} }$ is the related complex amplitude channel gain. For a fair comparison among fading channels with different distributions, the normalization constraint ${\mathbb{E}}\left\{ {g_{t,b} } \right\} = {\mathbb{E}}\left\{ {\left| {h_{t,b} } \right|^2 } \right\} = \Omega  = 1$ is assumed for every $b$ and for $t=1, 2, \ldots, T$.

The frameworks developed in the present paper are applicable to single--input--single--output transmission systems. In other words, BSs and ${\rm{MT}}_0$ are equipped with a single transmit and receive antenna, respectively. The generalization of the proposed analytical methodology to more advanced transmission technologies is currently under investigation, but it is beyond the scope of the present paper. The interested reader can, however, find preliminary results to the analysis of multi--antenna receivers and dual--hop relaying in \cite{MDR_TCOM2013} and \cite{MDR_WCNC2013}, respectively. The main limitation of \cite{MDR_TCOM2013} and \cite{MDR_WCNC2013} is that cell association is not considered and that the distance from serving BS to probe mobile terminal is assumed to be fixed. Finally, we mention that the average rate is computed under the same assumptions as in \cite[Sec. IV]{AndrewsNov2011}, \emph{i.e.}, the interference is treated as noise and the typical mobile terminal uses adaptive modulation/coding such that the Shannon bound, for the operating instantaneous SINR, can be achieved.
\subsection{Biased Long--Term Averaged Tier and BS Association} \label{TierAssociationPolicy}
We assume that the BSs of each tier operate in open access mode for ${\rm{MT}}_0$ \cite{AndrewsApr2012}. As a consequence, ${\rm{MT}}_0$ is allowed to access to any tiers without any restrictions. In a multi--tier cellular networks model, both tier and BS associations have to be properly defined. Similar to \cite[Sec. II--A]{AndrewsOct2012}, throughout this paper we consider a long--term averaged maximum biased--received--power association policy. Let $d_t  = \min \left\{ {d_{t,b} } \right\}$ for $t=1, 2, \ldots, T$ be the distance from ${\rm{MT}}_0$ to the \emph{nearest} BS of the $t$--th tier. Let ${\rm{BS}}_{t}$ for $t=1, 2, \ldots, T$ be the $T$ nearest BSs. Then, ${\rm{MT}}_0$ is associated (tagged) to the tier $t^*$ defined as follows:
\setcounter{equation}{0}
\begin{equation}
\label{Eq_1} t^*  = \mathop {\arg \max }\limits_{t = 1,2, \ldots ,T} \left\{ {P_t d_t^{ - \alpha _t } B_t } \right\}
\end{equation}
\noindent and the tagged (serving) BS is denoted by ${\rm{BS}}_{t^*} = {\rm{BS}}_{0}$.

In other words, ${\rm{MT}}_0$ is connected to the BS that offers the highest average received power to it. Accordingly, the ${\rm{BS}}_0$--to--${\rm{MT}}_0$ link is the useful signal, while all the other BSs in every tier act as interferers. Since the positions of the BSs are random, the ${\rm{BS}}_0$--to--${\rm{MT}}_0$ distance is a random variable as well \cite{AndrewsNov2011}.

The biasing factor, $B_t > 0$ for $t=1, 2, \ldots, T$, modifies the coverage range of each tier for a better offloading strategy. For example, if $B_t > 1$ the coverage range of the $t$--th tier is increased. Throughout this paper, we assume, similar to \cite{AndrewsOct2012}, that all the BSs are fully--loaded (\emph{i.e.}, their queues are full and, thus, they have always data to transmit). The analysis of heterogeneous cellular networks with partially--loaded BSs is postponed to future research, for example either using the conditionally thinning approach proposed in \cite{DhillonApr2012} or the recent results in \cite{AndrewsNov2012} and \cite{AndrewsFeb2013}.
\begin{figure*}[!t]
\setcounter{equation}{5}
\begin{equation}
\label{Eq_6}
\mathcal{\tilde R}_t  = 2\pi \lambda _t \int\nolimits_0^{ + \infty } {\xi \exp \left\{ { - \pi \sum\limits_{q = 1}^T {\left[ {\lambda _q \left( {\frac{{P_q }}{{P_t }}\frac{{B_q }}{{B_t }}} \right)^{\frac{2}{{\alpha _q }}} \xi ^{\frac{{2\alpha _t }}{{\alpha _q }}} } \right]} } \right\}{\mathbb{E}}\left\{ {\ln \left( {1 + \frac{{P_t g_{t,0} \xi ^{ - \alpha _t } }}{{\sigma _N^2  + I_{{\rm{agg}}} \left( \xi  \right)}}} \right)} \right\}d\xi }
\end{equation}
\normalsize \hrulefill \vspace*{0pt}
\end{figure*}
\begin{figure*}[!t]
\setcounter{equation}{7}
\begin{equation}
\label{Eq_8}
\left\{ \begin{split}
 & \mathcal{R} = \int\nolimits_0^{ + \infty } {\left[ {1 - \mathcal{M}_0 \left( {{\rm{SNR}}y} \right)} \right]\frac{{\mathcal{G}_I \left( y \right)}}{y}dy}  \\
 & \mathcal{G}_I \left( y \right) = \frac{1}{{\mathcal{Z}_I \left( {{\rm{SNR}}y} \right)}} - \frac{\alpha }{2}\frac{y}{{\mathcal{Z}_I \left( {{\rm{SNR}}y} \right)}}\int\nolimits_0^{ + \infty } {\xi ^{\frac{\alpha }{2} - 1} \exp \left\{ { - \pi \lambda\mathcal{Z}_I \left( {{\rm{SNR}}y} \right)\xi } \right\}\exp \left\{ { - y\xi ^{\frac{\alpha }{2}} } \right\}d\xi }  \\
 \end{split} \right.
\end{equation}
\normalsize \hrulefill \vspace*{0pt}
\end{figure*}
\subsection{Problem Statement} \label{ProblemStatement}
The main objective of this paper is to compute the average (ergodic) rate of a heterogeneous cellular network, which is modeled as the superposition of $T$ independent PPPs. According to \cite{AndrewsNov2011} and \cite{AndrewsOct2012}, the average rate can be written as follows:
\setcounter{equation}{1}
\begin{equation}
\label{Eq_2} \mathcal{R} = \sum\limits_{t = 1}^T {\mathcal{A}_t \mathcal{R}_t }
\end{equation}
\noindent where: i) $\mathcal{A}_t$ is the probability that ${\rm{MT}}_0$ is associated to the $t$--th tier; and ii) $\mathcal{R}_t$ is the average rate of ${\rm{MT}}_0$ conditioned on its association to the $t$--th tier. For the tier association policy introduced in Section \ref{TierAssociationPolicy}, $\mathcal{A}_t$ is available in \cite[Lemma 1]{AndrewsOct2012}. On the other hand, $\mathcal{R}_t$ is defined as follows \cite[Sec. IV]{AndrewsNov2011}, \cite[Eq. (46)]{AndrewsOct2012}:
\setcounter{equation}{2}
\begin{equation}
\label{Eq_3}
\mathcal{R}_t  = \int\nolimits_0^{ + \infty } {\mathcal{R}_t \left( \xi  \right) f_{d_{t,0} } \left( \xi  \right)d\xi }
\end{equation}
\noindent where: i) $d_{t,0}$ is the distance of ${\rm{MT}}_0$ from its serving BS by conditioning on ${\rm{MT}}_0$ being tagged to the $t$--th tier; ii) ${f_{d_{t,0} } \left( \cdot  \right)}$ is the PDF of the random distance $d_{t,0}$, which is given in \cite[Lemma 3]{AndrewsOct2012}:
\setcounter{equation}{3}
\begin{equation}
\label{Eq_4}
f_{d_{t,0} } \left( \xi  \right) = \frac{{2\pi \lambda _t }}{{\mathcal{A}_t }}\xi \exp \left\{ { - \pi \sum\limits_{q = 1}^T {\left[ {\lambda _q \left( {\frac{{P_q }}{{P_t }}\frac{{B_q }}{{B_t }}} \right)^{\frac{2}{{\alpha _q }}} \xi ^{\frac{{2\alpha _t }}{{\alpha _q }}} } \right]} } \right\}
\end{equation}
\noindent and iii) ${\mathcal{R}_t \left( \xi  \right)}$ is the average rate of ${\rm{MT}}_0$ conditioned on this terminal being tagged to the $t$--th tier and on $d_{t,0}$ being equal to $d_{t,0} = \xi$. From \cite[Sec. IV]{AndrewsNov2011} and \cite[Eq. (14)]{AndrewsOct2012}, ${\mathcal{R}_t \left( \cdot  \right)}$ can be written as follows:
\setcounter{equation}{4}
\begin{equation}
\label{Eq_5}
\left\{ \begin{array}{l}
\begin{split} \mathcal{R}_t \left( \xi  \right) &= {\mathbb{E}}\left\{ {\ln \left( {1 + {\rm{SINR}}_t \left( \xi  \right)} \right)} \right\} \\ & = {\mathbb{E}}\left\{ {\ln \left( {1 + \frac{{P_t g_{t,0} \xi ^{ - \alpha _t } }}{{\sigma _N^2  + I_{{\rm{agg}}} \left( \xi  \right)}}} \right)} \right\} \end{split} \\
I_{{\rm{agg}}} \left( \xi  \right) = \sum\limits_{q = 1}^T {\sum\limits_{b \in \Phi _q \left\{ {\backslash {\rm{BS}}_{t,0} \left( \xi  \right)} \right\}} {\left( {P_q g_{q,b} d_{q,b}^{ - \alpha _q } } \right)} } \\
 \end{array} \right.
\end{equation}
\noindent where: i) ${\sigma _N^2 }$ is the noise power; ii) ${{\rm{BS}}_{t,0} \left( \xi  \right)}$ is the serving BS at distance $d_{t,0} = \xi$ and $g_{t,0}$ is the ${{\rm{BS}}_{t,0} }$--to--${\rm{MT}}_0$ power channel gain; and iii) $I_{{\rm{agg}}} \left( \xi  \right)$ is the aggregate interference conditioned on $d_{t,0} = \xi$, which is generated by all BSs except the serving BS\footnote{Throughout this paper, the serving BS is denoted by  ${{\rm{BS}}_{t,0} \left( \xi  \right)}$ when used in equations, and by ${{\rm{BS}}_{t,0} }$ when used in the text.} ${{\rm{BS}}_{t,0} }$. From (\ref{Eq_2})--(\ref{Eq_5}), the average rate reduces to $\mathcal{R} = \sum\nolimits_{t = 1}^T {\mathcal{\tilde R}_t }$ with $\mathcal{\tilde R}_t$ given in (\ref{Eq_6}) at the top of this page.

The main objective of the next sections is to introduce a new MGF--based approach to efficiently compute $\mathcal{\tilde R}_t$ in (\ref{Eq_6}) for arbitrary fading channels. The main contribution is to avoid the computational complexity of the state--of--the--art Pcov--based approach \cite{BaccelliSep2009}, \cite{AndrewsNov2011}, \cite{AndrewsOct2012}. To this end, we introduce the simplified notation as follows, which originates from the assumption of identically distributed fading in each tier: i) $f_{t,0} \left(  \cdot  \right)$ and $\mathcal{M}_{t,0} \left(  \cdot  \right)$ are PDF and MGF of ${g_{t,0} }$ in (\ref{Eq_6}), respectively; ii) $f_{t,I} \left(  \cdot  \right)$ and $\mathcal{M}_{t,I} \left(  \cdot  \right)$ are PDF and MGF of ${g_{t,b} }$ in (\ref{Eq_5}), respectively; iii) $\mathcal{M}_{I_{{\rm{agg}}}} \left(  \cdot ; \xi \right)$ is the MGF of $I_{{\rm{agg}}} \left( \xi  \right)$ in (\ref{Eq_5}); and iv) $\mathcal{M}_{q,I_{{\rm{agg}}} } \left( { \cdot ;\xi } \right)$ is the MGF of ${I}_{q,{\rm{agg}}} \left( \xi  \right) = \sum\nolimits_{b \in \Phi _q \left\{ {\backslash {\rm{BS}}_{t,0} \left( \xi  \right)} \right\}} {\left( {P_q g_{q,b} d_{q,b}^{ - \alpha _q } } \right)}$, \emph{i.e.}, the per--tier aggregate MGF in (\ref{Eq_5}).

Finally, we mention that the average rate in (\ref{Eq_6}) provides an estimate of the mean data rate over a cell that is achievable by a typical mobile terminal \cite[Sec. IV]{AndrewsNov2011}. This interpretation immediately follows from the validation procedure of the PPP--based abstraction model against conventional grid--based abstraction models, as discussed in detail in \cite[Sec. V--A]{AndrewsNov2011}.
\begin{figure*}[!t]
\setcounter{equation}{10}
\begin{equation}
\label{Eq_10}
\left\{ \begin{array}{l}
 \left. {\mathcal{G}_I \left( y \right)} \right|_{\alpha  = 4} \mathop  = \limits^{\left( a \right)} \frac{1}{4}\sqrt {\frac{\pi }{y}} \exp \left\{ {\frac{{\left( {\pi \lambda } \right)^2 \mathcal{Z}_I^2 \left( {{\rm{SNR}}y} \right)}}{{4y}}} \right\}{\rm{erfc}}\left( {\frac{{\left( {\pi \lambda } \right) \mathcal{Z}_I \left( {{\rm{SNR}}y} \right)}}{{2\sqrt y }}} \right) \\
 \left. {\mathcal{G}_I \left( y \right)} \right|_{\alpha  = 6} \mathop  = \limits^{\left( b \right)} \frac{1}{4}\sqrt {\frac{{\left( {\pi \lambda } \right) \mathcal{Z}_I \left( {{\rm{SNR}}y} \right)}}{{27y}}} S_{0,{1 \mathord{\left/
 {\vphantom {1 3}} \right.
 \kern-\nulldelimiterspace} 3}} \left( {2\sqrt {\frac{{\left( {\pi \lambda } \right)^3 \mathcal{Z}_I^3 \left( {{\rm{SNR}}y} \right)}}{{27y}}} } \right) \\
 \left. {\mathcal{G}_I \left( y \right)} \right|_{{\alpha  \mathord{\left/
 {\vphantom {\alpha  2}} \right.
 \kern-\nulldelimiterspace} 2} = {{\alpha _N } \mathord{\left/
 {\vphantom {{\alpha _N } {\alpha _D }}} \right.
 \kern-\nulldelimiterspace} {\alpha _D }}} \mathop  = \limits^{\left( c \right)} \frac{1}{{\mathcal{Z}_I \left( {{\rm{SNR}}y} \right)}} - \frac{\alpha }{2}\frac{y}{{ \left( {\pi \lambda } \right)^{\nu _\alpha   + 1} \mathcal{Z}_I^{\nu _\alpha   + 2} \left( {{\rm{SNR}}y} \right)}}\frac{{\sqrt {\alpha _D } \alpha _N^{\nu _\alpha   + \frac{1}{2}} }}{{\left( {2\pi } \right)^{\frac{{\alpha _N  + \alpha _D }}{2} - 1} }}G_{\alpha _N ,\alpha _D }^{\alpha _D ,\alpha _N } \left( {\left. {\frac{{\alpha _N^{\alpha _N } y^{\alpha _D } }}{{\alpha _D^{\alpha _D } \left( {\pi \lambda } \right)^{\alpha _N } \mathcal{Z}_I^{\alpha _N } \left( {{\rm{SNR}}y} \right)}}} \right|\begin{array}{*{20}c}
   {\Delta \left( {\alpha _N , - \nu _\alpha  } \right)}  \\
   {\Delta \left( {\alpha _D ,0} \right)}  \\
\end{array}} \right) \\
 \end{array} \right.
\end{equation}
\normalsize \hrulefill \vspace*{0pt}
\end{figure*}
\section{Single--Tier Cellular Networks} \label{SingleTier}
To better introduce the proposed MGF--based analytical methodology to compute the average rate in (\ref{Eq_6}), we start by considering the single--tier reference scenario with $T=1$. In this case, (\ref{Eq_6}) simplifies as follows:
\setcounter{equation}{6}
\begin{equation}
\label{Eq_7}
\begin{split} \mathcal{R} &= 2\pi \lambda \\ &\times \int\nolimits_0^{ + \infty } {\xi \exp \left\{ { - \pi \lambda \xi ^2 } \right\}{\mathbb{E}}\left\{ {\ln \left( {1 + \frac{{Pg_0 \xi ^{ - \alpha } }}{{\sigma _N^2  + I_{{\rm{agg}}} \left( \xi  \right)}}} \right)} \right\}d\xi } \end{split}
\end{equation}
\noindent with $I_{{\rm{agg}}} \left( \xi  \right) = \sum\nolimits_{b \in \Phi \left\{ {\backslash {\rm{BS}}_0 \left( \xi  \right)} \right\}} {\left( {Pg_b d_b^{ - \alpha } } \right)}$. Since $T=1$, for ease of notation, in (\ref{Eq_7}) the subscript $t$ that denotes the tier is dropped. Likewise, the subscript $t$ is dropped in $f_{0} \left(  \cdot  \right)$, $\mathcal{M}_{0} \left(  \cdot  \right)$, $f_{I} \left(  \cdot  \right)$, and $\mathcal{M}_{I} \left(  \cdot  \right)$ as well.

By using the MGF--based approach, an integral closed--form expression of (\ref{Eq_7}) is given in \emph{Theorem} \ref{Theo_1}.
\begin{theorem} \label{Theo_1}
Let ${\rm{SNR}} = {P \mathord{\left/ {\vphantom {P {\sigma _N^2 }}} \right. \kern-\nulldelimiterspace} {\sigma _N^2 }}$ be the Signal--to--Noise--Ratio (SNR), then the average rate, $\mathcal{R}$, of a single--tier cellular network over generalized fading channels is given in (\ref{Eq_8}) at the top of this page, where:
\setcounter{equation}{8}
\begin{equation}
\label{Eq_9}
\left\{ \begin{array}{l}
 \mathcal{Z}_I \left( y \right) = \mathcal{M}_I \left( y \right) + \mathcal{T}_I \left( y \right) \\
 \mathcal{T}_I \left( y \right) = \Gamma \left( {1 - \frac{2}{\alpha }} \right)\sum\limits_{k = 0}^{ + \infty } {y^{k + 1} \mathcal{M}_I^{\left( k \right)} \left( y \right)\left[ {\Gamma \left( {2 - \frac{2}{\alpha } + k} \right)} \right]^{ - 1} }  \\
 \mathcal{M}_I^{\left( k \right)} \left( y \right) = {\mathbb{E}}\left\{ {g_b^{k + 1} \exp \left\{ { - yg_b } \right\}} \right\} \\
 \end{array} \right.
\end{equation}

\smallskip \emph{Proof}: See Appendix \ref{APP__Theo_1}. \hfill $\Box$
\end{theorem}

The framework in (\ref{Eq_8}) and (\ref{Eq_9}) is called MGF--based approach because $\mathcal{R}$ can be directly computed from the MGFs of useful and interference links. In fact, $\mathcal{M}_I^{\left( k \right)} \left( \cdot \right)$ can be obtained from the $(k+1)$--th derivative of $\mathcal{M}_I \left( \cdot \right)$, \emph{i.e.}, $\mathcal{M}_I^{\left( k \right)} \left( y \right) = \left( {{{ - d} \mathord{\left/ {\vphantom {{ - d} {dy}}} \right. \kern-\nulldelimiterspace} {dy}}} \right)^{k + 1} \mathcal{M}_I \left( y \right)$ \cite[Eq. (1.1.2.9)]{Prudnikov_Vol4}. In the sequel, we show that $\mathcal{M}_I^{\left( k \right)} \left( \cdot \right)$ can be explicitly computed in closed--form for many fading channel models. Furthermore, closed--form expressions of $\mathcal{M}_0 \left( \cdot \right)$ and $\mathcal{M}_I \left( \cdot \right)$ are available in \cite[Sec. 2.2]{SimonBook}, \cite[Tables II--IV]{AlouiniFading_A}, and \cite[Tables II--V]{AlouiniFading_B} for many fading channel models. Compared with the Pcov--based approach in, \emph{e.g.}, \cite{BaccelliSep2009}, \cite{AndrewsNov2011}, and \cite{AndrewsOct2012}, the framework in (\ref{Eq_8}) reduces the number of fold integrals to be computed from four to two.

By carefully looking at (\ref{Eq_9}), some important conclusions about the system behavior as a function of the BSs density, $\lambda$, can be drawn, as summarized in \emph{Remark} \ref{Remark_0}.
\begin{remark} \label{Remark_0}
Since the integrand function of $\mathcal{G}_I \left( \cdot \right)$ in (\ref{Eq_8}) is always greater than zero, it follows that $\mathcal{R}$ is a monotonically increasing function of $\lambda$. Furthermore, $\mathcal{R}$ is upper--bounded as follows:
\setcounter{equation}{9}
\begin{equation}
\label{Eq_9A}
\mathcal{R} \le \mathop {\lim }\limits_{\lambda  \to  + \infty } \mathcal{R}\left( \lambda  \right) = \mathcal{R}^{\left( {\lambda _\infty  } \right)}  = \int\nolimits_0^{ + \infty } {\frac{{1 - \mathcal{M}_0 \left( z \right)}}{{\mathcal{M}_I \left( z \right) + \mathcal{T}_I \left( z \right)}}\frac{{dz}}{z}}
\end{equation}

The analytical derivation of (\ref{Eq_9A}) is available in Appendix \ref{APP__Remark_0}. From (\ref{Eq_9A}), we observe that: i) $\mathcal{R}^{\left( {\lambda _\infty  } \right)}$ is independent of the ${\rm{SNR}} = {P \mathord{\left/ {\vphantom {P {\sigma _N^2 }}} \right. \kern-\nulldelimiterspace} {\sigma _N^2 }}$. Thus, for very dense BSs deployments increasing the transmit--power does not help in increasing the average rate; and ii) the existence of a finite upper--bound for increasing $\lambda$ confirms that the deployment of many BSs is not sufficient to achieve very high data rates but more advanced interference management techniques seem to be needed. \hfill $\Box$
\end{remark}

In the remainder of this section, we show that the two--fold integral in (\ref{Eq_8}) can often be reduced to a single integral, since closed--form expressions of $\mathcal{G}_I \left( \cdot \right)$ exist for many path--loss exponents $\alpha$. Also, we show that the infinite series in (\ref{Eq_9}) can be calculated for common fading distributions of the interference channels.

Let us consider the computation of $\mathcal{G}_I \left( \cdot \right)$ as a function of the path--loss exponent $\alpha$. The main result is summarized in \emph{Corollary} \ref{Cor_1}.
\begin{corollary} \label{Cor_1}
Let $\alpha=4$, $\alpha=6$, and ${\alpha  \mathord{\left/ {\vphantom {\alpha  2}} \right. \kern-\nulldelimiterspace} 2} = {{\alpha _N } \mathord{\left/ {\vphantom {{\alpha _N } {\alpha _D }}} \right. \kern-\nulldelimiterspace} {\alpha _D }}$ with $\alpha_N$ and $\alpha_D$ being two positive integer numbers, then $\mathcal{G}_I \left( \cdot \right)$ in (\ref{Eq_8}) has closed--form expression shown in (\ref{Eq_10}) at the top of this page, where $\nu _\alpha   = {\alpha  \mathord{\left/ {\vphantom {\alpha  2}} \right. \kern-\nulldelimiterspace} 2} - 1$.

\smallskip \emph{Proof}: Equation (\ref{Eq_10}) follows from some notable integrals. More specifically: (a) from \cite[Eq. (2.2.1.8)]{Prudnikov_Vol4}; (b) from \cite[Eq. (2.2.1.13), Eq. (2.2.1.14)]{Prudnikov_Vol4}; (c) from \cite[Eq. (2.2.1.22)]{Prudnikov_Vol4}. This concludes the proof. \hfill $\Box$
\end{corollary}

Since the case study ${\alpha  \mathord{\left/ {\vphantom {\alpha  2}} \right. \kern-\nulldelimiterspace} 2} = {{\alpha _N } \mathord{\left/ {\vphantom {{\alpha _N } {\alpha _D }}} \right. \kern-\nulldelimiterspace} {\alpha _D }}$ encompasses many scenarios of practical interest, when referring to \emph{Corollary} \ref{Cor_1}, we will implicitly assume the closed--form expression of $\mathcal{G}_I \left( \cdot \right)$ using the Meijer G--function.

The single--integral expression in \emph{Theorem} \ref{Theo_1} can be efficiently computed by using the Gauss--Chebyshev quadrature rule, as summarized in \emph{Remark} \ref{Remark_1} as follows.
\begin{remark} \label{Remark_1}
By using Gauss--Chebyshev integration, $\mathcal{R}$ in \emph{Theorem} \ref{Theo_1} can be computed as \cite[Eq. (25.4.39)]{AbramowitzStegun}:
\setcounter{equation}{11}
\begin{equation}
\label{Eq_11}
\mathcal{R} \approx \sum\limits_{n = 1}^{N_{{\rm{GCQ}}} } {\frac{{w_n }}{{s_n }}\left[ {1 - \mathcal{M}_0 \left( {{\rm{SNR}}s_n } \right)} \right]\mathcal{G}_I \left( {s_n } \right)}
\end{equation}
\noindent where $w_n$ and $s_n$ for $n=1,2,\ldots, N_{{\rm{GCQ}}}$ are weights and abscissas, respectively, of the quadrature rule \cite[Eq. (22) and Eq. (23)]{Yilmaz}:
\setcounter{equation}{12}
\begin{equation}
\label{Eq_12}
\left\{ \begin{split} & w_n  = \frac{{\pi ^2 \sin \left( {\frac{{2n - 1}}{{2N_{{\rm{GCQ}}} }}\pi } \right)}}{{4N_{{\rm{GCQ}}} \cos ^2 \left[ {\frac{\pi }{4}\cos \left( {\frac{{2n - 1}}{{2N_{{\rm{GCQ}}} }}\pi } \right) + \frac{\pi }{4}} \right]}} \\  & s_n  = \tan \left[ {\frac{\pi }{4}\cos \left( {\frac{{2n - 1}}{{2N_{{\rm{GCQ}}} }}\pi } \right) + \frac{\pi }{4}} \right] \end{split} \right. \end{equation} \hfill $\Box$
\end{remark}
\begin{figure*}[!t]
\setcounter{equation}{16}
\begin{equation}
\label{Eq_16}
\left\{ \begin{split}
 & \mathcal{T}_I \left( y \right) \approx \left( {1 + K} \right)\exp \left\{ { - K} \right\}\left( {1 - \frac{2}{\alpha }} \right)^{ - 1} y\frac{1}{{\sqrt \pi  }}\sum\limits_{n = 1}^{N_{{\rm{GHQ}}} } {\tilde w_n \tilde \omega _n \left[ {y + \left( {1 + K} \right)\tilde \omega _n } \right]^{ - 2} \mathcal{T}_I^{\left( n \right)} \left( y \right)}  \\
 & \mathcal{T}_I^{\left( n \right)} \left( y \right) = \sum\limits_{l = 0}^{ + \infty } {\frac{{l + 1}}{{\left( {l!} \right)}}\left[ {\frac{{K\left( {1 + K} \right)}}{{1 + K + \left( {{y \mathord{\left/
 {\vphantom {y {\tilde \omega _n }}} \right.
 \kern-\nulldelimiterspace} {\tilde \omega _n }}} \right)}}} \right]^l {}_2F_1 \left( {l + 2,1,2 - \frac{2}{\alpha },y\left[ {y + \left( {1 + K} \right)\tilde \omega _n } \right]^{ - 1} } \right)}  \\
 \end{split} \right.
\end{equation}
\normalsize \hrulefill \vspace*{0pt}
\end{figure*}
\subsection{Computation of $\mathcal{T}_I \left( \cdot \right)$ in (\ref{Eq_9}) for General Fading Channels} \label{Computation_TI}
\emph{Theorem} \ref{Theo_1} and \emph{Corollary} \ref{Cor_1} need the computation of $\mathcal{T}_I \left( \cdot \right)$, which depends on the fading distribution of the interference channels. As mentioned in \emph{Theorem} \ref{Theo_1}, $\mathcal{T}_I \left( \cdot \right)$ can, in principle, be computed from the derivatives of $\mathcal{M}_{I} \left(  \cdot  \right)$. However, closed--form expressions can be obtained for many fading channel models by also avoiding the computation of the infinite series in (\ref{Eq_9}). Some key case studies are analyzed in \emph{Propositions} \ref{Prop_1}--\ref{Prop_4} for Nakagami--\emph{m}, Log--Normal, composite Nakagami--\emph{m} and Log--Normal, and composite Rice and Log--Normal fading, respectively.
\begin{proposition} \label{Prop_1}
Let the interference links experience Nakagami--\emph{m} fading. Accordingly, $g_b$ follows a Gamma distribution with parameters $\left( {m,\Omega } \right)$, which we denote as $g_b  \sim {\rm{Gamma}}\left( {m,\Omega } \right)$ \cite[Sec. 2.2.1.4]{SimonBook}. Then, $\mathcal{T}_I \left( \cdot \right)$ in (\ref{Eq_9}) has closed--form expression as follows:
\setcounter{equation}{13}
\begin{equation}
\label{Eq_13}
\begin{split} \mathcal{T}_I \left( y \right) & = m\left( {\frac{m}{\Omega }} \right)^m \left( {1 - \frac{2}{\alpha }} \right)^{ - 1} y\left( {y + \frac{m}{\Omega }} \right)^{ - \left( {m + 1} \right)} \\ &\times {}_2F_1 \left( {m + 1,1,2 - \frac{2}{\alpha },y\left( {y + \frac{m}{\Omega }} \right)^{ - 1} } \right) \end{split}
\end{equation}

\smallskip \emph{Proof}: See Appendix \ref{APP__Prop_1}. \hfill $\Box$
\end{proposition}
\begin{proposition} \label{Prop_2}
Let the interference links experience Log--Normal fading. Accordingly, $g_b$ follows a Log--Normal distribution with parameters (in dB) $\left( {\mu,\sigma^2 } \right)$, which we denote as $g_b  \sim {\rm{LogN}}\left( {\mu,\sigma^2 } \right)$ \cite[Sec. 2.2.2]{SimonBook}. Then, $\mathcal{T}_I \left( \cdot \right)$ in (\ref{Eq_9}) has closed--form expression as follows:
\setcounter{equation}{14}
\begin{equation}
\label{Eq_14}
\begin{split} \mathcal{T}_I \left( y \right) & \approx \left( {1 - \frac{2}{\alpha }} \right)^{ - 1} y \frac{1}{{\sqrt \pi  }}\sum\limits_{n = 1}^{N_{{\rm{GHQ}}} } {\tilde w_n 10^{{{\left( {\sqrt 2 \sigma \tilde s_n  + \mu } \right)} \mathord{\left/
 {\vphantom {{\left( {\sqrt 2 \sigma \tilde s_n  + \mu } \right)} {10}}} \right.
 \kern-\nulldelimiterspace} {10}}}    } \\ &\times   \exp \left\{ { - 10^{{{\left( {\sqrt 2 \sigma \tilde s_n  + \mu } \right)} \mathord{\left/
 {\vphantom {{\left( {\sqrt 2 \sigma \tilde s_n  + \mu } \right)} {10}}} \right.
 \kern-\nulldelimiterspace} {10}}} y} \right\} \\ &\times   {}_1F_1 \left( {1,2 - \frac{2}{\alpha },10^{{{\left( {\sqrt 2 \sigma \tilde s_n  + \mu } \right)} \mathord{\left/
 {\vphantom {{\left( {\sqrt 2 \sigma \tilde s_n  + \mu } \right)} {10}}} \right.
 \kern-\nulldelimiterspace} {10}}} y} \right) \end{split}
\end{equation}
\noindent where $\tilde w_n$ and $\tilde s_n$ for $n=1,2,\ldots, N_{{\rm{GHQ}}}$ are weights and abscissas, respectively, of the Gauss--Hermite quadrature rule \cite[Eq. (25.4.46)]{AbramowitzStegun}.

\smallskip \emph{Proof}: See Appendix \ref{APP__Prop_2}. \hfill $\Box$
\end{proposition}
\begin{proposition} \label{Prop_3}
Let the interference links experience composite Nakagami--\emph{m} and Log--Normal fading. Accordingly, $g_b$ follows a Gamma distribution by conditioning on its mean power, which, in turn, follows a Log--Normal distribution. We denote this distribution as $g_b  \sim {\rm{Gamma/LogN}}\left( {m,\mu,\sigma^2 } \right)$ \cite[Sec. 2.2.3.1]{SimonBook}. Then, $\mathcal{T}_I \left( \cdot \right)$ in (\ref{Eq_9}) has closed--form expression as follows:
\setcounter{equation}{15}
\begin{equation}
\label{Eq_15}
\begin{split} \mathcal{T}_I \left( y \right) & \approx m^{m + 1} \left( {1 - \frac{2}{\alpha }} \right)^{ - 1} y\frac{1}{{\sqrt \pi  }} \\ & \times \sum\limits_{n = 1}^{N_{{\rm{GHQ}}} } {\tilde w_n \tilde \omega _n^m \left( {y + m\tilde \omega _n } \right)^{ - \left( {m + 1} \right)} } \\ &\times {}_2F_1 \left( {m + 1,1,2 - \frac{2}{\alpha },y\left( {y + m\tilde \omega _n } \right)^{ - 1} } \right)  \end{split}
\end{equation}
\noindent with $\tilde \omega _n  = 10^{{{ - \left( {\sqrt 2 \sigma \tilde s_n  + \mu } \right)} \mathord{\left/ {\vphantom {{ - \left( {\sqrt 2 \sigma \tilde s_n  + \mu } \right)} {10}}} \right. \kern-\nulldelimiterspace} {10}}}$.

\smallskip \emph{Proof}: See Appendix \ref{APP__Prop_3}. \hfill $\Box$
\end{proposition}
\begin{proposition} \label{Prop_4}
Let the interference links experience composite Rice and Log--Normal fading. Accordingly, $g_b$ follows a non--central Chi--Square distribution by conditioning on its mean power, which, in turn, follows a Log--Normal distribution. We denote this distribution as $g_b  \sim {\rm{ChiSquare/LogN}}\left( {K,\mu,\sigma^2 } \right)$ \cite[Eq. (6)]{Loo}, with $K$ being the Rice factor. If $K \ne 0$, $\mathcal{T}_I \left( \cdot \right)$ in (\ref{Eq_9}) has closed--form expression given in (\ref{Eq_16}) at the top of this page. If $K=0$, the composite Rice and Log--Normal fading reduces to the composite Nakagami--\emph{m} and Log--Normal fading with $m=1$ (Suzuki distribution \cite[Sec. 2.2.3.2]{SimonBook}) and \emph{Proposition} \ref{Prop_3} can be used.

\smallskip \emph{Proof}: See Appendix \ref{APP__Prop_4}. \hfill $\Box$
\end{proposition}

From (\ref{Eq_16}), we observe that, unlike the other fading distributions in \emph{Propositions} \ref{Prop_1}--\ref{Prop_3}, for composite Rice and Log--Normal fading we still need to calculate an infinite series to compute $\mathcal{T}_I \left( \cdot \right)$ in (\ref{Eq_9}). The computation of the series can be avoided as suggested in \emph{Remark} \ref{Remark_2} as follows.
\begin{remark} \label{Remark_2}
Using the mapping between the $m$ parameter of a Nakagami--\emph{m} distribution and the $K$ factor of a Rice distribution \cite[Eq. (2.26)]{SimonBook}, (\ref{Eq_16}) can be approximated by (\ref{Eq_15}) with $m = {{\left( {1 + K} \right)^2 } \mathord{\left/ {\vphantom {{\left( {1 + K} \right)^2 } {\left( {1 + 2K} \right)}}} \right.\kern-\nulldelimiterspace} {\left( {1 + 2K} \right)}}$. \hfill $\Box$
\end{remark}

Finally, we would like to emphasize that the fading distributions studied in \emph{Propositions} \ref{Prop_1}--\ref{Prop_4} are just some selected examples, which have been chosen because they are often used in theoretical analysis. However, our analytical methodology to compute $\mathcal{T}_I \left( \cdot \right)$ in (\ref{Eq_9}) is applicable to arbitrary fading distributions as described in \emph{Remark} \ref{Remark_3}.
\begin{remark} \label{Remark_3}
From (\ref{Eq_9}), we observe that $\mathcal{T}_I \left( \cdot \right)$ needs the computation of $\mathcal{M}_I^{\left( k \right)} \left( y \right) = {\mathbb{E}}\left\{ {g_b^{k + 1} \exp \left\{ { - yg_b } \right\}} \right\} = \int\nolimits_0^{ + \infty } {x^{k + 1} \exp \left\{ { - yx} \right\}f_{g_b } \left( x \right)dx}$. With the exception of the Log--Normal distribution, which is studied in \emph{Proposition} \ref{Prop_2}, from \cite[Sec. 2.2]{SimonBook}, \cite[Tables II--IV]{AlouiniFading_A}, and \cite[Tables II--V]{AlouiniFading_B} we note that two general situations can arise:
\begin{enumerate}
\item $\tilde f\left( {x;y} \right) = \exp \left\{ { - yx} \right\}f_{g_b } \left( x \right) = A\exp \left\{ { - B\left( y \right)x} \right\}$, where $A$ is a constant and $B\left(  \cdot  \right)$ is a function of $y$. In other words, $\tilde f\left( {\cdot;\cdot} \right)$ is still an exponential function in $x$. In this case, closed--form expressions of $\mathcal{T}_I \left( \cdot \right)$ can be obtained by using the same development as in \emph{Proposition} \ref{Prop_1} for Nakagami--\emph{m} fading.
\item $f_{g_b } \left( x \right) = Cx^\upsilon  G_{p,q}^{m,n} \left( {Dx\left| {\begin{array}{*{20}c} {\left( {a_p } \right)}  \\ {\left( {b_q } \right)}  \\ \end{array}} \right.} \right)$, where $C$, $D$, and $\upsilon$ are constants. In other words, the distribution of the power channel gain $g_b$ can be cast in terms of a Meijer G--function. Accordingly, $\mathcal{M}_I^{\left( k \right)} \left( \cdot \right)$ can be computed in closed--form as another Meijer G--function by using the Mellin--Barnes theorem and the notable integral in \cite[Eq. (2.24.3.1)]{Prudnikov_Vol3}. In general, in this case it is not possible to avoid the computation of the infinite series in (\ref{Eq_9}). \hfill $\Box$
\end{enumerate}
\end{remark}
\begin{figure*}[!t]
\setcounter{equation}{17}
\begin{equation}
\label{Eq_17}
\hspace{-0.25cm} \left\{ \begin{array}{l}
 \mathcal{G}_I \left( y \right) \approx \frac{1}{{\mathcal{Z}_I \left( {{\rm{SNR}}y} \right)}} - \frac{\alpha }{2}\frac{y}{{\left( {\pi \lambda } \right)^{\nu _\alpha   + 1} \mathcal{Z}_I^{\nu _\alpha   + 2} \left( {{\rm{SNR}}y} \right)}}\frac{{\sqrt {\alpha _D } \alpha _N^{\nu _\alpha   + \frac{1}{2}} }}{{\left( {2\pi } \right)^{\frac{{\alpha _N  + \alpha _D }}{2} - 1} }}\Upsilon _{\mathcal{H}} \left( {\frac{{\alpha _N^{\alpha _N } y^{\alpha _D } }}{{\alpha _D^{\alpha _D } \left( {\pi \lambda } \right)^{\alpha _N } \mathcal{Z}_I^{\alpha _N } \left( {{\rm{SNR}}y} \right)}}} \right) \\
  \Upsilon _{\mathcal{H}} \left( z \right) = \mathcal{U}\left( z \right) \mathcal{H}\left( {z - \varepsilon } \right) + \mathcal{U}^{\left( {{\rm{asymptote}}} \right)} \left( z \right)\left[ {1 - \mathcal{H}\left( {z - \varepsilon } \right)} \right] \\
 \mathcal{U}\left( z \right) = G_{\alpha _N ,\alpha _D }^{\alpha _D ,\alpha _N } \left( {z\left| {\begin{array}{*{20}c}
   {\Delta \left( {\alpha _N , - \nu _\alpha  } \right)}  \\
   {\Delta \left( {\alpha _D ,0} \right)}  \\
\end{array}} \right.} \right) \\
 \mathcal{U}^{\left( {{\rm{asymptote}}} \right)} \left( z \right) = \mathop {\lim }\limits_{z \to 0^ +  } \mathcal{U}\left( z \right) = \sum\limits_{q = 1}^{\alpha _D } {\left\{ {z^{\Delta _q \left( {\alpha _D ,0} \right)} \prod\limits_{\scriptstyle r = 1 \hfill \atop
  \scriptstyle r \ne q \hfill}^{\alpha _D } {\Gamma \left( {\Delta _r \left( {\alpha _D ,0} \right) - \Delta _q \left( {\alpha _D ,0} \right)} \right)} \prod\limits_{p = 1}^{\alpha _N } {\Gamma \left( {1 + \Delta _q \left( {\alpha _D ,0} \right) - \Delta _p \left( {\alpha _N , - \nu _\alpha  } \right)} \right)} } \right\}}  \\
 \end{array} \right.
\end{equation}
\normalsize \hrulefill \vspace*{0pt}
\end{figure*}
\begin{figure*}[!t]
\setcounter{equation}{18}
\begin{equation}
\label{Eq_18}
\begin{split}
\mathcal{G}_I \left( y \right) & \approx \frac{1}{{\mathcal{Z}_I \left( {{\rm{SNR}}y} \right)}} - \frac{\alpha }{2}\frac{y}{{ \left( {\pi \lambda } \right)^{\nu _\alpha   + 1} \mathcal{Z}_I^{\nu _\alpha   + 2} \left( {{\rm{SNR}}y} \right)}}\frac{{\sqrt {\alpha _D } \alpha _N^{\nu _\alpha   + \frac{1}{2}} }}{{\left( {2\pi } \right)^{\frac{{\alpha _N  + \alpha _D }}{2} - 1} }}G_{\alpha _N ,\alpha _D }^{\alpha _D ,\alpha _N } \left( {\frac{{\alpha _N^{\alpha _N } y^{\alpha _D } }}{{\alpha _D^{\alpha _D } \left( {\pi \lambda } \right)^{\alpha _N } \mathcal{Z}_I^{\alpha _N } \left( {{\rm{SNR}}y} \right)}}\left| {\begin{array}{*{20}c}
   {\Delta \left( {\alpha _N , - \nu _\alpha  } \right)}  \\
   {\Delta \left( {\alpha _D ,0} \right)}  \\
\end{array}} \right.} \right) \\ & \times \mathcal{H}\left( {\frac{{\alpha _N^{\alpha _N } y^{\alpha _D } }}{{\alpha _D^{\alpha _D } \left( {\pi \lambda } \right)^{\alpha _N } \mathcal{Z}_I^{\alpha _N } \left( {{\rm{SNR}}y} \right)}} - \varepsilon } \right)
\end{split}
\end{equation}
\normalsize \hrulefill \vspace*{0pt}
\end{figure*}
\subsection{Efficient Computation of the Meijer G--Function in (\ref{Eq_10})} \label{Computation_MeijerG}
The computation of the average rate in (\ref{Eq_7}) by using \emph{Theorem} \ref{Theo_1} and \emph{Corollary} \ref{Cor_1} needs, in general, the calculation of the Meijer G--function in (\ref{Eq_10}). This special function is commonly used in wireless communications theory, \emph{e.g.}, \cite{AlouiniFading_A}, \cite{AlouiniFading_B}, \cite{DiRenzoSep2009}--\cite{DiRenzoJan2010}, and it is available in several standard mathematical software packages. Thus, in general, its computation can be performed very efficiently. However, in (\ref{Eq_8}) the Meijer G--function must be calculated for all positive real values, and it is known that the numerical complexity and the numerical accuracy of common algorithms to compute the Meijer G--function increases and decreases, respectively, for small values of its argument, \emph{i.e.}, for $y \to 0$ in (\ref{Eq_10}), see, \emph{e.g.}, \cite{MeijerG_A} and \cite{MeijerG_B}. In order to provide a framework that is general and accurate but also simple and stable to compute, \emph{Corollary} \ref{Cor_2} provides a numerically efficient and stable solution to compute $\mathcal{G}_I \left( \cdot \right)$ in (\ref{Eq_10}), which exploits an asymptotic expansion of the Meijer G--function for large values of its argument.
\begin{corollary} \label{Cor_2}
Let ${\alpha  \mathord{\left/ {\vphantom {\alpha  2}} \right. \kern-\nulldelimiterspace} 2} = {{\alpha _N } \mathord{\left/ {\vphantom {{\alpha _N } {\alpha _D }}} \right. \kern-\nulldelimiterspace} {\alpha _D }}$ with $\alpha_N$ and $\alpha_D$ being two positive integer numbers, then $\mathcal{G}_I \left( \cdot \right)$ in (\ref{Eq_8}) can be efficiently computed as shown in (\ref{Eq_17}) at the top of the next page, where $\varepsilon$ is a small positive constant.

\smallskip \emph{Proof}: See Appendix \ref{APP__Cor_2}. \hfill $\Box$
\end{corollary}

The rationale behind \emph{Corollary} \ref{Cor_2} is to avoid the calculation of the Meijer G--function for small values of its argument, and to replace the Meijer G--function with an accurate, simple to compute, and numerically stable expansion formula. In other words, $\mathcal{G}_I \left( \cdot \right)$ is computed by using $\mathcal{U}\left( \cdot \right)$, \emph{i.e.}, the exact formula in \emph{Corollary} \ref{Cor_2}, as long as the argument of the Meijer G--function is no smaller than $\varepsilon$. On the other hand, when this occurs the asymptotical expansion $\mathcal{U}^{\left( {{\rm{asymptote}}} \right)} \left( \cdot \right)$ is used, which is simple and fast to be computed. This ``adaptive'' approach allows us to keep the desired accuracy without increasing the numerical complexity and without incurring in numerical instabilities. The key parameter for the efficient computation of $\mathcal{G}_I \left( \cdot \right)$ in (\ref{Eq_17}) is $\varepsilon$, which depends on the mathematical software package being used to compute the Meijer G--function. In practice, $\varepsilon$ is the smallest value of the argument of the Meijer G--function for which it can be efficiently computed. If $\varepsilon = 0$, \emph{Corollary} \ref{Cor_2} reduces to \emph{Corollary} \ref{Cor_1}.

Finally, we close this section with \emph{Remark} \ref{Remark_4} and \emph{Remark} \ref{Remark_4bis}.
\begin{remark} \label{Remark_4}
From (\ref{Eq_17}), we note that a very computationally efficient framework, which is accurate for sufficiently small values of $\varepsilon$, can be obtained by simply neglecting $\mathcal{U}^{\left( {{\rm{asymptote}}} \right)} \left( \cdot \right)$, as shown in (\ref{Eq_18}) at the top of this page \hfill $\Box$.
\end{remark}
\begin{remark} \label{Remark_4bis}
The integral in $\mathcal{G}_I \left( \cdot \right)$ belongs to the so--called ``Weibull--type'' integrals, since it coincides with the integral to be computed to obtain the MGF of the Weibull distribution \cite[Eq. (2)]{WeibullMGF_1} and \cite[Eq. (2)]{WeibullMGF_2}. Similar to \emph{Corollary} \ref{Cor_1}, it can be computed in terms of Meijer G--function \cite{WeibullMGF_1} and \cite{WeibullMGF_3} or in terms of generalized hypergeometric function \cite{WeibullMGF_2}. Furthermore, various closed--form approximations are available in the literature, such as \cite{WeibullMGF_4}--\cite{WeibullMGF_7} and references therein. The interested reader can consult these papers and the references therein to identify alternative ways of computing $\mathcal{G}_I \left( \cdot \right)$ that avoid special functions, such as the Meijer G--function. On the other hand, to the best of the authors knowledge, the approach proposed in \emph{Corollary} \ref{Cor_2} is not available in the literature. \hfill $\Box$.
\end{remark}
\subsection{Interference--Limited Scenario} \label{InterferenceLimited_1Tier}
In many practical situations of interest, the background noise is often negligible compared to the aggregate interference \cite{AndrewsNov2011} and \cite{AndrewsOct2012}. In this case, \emph{Theorem} \ref{Theo_1} simplifies as shown in \emph{Corollary} \ref{Cor_3}.
\begin{corollary} \label{Cor_3}
Let $\sigma _N^2  = 0$, then the average rate, $\mathcal{R}$, in (\ref{Eq_7}) simplifies as follows:
\setcounter{equation}{19}
\begin{equation}
\label{Eq_19}
\left. \mathcal{R} \right|_{\sigma _N^2  = 0}  = \mathcal{R}^{\left( {{\rm{SNR}}_\infty  } \right)}  = \int\nolimits_0^{ + \infty } {\frac{{1 - \mathcal{M}_0 \left( z \right)}}{{\mathcal{M}_I \left( z \right) + \mathcal{T}_I \left( z \right)}}\frac{{dz}}{z}}
\end{equation}

\smallskip \emph{Proof}: By using the change of variable ${\rm{SNR}}y = z$ in (\ref{Eq_8}), it follows that $\mathcal{G}_I \left( {{\rm{SNR}}^{ - 1} z} \right) = {1 \mathord{\left/ {\vphantom {1 {Z_I \left( z \right)}}} \right. \kern-\nulldelimiterspace} {\mathcal{Z}_I \left( z \right)}}$ since ${\rm{SNR}} \to \infty$ if $\sigma _N^2  = 0$. This concludes the proof. \hfill $\Box$
\end{corollary}

From (\ref{Eq_19}), interesting considerations about the average rate can be made, as summarized in \emph{Remark} \ref{Remark_5}.
\begin{remark} \label{Remark_5}
Similar to \cite[Eq. (8)]{AndrewsNov2011}, (\ref{Eq_19}) confirms that for interference--limited cellular networks the average rate is independent of the density of BSs as well as of the transmit--power. Thus, increasing either the BSs density or the transmit--power are not effective solutions to increase the average rate. More advanced interference management mechanisms are needed. Furthermore, our framework shows that these two trends hold regardless of the fading channel model, and that they seem to be mainly related to the PPP spatial model of the BSs. Finally, by comparing (\ref{Eq_19}) with (\ref{Eq_9A}) we observe that $\mathcal{R} \le \mathcal{R}^{\left( {\lambda _\infty  } \right)}  = \mathcal{R}^{\left( {{\rm{SNR}}_\infty  } \right)}$. This implies that the average rate of a cellular system with unbounded BSs density and finite transmit--power is the same as the average rate of a cellular system with unbounded transmit--power and finite BSs density. \hfill $\Box$
\end{remark}
\subsection{High--SNR Scenario} \label{HighSNR_1Tier}
In \emph{Corollary} \ref{Cor_3}, we have studied the average rate in the absence of background noise. In \emph{Corollary} \ref{Cor_4}, we study the scenario with small but non--zero noise, \emph{i.e.}, the high--SNR setup.
\begin{corollary} \label{Cor_4}
As a function of the SNR, the average rate, $\mathcal{R}$, in (\ref{Eq_7}) is upper-- and lower--bounded as follows:
\setcounter{equation}{20}
\begin{equation}
\label{Eq_20}
\begin{split} \mathcal{R}^{\left( {{\rm{SNR}} \gg 1} \right)}  &= \mathcal{R}^{\left( {{\rm{SNR}}_\infty  } \right)}  - \left( {\pi \lambda } \right)^{ - {\alpha  \mathord{\left/
 {\vphantom {\alpha  2}} \right.
 \kern-\nulldelimiterspace} 2}} \Gamma \left( {1 + \frac{\alpha }{2}} \right)\frac{1}{{{\rm{SNR}}}} \\ &\times \int\nolimits_0^{ + \infty } {\frac{{1 - \mathcal{M}_0 \left( z \right)}}{{\left[ {\mathcal{M}_I \left( z \right) + \mathcal{T}_I \left( z \right)} \right]^{1 + \left( {{\alpha  \mathord{\left/
 {\vphantom {\alpha  2}} \right. \kern-\nulldelimiterspace} 2}} \right)} }}dz} \\ & \le \mathcal{R}\left( {{\rm{SNR}}} \right) \le \mathcal{R}^{\left( {{\rm{SNR}}_\infty  } \right)} \end{split}
\end{equation}

\smallskip \emph{Proof}: Equation (\ref{Eq_20}) immediately follows, with the same analytical steps, from (\ref{APP1A__Eq_3}) in Appendix \ref{APP__Remark_0}, and by using the identity $\left( {{\alpha  \mathord{\left/{\vphantom {\alpha  2}} \right.\kern-\nulldelimiterspace} 2}} \right)\Gamma \left( {{\alpha  \mathord{\left/{\vphantom {\alpha  2}} \right.\kern-\nulldelimiterspace} 2}} \right) = \Gamma \left( {1 + {\alpha  \mathord{\left/{\vphantom {\alpha  2}} \right.\kern-\nulldelimiterspace} 2}} \right)$ in (\ref{APP1A__Eq_4}). This concludes the proof. \hfill $\Box$
\end{corollary}

From (\ref{Eq_20}), interesting considerations about the average rate can be made, as summarized in \emph{Remark} \ref{Remark_6}.
\begin{remark} \label{Remark_6}
By direct inspection of (\ref{Eq_20}), we observe that: i) the lower--bound, $\mathcal{R}^{\left( {{\rm{SNR}} \gg 1} \right)}$, is the high--SNR approximation of the average rate since $\mathop {\lim }\nolimits_{{\rm{SNR}} \to  + \infty } \mathcal{R}^{\left( {{\rm{SNR}} \gg 1} \right)}  = \mathcal{R}^{\left( {{\rm{SNR}}_\infty  } \right)}$; ii) the average rate increases with the SNR by approaching the upper--bound $\mathcal{R}^{\left( {{\rm{SNR}}_\infty  } \right)}$ with linear convergence rate; iii) the average rate increases with the BSs density by approaching the upper--bound $\mathcal{R}^{\left( {{\rm{SNR}}_\infty  } \right)} = \mathcal{R}^{\left( {\lambda _\infty  } \right)}$ with ${\left( {{\alpha  \mathord{\left/ {\vphantom {\alpha  2}} \right. \kern-\nulldelimiterspace} 2}} \right)}$--order convergence rate; and iv) the larger the path--loss exponent, $\alpha$, the faster the convergence speed to $\mathcal{R}^{\left( {{\rm{SNR}}_\infty  } \right)} = \mathcal{R}^{\left( {\lambda _\infty  } \right)}$ as a function of $\lambda$. \hfill $\Box$
\end{remark}
\subsection{Frequency Reuse} \label{FrequencyReuse_1Tier}
In this section, we study the impact of frequency reuse on the average rate. In particular, we consider a cellular network with $F_B \ge 1$ frequency bands. The setup with $F_B=1$ corresponds to the universal frequency reuse case studied in \emph{Theorem} \ref{Theo_1}. Also, similar to \cite{AndrewsNov2011}, we assume that each interfering BS picks at random one of the $F_B$ frequency bands when transmitting. The average rate is given in \emph{Corollary} \ref{Cor_5}.
\begin{corollary} \label{Cor_5}
The average rate, $\mathcal{R}^{\left( {F_B } \right)}$, of a single--tier cellular network with $F_B \ge 1$ available frequency bands and random frequency reuse coincides with $\mathcal{R}$ in (\ref{Eq_8}), (\ref{Eq_9A}), (\ref{Eq_19}), and (\ref{Eq_20}) by replacing ${\mathcal{Z}_I \left( \cdot \right)}$ in (\ref{Eq_9}) with $\mathcal{Z}_I^{\left( {F_B } \right)} \left( z \right) = \left( {F_B  - 1} \right) + \mathcal{M}_I \left( z \right) + \mathcal{T}_I \left( z \right)$, and $\lambda$ (when available) with ${{\lambda ^{\left( {F_B } \right)}  = \lambda } \mathord{\left/ {\vphantom {{\lambda ^{\left( {F_B } \right)}  = \lambda } {F_B }}} \right. \kern-\nulldelimiterspace} {F_B }}$.

\smallskip \emph{Proof}: The proof follows by taking into account that for $F_B \ge 1$: i) the average rate in (\ref{Eq_5}) becomes $\mathcal{R}^{\left( {F_B } \right)}\left( \xi  \right) = \left( {{1 \mathord{\left/ {\vphantom {1 {F_B }}} \right. \kern-\nulldelimiterspace} {F_B }}} \right){\mathbb{E}}\left\{ {\ln \left( {1 + {\rm{SINR}}\left( \xi  \right)} \right)} \right\}$; ii) on average, the interference originates from a PPP with BSs density equal to $\lambda ^{\left( {F_B } \right)}  = {\lambda  \mathord{\left/ {\vphantom {\lambda  {F_B }}} \right. \kern-\nulldelimiterspace} {F_B }}$; and iii) the tier and BS association PDF is independent of $F_B$ \cite[Sec. VI--A]{AndrewsNov2011}. Accordingly, the proof proceeds along the same lines as \emph{Theorem} \ref{Theo_1}. This concludes the proof. \hfill $\Box$
\end{corollary}

From \emph{Corollary} \ref{Cor_5}, interesting considerations about the average rate can be made, as given in \emph{Remark} \ref{Remark_7}.
\begin{remark} \label{Remark_7}
By direct inspection of $\mathcal{R}^{\left( {F_B } \right)}$ in \emph{Corollary} \ref{Cor_5}, it follows that the average rate is maximized for $F_B=1$, \emph{i.e.}, for universal frequency reuse. Also, for densely deployed BSs (\ref{Eq_9A}) and for interference--limited cellular networks (\ref{Eq_19}), the average rate linearly decreases with the number of available frequency bands $F_B$. These trends are in agreement with \cite[Sec. VI--B]{AndrewsNov2011} and hold for general fading channel models. \hfill $\Box$
\end{remark}
\subsection{Correlated Log--Normal Shadowing} \label{Correlated_LogN}
The MGF--based approach introduced so far can be applied to a wide variety of channel conditions, notably composite fading channels that account for Log--Normal shadowing (see \emph{Proposition} \ref{Prop_3} and \emph{Proposition} \ref{Prop_4}). The average rate in \emph{Theorem} \ref{Theo_1} is applicable, however, only to i.i.d. fast--fading and Log--Normal shadowing. It is well--known, on the other hand, that shadowing correlation severely affects the performance of cellular networks \cite{NewportSep2007}--\cite{BaszczyszynJune2012}. In this section, we provide a simple methodology to extend the framework in \emph{Theorem} \ref{Theo_1} to c.i.d. fading channels. The reason of restricting the analysis to equi--correlated fading originates from the stochastic geometry approach for other--cell interference modeling used in the present paper, which is applicable only to identically distributed fading. The methodology used to obtain equi--correlated Log--Normal random variables exploits the Owen and Steck method for the generation of equi--correlated multivariate Normal distributions \cite{OwenSteck1962}.

As an illustrative example, let the generic downlink channel experience composite Nakagami--\emph{m} fast--fading and Log--Normal shadowing, as described in \emph{Proposition} \ref{Prop_3}. The proposed methodology is readily applicable to other fading channel models with correlated Log--Normal shadowing, as well as to multi--tier cellular networks by applying the same methodology to the framework discussed in Section \ref{MultiTier}. More specifically, we assume i.i.d. fast--fading and c.i.d. shadowing. Accordingly, ${\rm{card}}\left\{ \Phi  \right\}$ channel power gains $g_b$ for $b \in \Phi$ with correlation coefficient $\rho$ and parameters $(m,\mu,\sigma^2)$ can be obtained as follows \cite{OwenSteck1962}:
\begin{description}
\item[\emph{Step 1}]: Generate ${\rm{card}}\left\{ \Phi  \right\}$ equi--correlated Normal random variables as $X_b  = \sigma \sqrt \rho  \bar S  + \sigma \sqrt {1 - \rho } S_b  + \mu$ for $b \in \Phi$, where $\bar S$ and $S_b$ for $b \in \Phi$ are a set of i.i.d. Normal random variables with zero mean and unit variance. The ${\rm{card}}\left\{ \Phi  \right\}$ random variables $X_b$ have mean $\mu$ and variance $\sigma^2$, for $b \in \Phi$, regardless of the correlation coefficient $\rho$.
\item[\emph{Step 2}]: Convert the set of ${\rm{card}}\left\{ \Phi  \right\}$ equi--correlated Normal random variables into a set of ${\rm{card}}\left\{ \Phi  \right\}$ equi--correlated Log--Normal random variables as $Y_b  = 10^{{{X_b } \mathord{\left/ {\vphantom {{X_b } {10}}} \right. \kern-\nulldelimiterspace} {10}}}$ for $b \in \Phi$.
\item[\emph{Step 3}]: Generate ${\rm{card}}\left\{ \Phi  \right\}$ independent Gamma random variables $g_b$ with fading severity $m$ and mean value $Y_b$ for $b \in \Phi$.
\end{description}

From the above generation mechanism of c.i.d. composite Nakagami--\emph{m} and Log--Normal fading, it follows that the ${\rm{card}}\left\{ \Phi  \right\}$ random variables $g_b$ are, by conditioning upon the random variable $\bar S$, i.i.d. with parameters $\left( {m,\mu  + \sigma \sqrt \rho  \bar S ,\sigma ^2 \left( {1 - \rho } \right)} \right)$. Accordingly, we propose the following approach to compute the downlink average rate of cellular networks:
\begin{description}
\item[\emph{Step 1}]: The framework in \emph{Theorem} \ref{Theo_1} is applied by conditioning upon the random variable $\bar S$  and by substituting $\mu  \mapsto \mu  + \sigma \sqrt \rho  \bar S$ and $\sigma  \mapsto \sigma \sqrt {1 - \rho }$. The resulting average rate is denoted by $\mathcal{R}\left( {\bar S} \right)$.
\item[\emph{Step 2}]: The conditioning upon the standard Normal random variable $\bar S$ is removed by averaging over its PDF $f_{\bar S } \left( x  \right) = \left( {{1 \mathord{\left/ {\vphantom {1 {\sqrt {2\pi } }}} \right. \kern-\nulldelimiterspace} {\sqrt {2\pi } }}} \right)\exp \left\{ { - {{x ^2 } \mathord{\left/ {\vphantom {{x ^2 } 2}} \right. \kern-\nulldelimiterspace} 2}} \right\}$.
\end{description}

In formulas, the downlink average rate over c.i.d. fading channels can be computed as follows:
\setcounter{equation}{21}
\begin{equation}
\label{Eq_20bis}
 \mathcal{R} = \int\nolimits_{ - \infty }^{ + \infty } {\mathcal{R}\left( x \right)f_{\bar S } \left( x \right)dx}
 \mathop  \approx \limits^{\left( a \right)} \frac{1}{{\sqrt \pi  }} \sum\limits_{\eta = 1}^{N_{{\rm{GHQ}}} } {\tilde w_{\eta} \mathcal{R}\left( {\sqrt 2 \tilde s_{\eta} } \right)}
\end{equation}
\noindent where $(a)$ is obtained by applying Gauss--Hermite quadratures, and ${\mathcal{R}\left( x \right)}$ is the average rate in \emph{Theorem} \ref{Theo_1} with fading parameters $m\left( x \right) = m$, $\mu \left( x \right) = \mu  + \sigma \sqrt \rho  x$, and $\sigma \left( x \right) = \sigma \sqrt {1 - \rho }$.

In summary, the rationale behind the proposed approach to deal with shadowing correlation consists in: i) first, generating a set of correlated Log--Normal random variables that are conditionally independent and, thus, applying the framework for independent shadowing; and ii) then, removing the conditioning via a single numerical integration. Accordingly, shadowing correlation can be taken into account with only a single extra numerical integral, which can be efficiently computed using Gauss--Hermite quadratures as shown in (\ref{Eq_20bis}).
\begin{figure*}[!t]
\setcounter{equation}{22}
\begin{equation}
\label{Eq_20A}
\left\{ \begin{split}
 & \mathcal{R}\mathop  = \limits^{\left( a \right)} 2\pi \lambda \int\nolimits_0^{ + \infty } {\int\nolimits_0^{ + \infty } {r\exp \left\{ { - \pi \lambda r^2 } \right\}\bar T_1 \left( {r,t} \right)drdt} }  \\
 & \bar T_1 \left( {r,t} \right)\mathop  = \limits^{\left( b \right)} \int\nolimits_{ - \infty }^{ + \infty } {\exp \left\{ { - 2\pi \sigma _N^2 js} \right\}\left( {2\pi js} \right)^{ - 1} \left[ {\mathcal{M}_0 \left( { - 2\pi r^{ - \alpha } \left( {e^t  - 1} \right)^{ - 1} js} \right) - 1} \right]\bar T_2 \left( {r,s} \right)ds}  \\
 & \bar T_2 \left( {r,s} \right)\mathop  = \limits^{\left( c \right)} \exp \left\{ {\pi \lambda r^2  - 2\pi \lambda \alpha ^{ - 1} \left( {2\pi js} \right)^{{2 \mathord{\left/
 {\vphantom {2 \alpha }} \right.
 \kern-\nulldelimiterspace} \alpha }} \int\nolimits_0^{ + \infty } {x^{{2 \mathord{\left/
 {\vphantom {2 \alpha }} \right.
 \kern-\nulldelimiterspace} \alpha }} \left[ {\Gamma \left( { - {2 \mathord{\left/
 {\vphantom {2 \alpha }} \right.
 \kern-\nulldelimiterspace} \alpha },2\pi jsr^{ - \alpha } x} \right) - \Gamma \left( { - {2 \mathord{\left/
 {\vphantom {2 \alpha }} \right.
 \kern-\nulldelimiterspace} \alpha }} \right)} \right]f_I \left( x \right)dx} } \right\} \\
 \end{split} \right.
\end{equation}
\normalsize \hrulefill \vspace*{0pt}
\end{figure*}
\subsection{Pcov-- vs. MGF--based Approach: A Comparison} \label{Pcov_vs_MGF}
In Section \ref{Intro_SOTA}, we have stated that both Pcov-- and MGF--based approaches can be applied to general fading distributions. However, Pcov-- and MGF--based approaches need, in general, the computation of a four-- and a two--fold numerical integral, respectively. In both cases, the integrals may involve the computation of special functions, which, however, are efficiently implemented in commercially available software packages. Due to the reduction of the number of fold integrals to be computed, the MGF--based approach is expected to be more computationally efficient. The aim of this section is to better compare strengths, weaknesses, and computational complexity of these two approaches.

To better conduct this comparison, we summarize in (\ref{Eq_20A}), shown at the top of this page, the four--fold integral expression of the average rate that is obtained from the Pcov--based approach. More specifically, (\ref{Eq_20A}) is obtained from \cite{AndrewsNov2011} as follows: (a) originates from \cite[Appendix C]{AndrewsNov2011}; (b) originates from \cite[Appendix B]{AndrewsNov2011}; and (c) originates from \cite[Theorem 4, Eq. (4)]{AndrewsNov2011}. For consistency and ease of comparison, the same notation as for the MGF--based approach is used.

By comparing the MGF--based approach in (\ref{Eq_8}) with the Pcov--based approach in (\ref{Eq_20A}) the following comments can be made:
\begin{itemize}
\item Both approaches may need the computation of some special functions. More specifically, the MGF--based approach involves the computation of hypergeometric functions in $\mathcal{T}_I \left( {\cdot} \right)$, and the Pcov--based approach involves the computation of the incomplete Gamma function in $\bar T_2 \left( {\cdot,\cdot} \right)$.
\item Both approaches may need to use Gauss--Hermite quadratures to compute $f_{0} \left(  \cdot  \right)$, $\mathcal{M}_{0} \left(  \cdot  \right)$, $f_{I} \left(  \cdot  \right)$, and $\mathcal{M}_{I} \left(  \cdot  \right)$ for composite channel models. This need originates from the analytical intractability of Log--Normal shadowing and it is independent of either the Pcov-- or the MGF--based approach being used.
\item By using \emph{Corollary} \ref{Cor_1}, the two--fold integral in (\ref{Eq_8}) may be reduced to a single--integral for some path--loss exponents. Likewise, by using the Mellin--Barnes theorem in \cite[Eq. (2.24.2.1)]{Prudnikov_Vol3} and the Meijer G--function representation of the upper--incomplete Gamma function in \cite[Ea. (8.4.16.2)]{Prudnikov_Vol3}, a closed--form expression of $\bar T_2 \left( {\cdot,\cdot} \right)$ in (\ref{Eq_20A}) may be obtained. As a consequence, the MGF--based approach reduces the number of fold integrals to be computed and avoids the computation of complex integrals.
\item In interference--limited scenarios, the MGF--based approach in \emph{Corollary} \ref{Cor_3} offers a significant reduction of the computational complexity and the average rate can be calculated from the simple single integral in (\ref{Eq_19}). On the other hand, the computational complexity of the Pcov--based approach is not significantly affected in this scenario. In fact, the only simplification in (\ref{Eq_20A}) is $\exp \left\{ { - 2\pi \sigma _N^2 js} \right\} = 1$ in $\bar T_1 \left( {\cdot,\cdot} \right)$, which does not lead to further reduction of the number of fold integrals to be computed.
\item The desired form of the average rate offered by the MGF--based approach leads to simple and intuitive understanding of the performance of cellular networks for a variety of special operating scenarios, such as dense cellular networks deployments (\emph{Remark} \ref{Remark_0}), interference--dominated environments (\emph{Remark} \ref{Remark_5} and \emph{Remark} \ref{Remark_6}), frequency reuse strategies (\emph{Remark} \ref{Remark_7}). On the other hand, little insight can be gained from (\ref{Eq_20A}) for general fading distributions. However, (\ref{Eq_20A}) can be significantly simplified for Rayleigh fading channels and interesting design guidelines can be inferred from it \cite{AndrewsNov2011}.
\end{itemize}

The considerations above originate from the direct inspection of (\ref{Eq_8}) and (\ref{Eq_20A}), and provide a qualitative comparison of the reduction of computational complexity that can be expected by using the MGF-- instead of the Pcov--based approach. To better understand the advantages of the MGF--based approach, we have also conducted some numerical tests with the goal of providing a more quantitative assessment of the computational complexity. The conventional approach that is often used to conduct these tests it to consider a case study for which, with further analytical manipulations, the integral expressions in (\ref{Eq_8}) and (\ref{Eq_20A}) can be simplified or even computed in closed--form, and to compare their accuracy and computational time without applying any mathematical manipulations. Following this line of thought, we have considered Rayleigh fading as a benchmark and have implemented in Mathematica the formulas in (\ref{Eq_8}) and (\ref{Eq_20A}) as they appear in the present paper. In fact, simple closed--form expressions for Rayleigh fading are available in \cite{AndrewsNov2011}. As far as the MGF--based approach is concerned, the outer integral in (\ref{Eq_8}) is computed using (\ref{Eq_11}) with $N_{{\rm{GCQ}}} = 2000$. The high value of $N_{{\rm{GCQ}}}$ is chosen as a worst case setup for the MGF--based approach. Various combinations of path--loss exponents, $\alpha  = \left\{ {2.05,2.2,2.5,3,4,5} \right\}$, and densities of BSs, $\lambda  = \left\{ {10^{ - 6} ,10^{ - 4} ,10^{ - 2} ,10^{ - 1} } \right\}$, have been considered. The chosen path--loss exponents cover typical propagation environments for cellular applications \cite[Table 2.2]{GoldsmithBook}, \cite[Ch. 2, Sec. 5]{StuberBook}, and the chosen densities of BSs cover sparse, normal, and dense cellular deployments \cite{AndrewsNov2011}, \cite{HeathTSP2012}. The tests have been executed in a laptop computer. In all the analyzed scenarios, the MGF--based approach in (\ref{Eq_8}) has been able to provide accurate estimates of the average rate in less than five/six seconds for each ${\rm{SNR}}$ point to be computed. On the other hand, the Pcov--based approach in (\ref{Eq_20A}) has not been able to provide any numerical estimates after five minutes of computation. In interference--limited scenarios, \emph{i.e.}, $\sigma _N^2  = 0$, the computational complexity of the MGF--based approach is further reduced, while the computational complexity of the Pcov--based approach is not affected. These outcomes confirm the advantages of the proposed MGF--based approach for analysis and design of cellular networks.
\begin{figure*}[!t]
\setcounter{equation}{23}
\begin{equation}
\label{Eq_21}
\left\{ \begin{split}
 & \mathcal{\tilde R}_t  = 2\pi \lambda _t \int\nolimits_0^{ + \infty } {\left[ {1 - \mathcal{M}_{t,0} \left( {{\rm{SNR}}_t y} \right)} \right]\frac{{\mathcal{\tilde G}_I^{\left( t \right)} \left( y \right)}}{y}dy}  \\
 & \mathcal{\tilde G}_I^{\left( t \right)} \left( y \right) = \int\nolimits_0^{ + \infty } {\xi \exp \left\{ { - \pi \sum\limits_{q = 1}^T {\lambda _q \mathcal{\tilde Z}_I^{\left( {t,q} \right)} \left( y \right)\xi ^{2\frac{{\alpha _t }}{{\alpha _q }}} } } \right\}\exp \left\{ { - y\xi ^{\alpha _t } } \right\}d\xi }  \\
 \end{split} \right.
\end{equation}
\normalsize \hrulefill \vspace*{0pt}
\end{figure*}
\begin{figure*}[!t]
\setcounter{equation}{26}
\begin{equation}
\label{Eq_24}
\left\{ \begin{split}
 &\mathcal{\tilde R}_t  = \int\nolimits_0^{ + \infty } {\left[ {1 - \mathcal{M}_{t,0} \left( {{\rm{SNR}}_t y} \right)} \right]\frac{{\mathcal{\tilde G}_I^{\left( t \right)} \left( y \right)}}{y}dy}  \\
 &\mathcal{\tilde G}_I^{\left( t \right)} \left( y \right) = \frac{1}{{\mathcal{\tilde Z}_I^{\left( t \right)} \left( y \right)}} - \frac{\alpha }{2}\frac{y}{{\mathcal{\tilde Z}_I^{\left( t \right)} \left( y \right)}}\int\nolimits_0^{ + \infty } {\xi ^{\frac{\alpha }{2} - 1} \exp \left\{ { - \pi \lambda _t \mathcal{\tilde Z}_I^{\left( t \right)} \left( y \right)\xi } \right\}\exp \left\{ { - y\xi ^{\frac{\alpha }{2}} } \right\}d\xi }  \\
 \end{split} \right.
\end{equation}
\normalsize \hrulefill \vspace*{0pt}
\end{figure*}
\section{Multi--Tier Cellular Networks} \label{MultiTier}
In this section, we extend the analytical framework to generic multi--tier cellular networks. The analytical development is, in many ways, similar to Section \ref{SingleTier}. Thus, only the most important analytical details are reported in what follows. The departing point is (\ref{Eq_6}) and the main result is summarized in \emph{Theorem} \ref{Theo_2}.
\begin{theorem} \label{Theo_2}
Let $\mathcal{\tilde R}_t$ for $t=1, 2, \ldots, T$ in (\ref{Eq_6}). Let ${\rm{SNR}}_t = {P_t \mathord{\left/ {\vphantom {P_t {\sigma _N^2 }}} \right. \kern-\nulldelimiterspace} {\sigma _N^2 }}$ be the SNR of the $t$--th tier, an explicit closed--form expression of  $\mathcal{\tilde R}_t$ for arbitrary fading channels is given in (\ref{Eq_21}) at the top of this page, where:
\setcounter{equation}{24}
\begin{equation}
\label{Eq_22}
\begin{split} \mathcal{\tilde Z}_I^{\left( {t,q} \right)} \left( y \right) &= \left( {\frac{{P_q }}{{P_t }}\frac{{B_q }}{{B_t }}} \right)^{{2 \mathord{\left/
 {\vphantom {2 {\alpha _q }}} \right.
 \kern-\nulldelimiterspace} {\alpha _q }}} \mathcal{M}_{q,I} \left( {\left( {\frac{{P_q }}{{P_t }}\frac{{B_q }}{{B_t }}} \right)^{ - 1} {\rm{SNR}}_q y} \right) \\ &+ \left( {\frac{{P_q }}{{P_t }}\frac{{B_q }}{{B_t }}} \right)^{{2 \mathord{\left/
 {\vphantom {2 {\alpha _q }}} \right.
 \kern-\nulldelimiterspace} {\alpha _q }}} \mathcal{T}_{q,I} \left( {\left( {\frac{{P_q }}{{P_t }}\frac{{B_q }}{{B_t }}} \right)^{ - 1} {\rm{SNR}}_q y} \right) \end{split}
\end{equation}
\setcounter{equation}{25}
\begin{equation}
\label{Eq_23}
\begin{split} \mathcal{T}_{q,I} \left( y \right) & = \Gamma \left( {1 - \frac{2}{{\alpha _q }}} \right) \\ &\times \sum\limits_{k = 0}^{ + \infty } {y^{k + 1} \mathcal{M}_{q,I}^{\left( k \right)} \left( y \right)\left[ {\Gamma \left( {2 - \frac{2}{{\alpha _q }} + k} \right)} \right]^{ - 1} } \end{split}
\end{equation}
\noindent and $\mathcal{M}_{q,I}^{\left( k \right)} \left( s \right) = {\mathbb{E}}\left\{ {g_{q,b}^{k + 1} \exp \left\{ { - sg_{q,b} } \right\}} \right\}$.

\smallskip \emph{Proof}: The proof follows by using the same steps as in Appendix \ref{APP__Theo_1} and by taking into account that, thanks to the spatial and channel independence of the PPPs, the MGF of the aggregate interference for the generic $t$--th tier is $\mathcal{M}_{I_{{\rm{agg}}} } \left( {s;\xi } \right) = \prod\nolimits_{q = 1}^T {\mathcal{M}_{q,I_{{\rm{agg}}} } \left( {s;\xi_q } \right)}$ where $\mathcal{M}_{q,I_{{\rm{agg}}} } \left( { \cdot ;\xi_q } \right)$ is given in (\ref{APP1__Eq_4}) and can be computed in closed--form from $\lambda_q$, $\mathcal{M}_{q,I} \left(  \cdot  \right)$, $\mathcal{T}_{q,I} \left(  \cdot  \right)$, and $\xi _q  = \left( {{{P_q } \mathord{\left/ {\vphantom {{P_q } {P_t }}} \right. \kern-\nulldelimiterspace} {P_t }}} \right)^{{1 \mathord{\left/ {\vphantom {1 {\alpha _q }}} \right. \kern-\nulldelimiterspace} {\alpha _q }}} \left( {{{B_q } \mathord{\left/ {\vphantom {{B_q } {B_t }}} \right. \kern-\nulldelimiterspace} {B_t }}} \right)^{{1 \mathord{\left/ {\vphantom {1 {\alpha _q }}} \right. \kern-\nulldelimiterspace} {\alpha _q }}} \xi ^{{{\alpha _t } \mathord{\left/ {\vphantom {{\alpha _t } {\alpha _q }}} \right. \kern-\nulldelimiterspace} {\alpha _q }}}$ \cite[Eq. (42)]{AndrewsOct2012}. This concludes the proof. \hfill $\Box$
\end{theorem}

\emph{Theorem} \ref{Theo_2} provides a very general expression for the average rate of multi--tier cellular networks that, in general, needs the computation of a two--fold numerical integral but is applicable to tiers having different path--loss exponents and fading distributions. Furthermore, closed--form expressions for $\mathcal{T}_{q,I} \left(  \cdot  \right)$ in (\ref{Eq_23}) can be obtained from \emph{Propositions} \ref{Prop_1}--\ref{Prop_4}, similar to the single--tier case. The extension to correlated Log--Normal shadowing follows immediately from Section \ref{Correlated_LogN}.

Even though general, \emph{Theorem} \ref{Theo_2} provides a framework that is less analytically tractable than \emph{Theorem} \ref{Theo_1} and \emph{Corollary} \ref{Cor_1}. A simpler and more insightful analytical framework can be obtained by assuming that all the tiers have the same path--loss exponent, \emph{i.e.}, $\alpha _t  = \alpha$ for $t=1, 2, \ldots, T$ while still keeping the assumption that the per--tier fading distribution is different. The related framework is given in \emph{Corollary} \ref{Cor_6}.
\begin{corollary} \label{Cor_6}
Let $\alpha _t  = \alpha$ for $t=1, 2, \ldots, T$, then $\mathcal{\tilde R}_t$ in (\ref{Eq_6}) can be explicitly computed as shown in (\ref{Eq_24}) at the top of this page, where:
\setcounter{equation}{27}
\begin{equation}
\label{Eq_25}
\left\{ \begin{split} & \mathcal{\tilde Z}_I^{\left( t \right)} \left( y \right) = \sum\limits_{q = 1}^T {\left[ {\frac{{\lambda _q }}{{\lambda _t }} \mathcal{\tilde Z}_I^{\left( {t,q} \right)} \left( y \right)} \right]} \\
& \mathcal{\tilde Z}_I^{\left( {t,q} \right)} \left( y \right) = \left( {\frac{{P_q }}{{P_t }}\frac{{B_q }}{{B_t }}} \right)^{{2 \mathord{\left/
 {\vphantom {2 \alpha }} \right.
 \kern-\nulldelimiterspace} \alpha }} \mathcal{M}_{q,I} \left( {\left( {\frac{{P_q }}{{P_t }}\frac{{B_q }}{{B_t }}} \right)^{ - 1} {\rm{SNR}}_q y} \right) \\
 & \hspace{1.35cm} + \left( {\frac{{P_q }}{{P_t }}\frac{{B_q }}{{B_t }}} \right)^{{2 \mathord{\left/
 {\vphantom {2 \alpha }} \right.
 \kern-\nulldelimiterspace} \alpha }} \mathcal{T}_{q,I} \left( {\left( {\frac{{P_q }}{{P_t }}\frac{{B_q }}{{B_t }}} \right)^{ - 1} {\rm{SNR}}_q y} \right)
 \end{split} \right.
\end{equation}

\smallskip \emph{Proof}: The proof follows directly from (\ref{Eq_21}) and from some algebraic manipulations similar to Appendix \ref{APP__Theo_1}. This concludes the proof. \hfill $\Box$
\end{corollary}
\begin{figure*}[!t]
\setcounter{equation}{29}
\begin{equation} \footnotesize
\label{Eq_26}
\begin{split}
\mathcal{\tilde R}_t  & \le \mathop {\lim }\limits_{\lambda  \to  + \infty } \mathcal{\tilde R}_t \left( \lambda  \right) = \mathcal{\tilde R}_t^{\left( {\lambda _\infty  } \right)} \\ &  = \int\nolimits_0^{ + \infty } {\frac{{1 - \mathcal{M}_{t,0} \left( z \right)}}{{\sum\limits_{q = 1}^T {\left\{ {\left( {\frac{{\kappa _q }}{{F_B^{\left( q \right)} }}\frac{{F_B^{\left( t \right)} }}{{\kappa _t }}} \right)\left( {\frac{{P_q }}{{P_t }}\frac{{B_q }}{{B_t }}} \right)^{{2 \mathord{\left/ {\vphantom {2 \alpha }} \right. \kern-\nulldelimiterspace} \alpha }} \left[ {\left( {F_B^{\left( q \right)}  - 1} \right) + \mathcal{M}_{q,I} \left( {\left( {\frac{{P_q }}{{P_t }}\frac{{B_q }}{{B_t }}} \right)^{ - 1} {\rm{SNR}}_q z} \right) + \mathcal{T}_{q,I} \left( {\left( {\frac{{P_q }}{{P_t }}\frac{{B_q }}{{B_t }}} \right)^{ - 1} {\rm{SNR}}_q z} \right)} \right]} \right\}} }}\frac{{dz}}{z}} \end{split}
\end{equation}
\normalsize \hrulefill \vspace*{0pt}
\end{figure*}
\begin{figure*}[!t]
\setcounter{equation}{30}
\begin{equation} \footnotesize
\label{Eq_27}
\hspace{-0.25cm} \begin{split}
\mathcal{\tilde R}_t^{\left( {{\rm{SNR}} \gg 1} \right)}  & \approx \mathcal{\tilde R}_t^{\left( {{\rm{SNR}}_\infty  } \right)}  - \left( {\pi \frac{{\lambda _t }}{{F_B^{\left( t \right)} }}} \right)^{ - {\alpha  \mathord{\left/
 {\vphantom {\alpha  2}} \right.
 \kern-\nulldelimiterspace} 2}} \Gamma \left( {1 + \frac{\alpha }{2}} \right)\frac{1}{{{\rm{SNR}}_t }} \\ & \times \int\nolimits_0^{ + \infty } {\frac{{1 - \mathcal{M}_{t,0} \left( z \right)}}{{\left( {\sum\limits_{q = 1}^T {\left\{ {\left( {\frac{{\kappa _q }}{{F_B^{\left( q \right)} }}\frac{{F_B^{\left( t \right)} }}{{\kappa _t }}} \right)\left( {\frac{{P_q }}{{P_t }}\frac{{B_q }}{{B_t }}} \right)^{{2 \mathord{\left/
 {\vphantom {2 \alpha }} \right.
 \kern-\nulldelimiterspace} \alpha }} \left[ {\left( {F_B^{\left( q \right)}  - 1} \right) + \mathcal{M}_{q,I} \left( {\left( {\frac{{P_q }}{{P_t }}\frac{{B_q }}{{B_t }}} \right)^{ - 1} {\rm{SNR}}_q z} \right) + \mathcal{T}_{q,I} \left( {\left( {\frac{{P_q }}{{P_t }}\frac{{B_q }}{{B_t }}} \right)^{ - 1} {\rm{SNR}}_q z} \right)} \right]} \right\}} } \right)^{1 + \left( {{\alpha  \mathord{\left/
 {\vphantom {\alpha  2}} \right.
 \kern-\nulldelimiterspace} 2}} \right)} }}dz} \end{split}
\end{equation}
\normalsize \hrulefill \vspace*{0pt}
\end{figure*}
From \emph{Corollary} \ref{Cor_6}, the following important remark can be made.
\begin{remark} \label{Remark_8}
By comparing \emph{Corollary} \ref{Cor_6} and \emph{Corollary} \ref{Cor_1}, we observe that the two formulas have the same structure. More specifically, $\mathcal{\tilde R}_t$ in (\ref{Eq_24}) can be obtained from $\mathcal{R}$ in (\ref{Eq_8}) by simply replacing ${\mathcal{Z}_I \left( \cdot \right)}$ with $\mathcal{\tilde Z}_I^{\left( t \right)} \left( \cdot \right)$. Furthermore, \emph{Theorem} \ref{Theo_1} reduces to \emph{Corollary} \ref{Cor_6}, as expected, for $T=1$. \hfill $\Box$
\end{remark}

\emph{Remark} \ref{Remark_8} allows us to easily generalize many important results obtained for the single--tier setup to the multi--tier case. In particular, \emph{Corollary} \ref{Cor_7} generalizes \emph{Corollary} \ref{Cor_5} for multi--tier cellular networks with random frequency reuse; \emph{Corollary} \ref{Cor_8} generalizes \emph{Remark} \ref{Remark_0} by investigating the impact of dense BSs deployments; and \emph{Corollary} \ref{Cor_9} generalizes \emph{Corollary} \ref{Cor_3} and \emph{Corollary} \ref{Cor_4} by studying interference--limited multi--tier cellular systems and high--SNR operating conditions.
\begin{corollary} \label{Cor_7}
The average rate of the $t$--th tier of a multi--tier cellular network with $\alpha _t  = \alpha$ and with $F_B^{\left( t \right)} \ge 1$ available frequency bands and random frequency reuse for every tier $t=1, 2, \ldots, T$ can be obtained from (\ref{Eq_24}) by replacing $\mathcal{\tilde Z}_I^{\left( {t,q} \right)} \left(  \cdot  \right)$ in (\ref{Eq_25}) with:
\setcounter{equation}{28}
\begin{equation}
\label{Eq_25bis}
\begin{split} \mathcal{\tilde Z}_I^{\left( {t,q} \right)} \left( y \right) & = \left( {\frac{{P_q }}{{P_t }}\frac{{B_q }}{{B_t }}} \right)^{{2 \mathord{\left/
 {\vphantom {2 \alpha }} \right. \kern-\nulldelimiterspace} \alpha }} \left( {F_B^{\left( q \right)}  - 1} \right) \\
 & + \left( {\frac{{P_q }}{{P_t }}\frac{{B_q }}{{B_t }}} \right)^{{2 \mathord{\left/
 {\vphantom {2 \alpha }} \right. \kern-\nulldelimiterspace} \alpha }}  \mathcal{M}_{q,I} \left( {\left( {\frac{{P_q }}{{P_t }}\frac{{B_q }}{{B_t }}} \right)^{ - 1} {\rm{SNR}}_q y} \right) \\
 & + \left( {\frac{{P_q }}{{P_t }}\frac{{B_q }}{{B_t }}} \right)^{{2 \mathord{\left/
 {\vphantom {2 \alpha }} \right. \kern-\nulldelimiterspace} \alpha }}  \mathcal{T}_{q,I} \left( {\left( {\frac{{P_q }}{{P_t }}\frac{{B_q }}{{B_t }}} \right)^{ - 1} {\rm{SNR}}_q y} \right) \end{split}
\end{equation}

\noindent for $t,q=1, 2, \ldots, T$, and ${\lambda _t }$ with $\lambda _t^{\left( {F_B^{\left( t \right)} } \right)}  = \lambda _t / {F_B^{\left( t \right)} }$ for $t=1, 2, \ldots, T$.

\smallskip \emph{Proof}: It follows from \emph{Corollary} \ref{Cor_5} and \emph{Remark} \ref{Remark_8}. This concludes the proof. \hfill $\Box$
\end{corollary}
\begin{corollary} \label{Cor_8}
Let us consider a multi--tier cellular network with $\alpha _t  = \alpha$ and with $F_B^{\left( t \right)} \ge 1$ available frequency bands and random frequency reuse for every tier $t=1, 2, \ldots, T$. Also, let $\lambda _t  = \kappa _t \lambda$ and ${\rm{SNR}}_t  = \chi _t {\rm{SNR}}$ for $t=1, 2, \ldots, T$. If $ \lambda  \to  + \infty$, then the average rate in \emph{Corollary} \ref{Cor_7} is upper--bounded as shown in (\ref{Eq_26}) at the top of this page.

\smallskip \emph{Proof}: The proof follows immediately from \emph{Remark} \ref{Remark_0} and \emph{Remark} \ref{Remark_8}. This concludes the proof. \hfill $\Box$
\end{corollary}
\begin{corollary} \label{Cor_9}
Let us consider a multi--tier cellular network with $\alpha _t  = \alpha$ and with $F_B^{\left( t \right)} \ge 1$ available frequency bands and random frequency reuse for every tier $t=1, 2, \ldots, T$. Also, let $\lambda _t  = \kappa _t \lambda$ and ${\rm{SNR}}_t  = \chi _t {\rm{SNR}}$ for $t=1, 2, \ldots, T$. If $\sigma _N^2  = 0$ (\emph{i.e.}, ${\rm{SNR}} \to  + \infty$), which implies that the system is interference--limited, then the average rate in \emph{Corollary} \ref{Cor_7} is upper--bounded as $\mathcal{\tilde R}_t  \le \mathcal{\tilde R}_t^{\left( {{\rm{SNR}}_\infty  } \right)}  = \mathcal{\tilde R}_t^{\left( {\lambda _\infty  } \right)}$, where $\mathcal{\tilde R}_t^{\left( {\lambda _\infty  } \right)}$ is given in (\ref{Eq_26}). Furthermore, if $\sigma _N^2$ is small but non--zero, \emph{i.e.}, ${\rm{SNR}} \gg 1$, then the average rate in \emph{Corollary} \ref{Cor_7} can be approximated as shown in (\ref{Eq_27}) at the top of this page.

\smallskip \emph{Proof}: The proof follows from \emph{Corollary} \ref{Cor_3}, \emph{Corollary} \ref{Cor_4}, and \emph{Remark} \ref{Remark_8}. This concludes the proof. \hfill $\Box$
\end{corollary}
\section{Numerical and Simulation Results} \label{NumericalResults}
In this section, we show some numerical examples in order to verify the accuracy of the proposed analytical methodology against Monte Carlo simulations, as well as to show the impact of different fading parameters and distributions on the average rate. For a fair comparison among different fading distributions, the mean square value of each fading distribution is normalized and set equal to one. This implies $\Omega = 1$ for Rayleigh and Nakagami--\emph{m} distributions, and $\mu  =  - {{\ln \left( {10} \right)\sigma ^2 } \mathord{\left/ {\vphantom {{\ln \left( {10} \right)\sigma ^2 } {20}}} \right. \kern-\nulldelimiterspace} {20}}$ for Log--Normal, composite Nakagami--\emph{m} and Log--Normal, and composite Rice and Log--Normal distributions. Furthermore, useful and interference links are assumed to have the same fading distribution. The analytical framework is implemented as described in the captions of each figure. As far as the composite Rice and Log--Normal fading model is concerned, both frameworks in \emph{Proposition} \ref{Prop_4} and \emph{Remark} \ref{Remark_2} are implemented. We have verified that both frameworks provide the same accuracy. Thus, the application of \emph{Remark} \ref{Remark_2} is recommended since it is simpler to compute.
\paragraph{Monte Carlo Simulations} As far as Monte Carlo simulations are concerned, we have used the following methodology \cite[Appendix F]{BrownGLOBE2012}.
\begin{description}
  \item[\emph{Step 1}]: A finite circular area of (normalized) radius $R_A$ around the origin, \emph{i.e.}, where the probe mobile terminal is located, is considered. The radius is chosen sufficiently large to minimize the error committed in simulating the infinite bi--dimensional plane. In the considered setup, the radius $R_A$ is chosen such that $\lambda _{\min } R_A^2  \ge 100$, where $\lambda _{\min }  = \min \left\{ {\lambda _1 ,\lambda _2 , \ldots ,\lambda _T } \right\}$. For example, $R_A  = 100$ if $\lambda _{\min }  = 10^{ - 1}$ and $\lambda _{\min }  = 10^{ - 2}$, $R_A  = 1000$ if $\lambda _{\min }  = 10^{ - 4}$, and $R_A  = 10000$ if $\lambda _{\min }  = 10^{ - 6}$.
  \item[\emph{Step 2}]: For each tier, the number of BSs is generated following a Poisson distribution with density $\lambda_t$ and area $\pi R_A^2$.
  \item[\emph{Step 3}]: The BSs of each tier are distributed following a uniform distribution over the circular region of area $\pi R_A^2$.
  \item[\emph{Step 4}]: Independent channel gains are generated for each BS of every tier.
  \item[\emph{Step 5}]: The tier and BS association policy described in Section \ref{TierAssociationPolicy} is applied, and useful and interference links are identified.
  \item[\emph{Step 6}]: Given the associated tier and its tagged BS, the SINR is computed as shown in (\ref{Eq_5}).
  \item[\emph{Step 7}]: The rate of the generic Monte Carlo trial is computed as $\mathcal{R}_{{\rm{mc}}}  = (1/{F_B^{\left( {t^ *  } \right)} })\ln \left( {1 + {\rm{SINR}}_{t^ *  } } \right)$, where $t^*$ is the tagged tier.
  \item[\emph{Step 8}]: Finally, the average rate is computed by repeating \emph{Step 1}--\emph{Step 7} for $N_{{\rm{mc}}}$ times and eventually calculating $\mathcal{R} = \left( {{1 \mathord{\left/ {\vphantom {1 {N_{\rm{mc}} }}} \right. \kern-\nulldelimiterspace} {N_{\rm{mc}} }}} \right)\sum\nolimits_{{\rm{mc}} = 1}^{N_{{\rm{mc}}} } {\mathcal{R}_{{\rm{mc}}} }$. In our simulations, we have considered $N_{{\rm{mc}}}  = 10^6$.
\end{description}

In Section \ref{Pcov_vs_MGF}, we have compared the computational complexity of Pcov-- and MGF--based approaches, and we have shown that the proposed analytical methodology turns out to be more computational efficient for general fading distributions. As far as the computational complexity comparison with Monte Carlo simulations is concerned, our experiments have revealed that each simulation curve shown in this section can be obtained in a computation time of the order of a few minutes (five to ten minutes depending on the setup) by using the MGF--based approach. On the other hand, the same curve can be obtained in tens of hours (ten to sixty depending on the setup) of computation time by using Monte Carlo simulations. In addition to the longer simulation time and to the more resources for the computation, it is important to mention that Monte Carlo simulations tend to be less accurate for: i) sparse cellular networks; ii) low path--loss exponents; and iii) high--SNR. The reason is that in these operating scenarios $R_A$ and $N_{{\rm{mc}}}$ must be increased in order to account for the interfering BSs that are far from the probe mobile terminal, and which, in these cases, can no longer be neglected.

\begin{figure}[!t]
\centering
\includegraphics [width=\columnwidth] {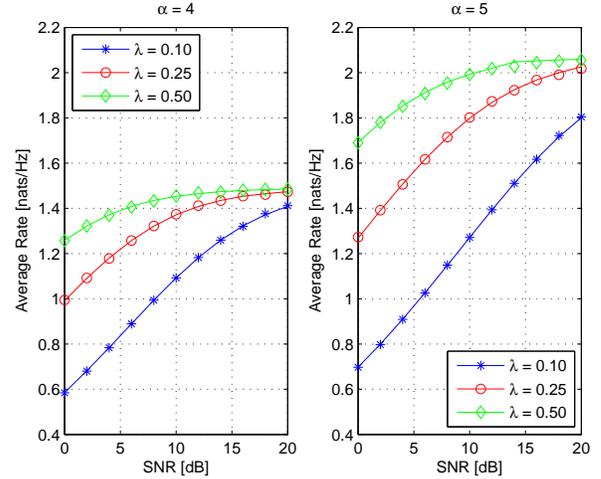}
\vspace{-0.60cm} \caption{Average rate of a single--tier cellular network over Rayleigh fading ($F_B=1$). Markers show Monte Carlo simulations. Solid lines show the analytical framework, which is computed by using \emph{Theorem} \ref{Theo_1}, \emph{Corollary} \ref{Cor_2} with $\varepsilon  = 0.05$, \emph{Proposition} \ref{Prop_1} with ($m=1$, $\Omega = 1$), and \emph{Remark} \ref{Remark_1} with $N_{{\rm{GCQ}}}  = 2000$. Furthermore, $\mathcal{M}_0 \left(  \cdot  \right) = \mathcal{M}_I \left(  \cdot  \right)$ are obtained from \cite[Eq. (2.8)]{SimonBook}.} \label{Fig_Ray}
\end{figure}
\begin{figure}[!t]
\centering
\includegraphics [width=\columnwidth] {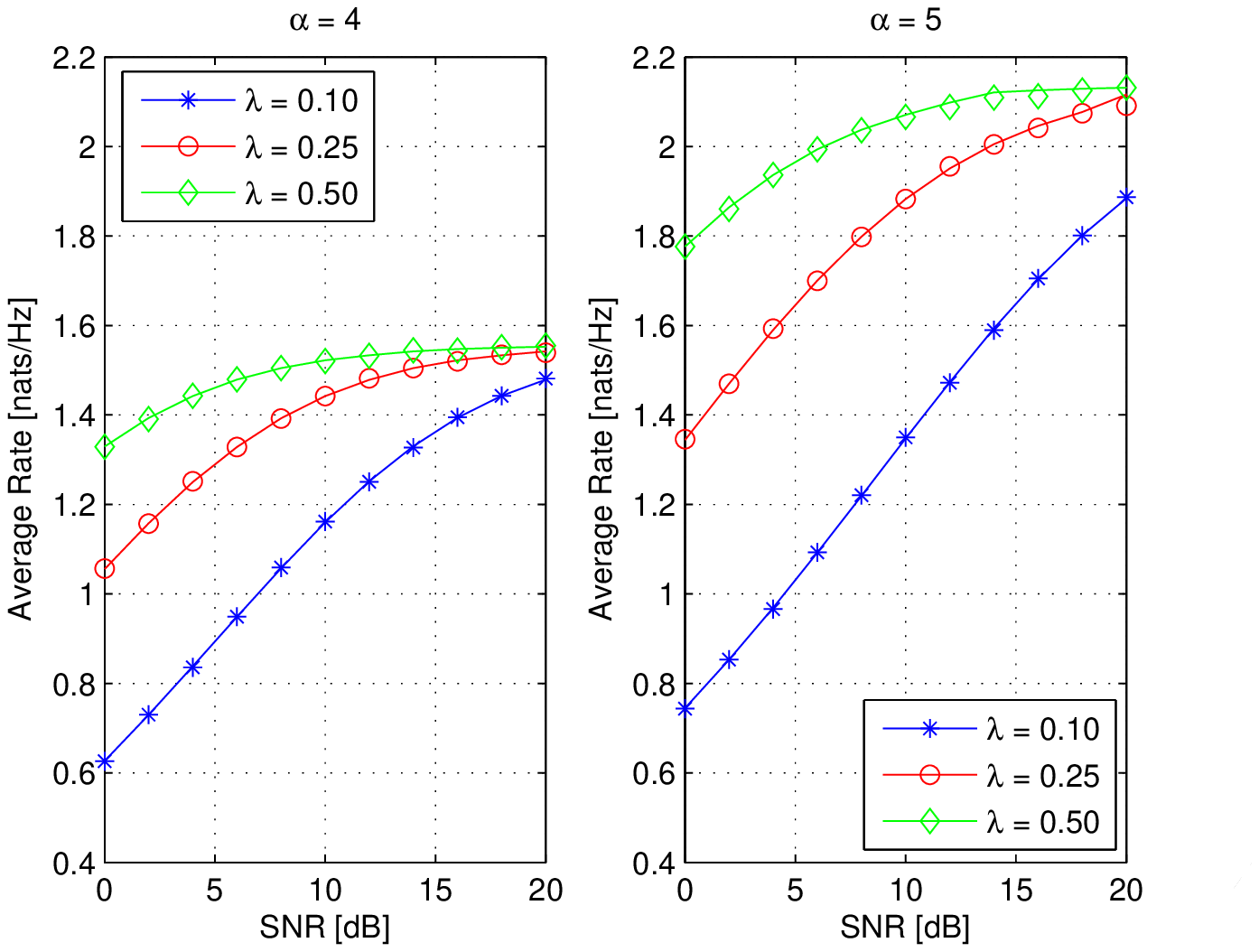}
\vspace{-0.60cm} \caption{Average rate of a single--tier cellular network over Nakagami--\emph{m} fading ($F_B=1$). Markers show Monte Carlo simulations. Solid lines show the analytical framework, which is computed by using \emph{Theorem} \ref{Theo_1}, \emph{Corollary} \ref{Cor_2} with $\varepsilon  = 0.05$, \emph{Proposition} \ref{Prop_1} with ($m=2.5$, $\Omega = 1$), and \emph{Remark} \ref{Remark_1} with $N_{{\rm{GCQ}}}  = 2000$. Furthermore, $\mathcal{M}_0 \left(  \cdot  \right) = \mathcal{M}_I \left(  \cdot  \right)$ are obtained from \cite[Eq. (2.22)]{SimonBook}.} \label{Fig_NakM}
\end{figure}
\begin{figure}[!t]
\centering
\includegraphics [width=\columnwidth] {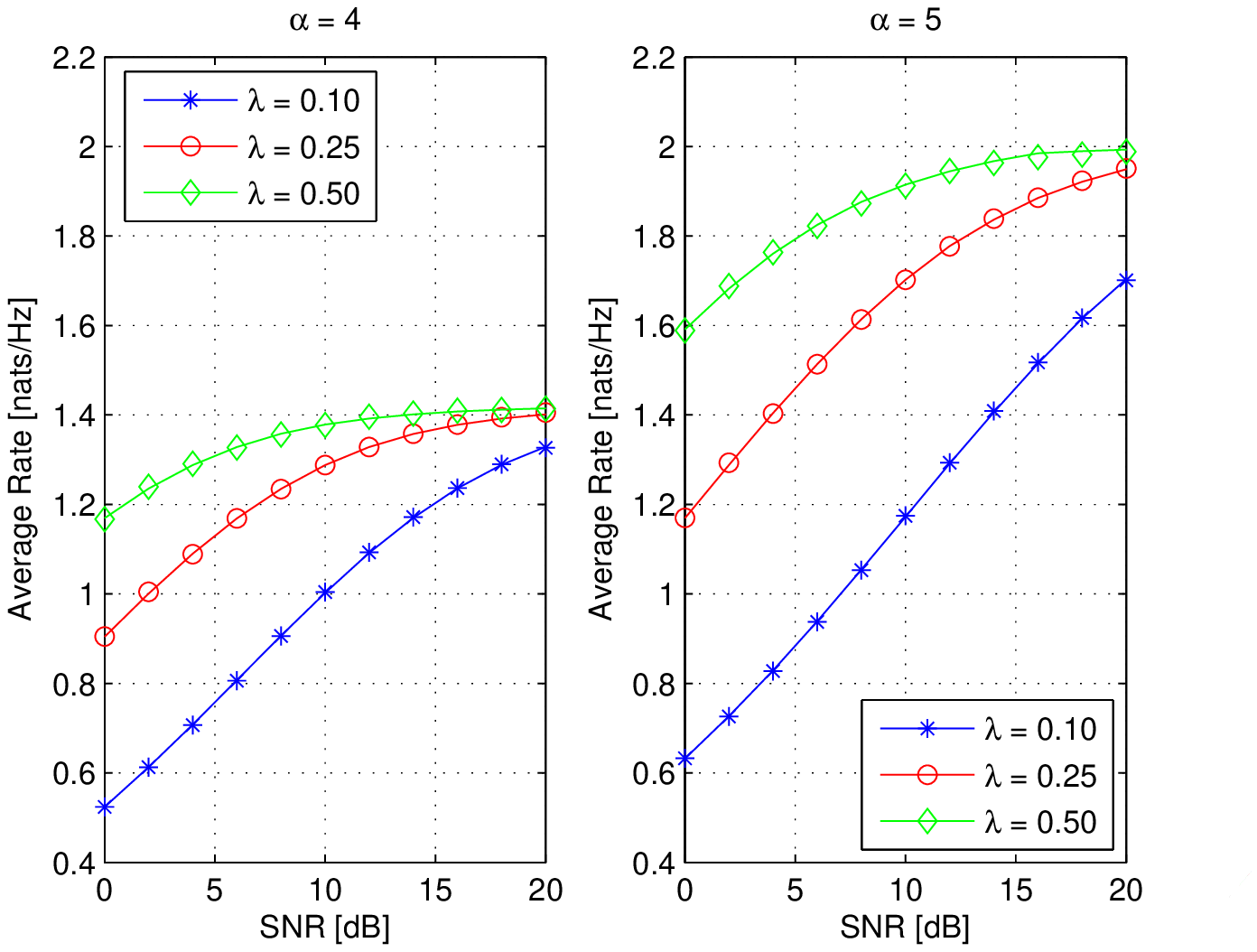}
\vspace{-0.60cm} \caption{Average rate of a single--tier cellular network over Log--Normal fading ($F_B=1$). Markers show Monte Carlo simulations. Solid lines show the analytical framework, which is computed by using \emph{Theorem} \ref{Theo_1}, \emph{Corollary} \ref{Cor_2} with $\varepsilon  = 0.05$, \emph{Proposition} \ref{Prop_2} with ($\sigma=6 \, {\rm{dB}}$, $\mu  =  - \ln \left( {10} \right)\sigma ^2 /20 \, {\rm{dB}}$) and $N_{{\rm{GHQ}}}  = 5$, and \emph{Remark} \ref{Remark_1} with $N_{{\rm{GCQ}}}  = 2000$. Furthermore, $\mathcal{M}_0 \left(  \cdot  \right) = \mathcal{M}_I \left(  \cdot  \right)$ are obtained from \cite[Eq. (2.54)]{SimonBook} with $N_{{\rm{GHQ}}}  = 5$.} \label{Fig_LogN}
\end{figure}
\begin{figure}[!t]
\centering
\includegraphics [width=\columnwidth] {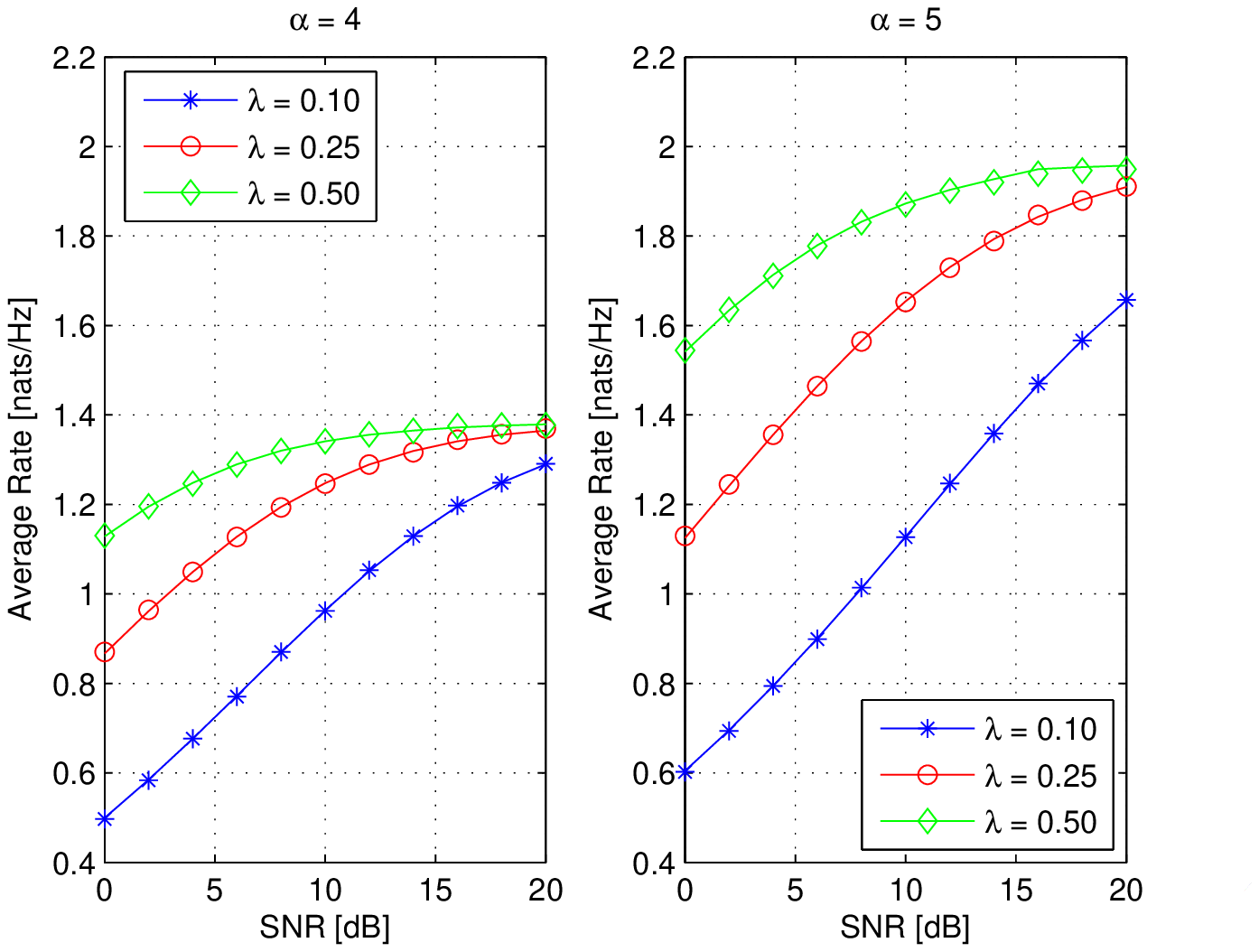}
\vspace{-0.60cm} \caption{Average rate of a single--tier cellular network over composite Nakagami--\emph{m} and Log--Normal fading ($F_B=1$). Markers show Monte Carlo simulations. Solid lines show the analytical framework, which is computed by using \emph{Theorem} \ref{Theo_1}, \emph{Corollary} \ref{Cor_2} with $\varepsilon  = 0.05$, \emph{Proposition} \ref{Prop_3} with ($m=2.5$, $\sigma=6 \, {\rm{dB}}$, $\mu  =  - \ln \left( {10} \right)\sigma ^2 /20 \, {\rm{dB}}$) and $N_{{\rm{GHQ}}}  = 5$, and \emph{Remark} \ref{Remark_1} with $N_{{\rm{GCQ}}}  = 2000$. Furthermore, $\mathcal{M}_0 \left(  \cdot  \right) = \mathcal{M}_I \left(  \cdot  \right)$ are obtained from \cite[Eq. (2.58)]{SimonBook} with $N_{{\rm{GHQ}}}  = 5$.} \label{Fig_NakMLogN}
\end{figure}
\begin{figure}[!t]
\centering
\includegraphics [width=\columnwidth] {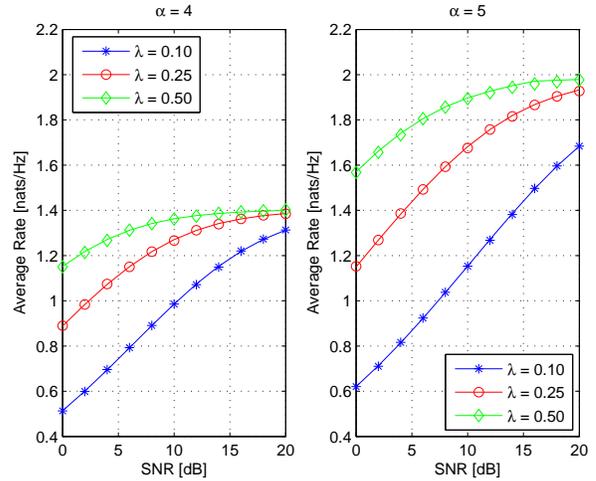}
\vspace{-0.60cm} \caption{Average rate of a single--tier cellular network over composite Rice and Log--Normal fading ($F_B=1$). Markers show Monte Carlo simulations. Solid lines show the analytical framework, which is computed by using \emph{Theorem} \ref{Theo_1}, \emph{Corollary} \ref{Cor_2} with $\varepsilon  = 0.05$, \emph{Proposition} \ref{Prop_4} with ($K=10$, $\sigma=6 \, {\rm{dB}}$, $\mu  =  - \ln \left( {10} \right)\sigma ^2 /20 \, {\rm{dB}}$) and $N_{{\rm{GHQ}}}  = 5$, and \emph{Remark} \ref{Remark_1} with $N_{{\rm{GCQ}}}  = 2000$. In \emph{Proposition} \ref{Prop_4}, the series in (\ref{Eq_16}) is truncated to the first 100 terms. The application of \emph{Remark} \ref{Remark_2} provides the same result and accuracy, but with less computational complexity. Furthermore, $\mathcal{M}_0 \left(  \cdot  \right) = \mathcal{M}_I \left(  \cdot  \right)$ are obtained from (\ref{APP5__Eq_1}) with $N_{{\rm{GHQ}}}  = 5$.} \label{Fig_RiceLogN}
\end{figure}
\paragraph{Framework Validation for Single--Tier Cellular Networks} In Figs. \ref{Fig_Ray}--\ref{Fig_RiceLogN}, the average rate of Rayleigh, Nakagami--\emph{m}, Log--Normal, composite Nakagami--\emph{m} and Log--Normal, and composite Rice and Log--Normal fading is shown, respectively, for a single--tier cellular network. Overall, we observe a very good accuracy of the proposed MGF--based approach. Furthermore, we observe, as expected, that the average rate: i) increases with the BSs density; ii) depends on the fading distribution; and iii) increases with the path--loss exponent.

\begin{figure}[!t]
\centering
\includegraphics [width=\columnwidth] {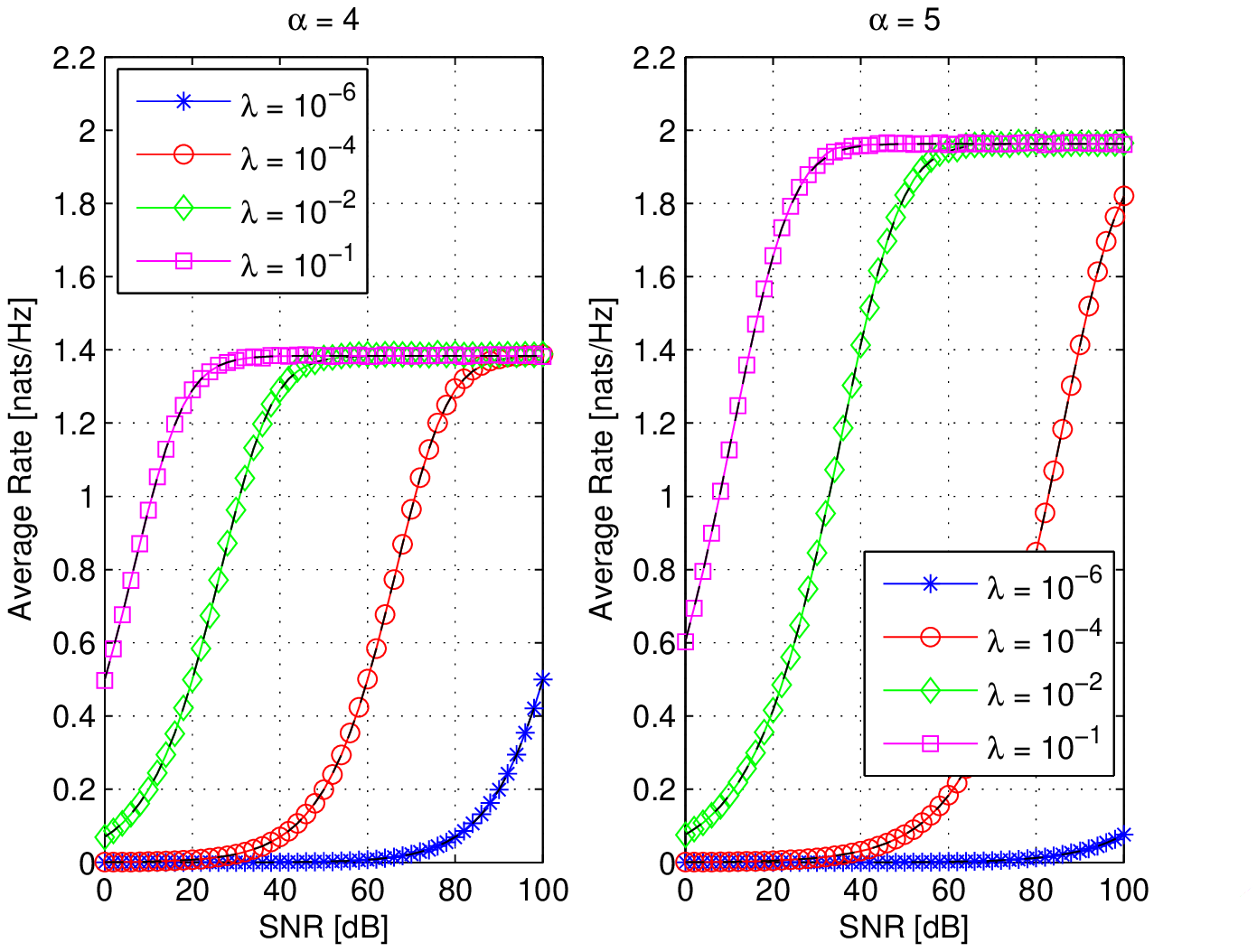}
\vspace{-0.60cm} \caption{Average rate of a single--tier cellular network over composite Nakagami--\emph{m} and Log--Normal fading ($F_B=1$). Markers show Monte Carlo simulations. Solid lines show the analytical framework, which is computed by using \emph{Theorem} \ref{Theo_1}, \emph{Corollary} \ref{Cor_2} with $\varepsilon  = 0.05$, \emph{Proposition} \ref{Prop_3} with ($m=2.5$, $\sigma=6 \, {\rm{dB}}$, $\mu  =  - \ln \left( {10} \right)\sigma ^2 /20 \, {\rm{dB}}$) and $N_{{\rm{GHQ}}}  = 5$, and \emph{Remark} \ref{Remark_1} with $N_{{\rm{GCQ}}}  = 2000$. Furthermore, $\mathcal{M}_0 \left(  \cdot  \right) = \mathcal{M}_I \left(  \cdot  \right)$ are obtained from \cite[Eq. (2.58)]{SimonBook} with $N_{{\rm{GHQ}}}  = 5$. The black dashed lines are obtained by directly computing the two--fold integral in \emph{Theorem} \ref{Theo_1} without using the Meijer G--function in \emph{Corollary} \ref{Cor_2}.} \label{Fig_Density45}
\end{figure}

\begin{figure}[!t]
\centering
\includegraphics [width=\columnwidth] {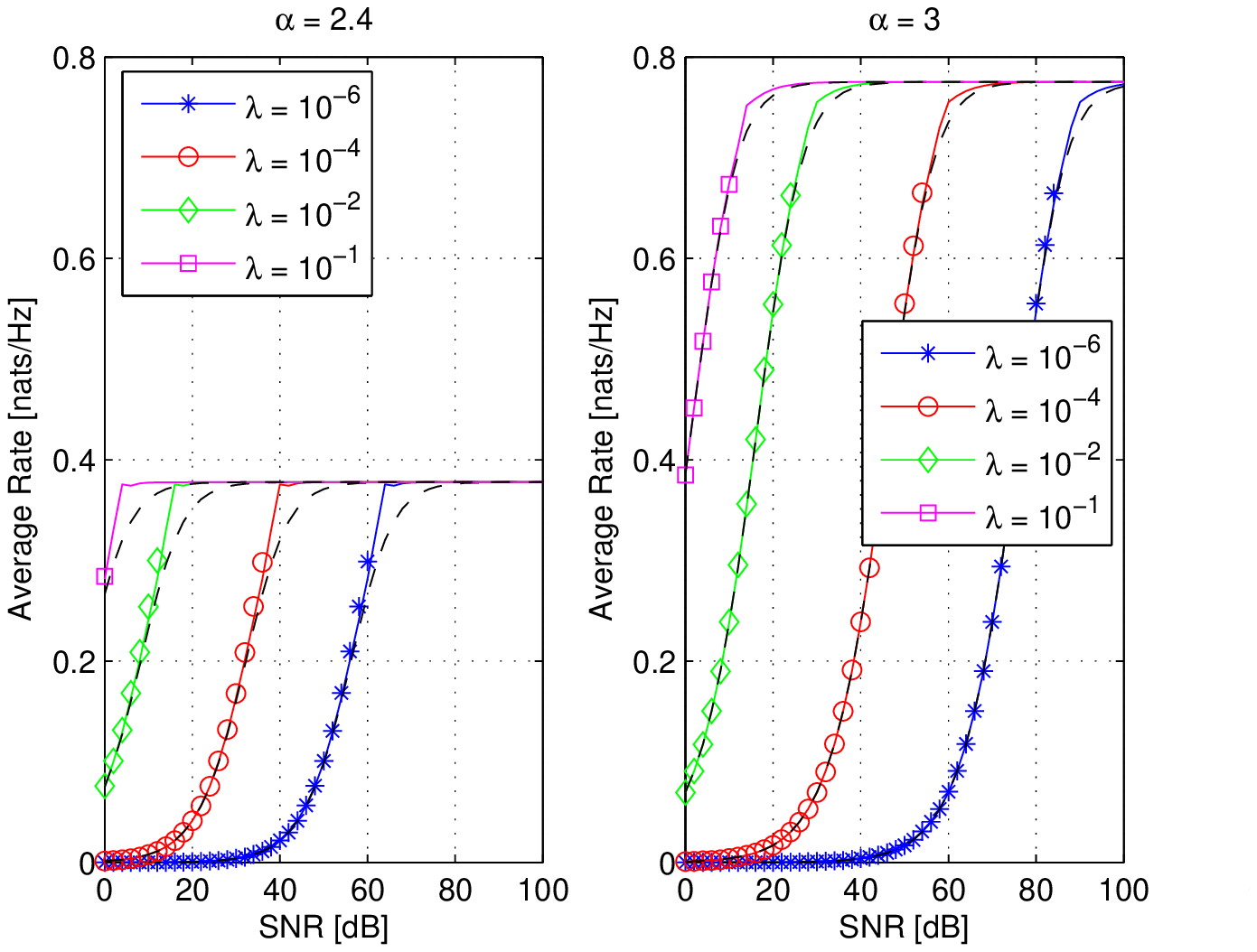}
\vspace{-0.60cm} \caption{Average rate of a single--tier cellular network over composite Nakagami--\emph{m} and Log--Normal fading ($F_B=1$). Markers show Monte Carlo simulations. Solid lines show the analytical framework, which is computed by using \emph{Theorem} \ref{Theo_1}, \emph{Corollary} \ref{Cor_2} with $\varepsilon  = 0.05$, \emph{Proposition} \ref{Prop_3} with ($m=2.5$, $\sigma=6 \, {\rm{dB}}$, $\mu  =  - \ln \left( {10} \right)\sigma ^2 /20 \, {\rm{dB}}$) and $N_{{\rm{GHQ}}}  = 5$, and \emph{Remark} \ref{Remark_1} with $N_{{\rm{GCQ}}}  = 2000$. Furthermore, $\mathcal{M}_0 \left(  \cdot  \right) = \mathcal{M}_I \left(  \cdot  \right)$ are obtained from \cite[Eq. (2.58)]{SimonBook} with $N_{{\rm{GHQ}}}  = 5$. The black dashed lines are obtained by directly computing the two--fold integral in \emph{Theorem} \ref{Theo_1} without using the Meijer G--function in \emph{Corollary} \ref{Cor_2}.} \label{Fig_Density23}
\end{figure}

\begin{figure}[!t]
\centering
\includegraphics [width=\columnwidth] {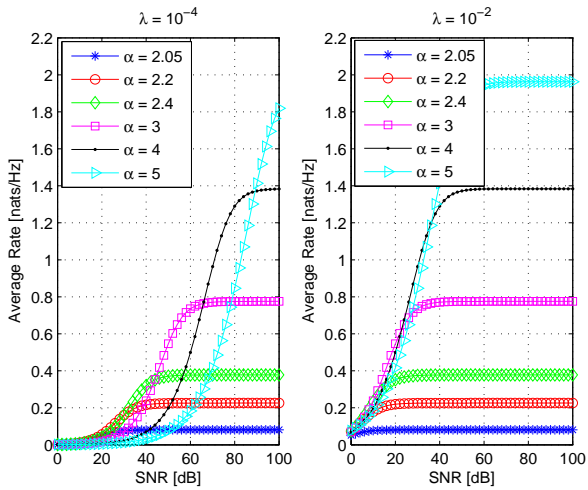}
\vspace{-0.60cm} \caption{Average rate of a single--tier cellular network over composite Nakagami--\emph{m} and Log--Normal fading ($F_B=1$). The curves are obtained by using the analytical framework and Monte Carlo simulations are not shown. More specifically, the curves are obtained by computing the two--fold integral in \emph{Theorem} \ref{Theo_1}, \emph{Proposition} \ref{Prop_3} with ($m=2.5$, $\sigma=6 \, {\rm{dB}}$, $\mu  =  - \ln \left( {10} \right)\sigma ^2 /20 \, {\rm{dB}}$) and $N_{{\rm{GHQ}}}  = 5$, and \emph{Remark} \ref{Remark_1} with $N_{{\rm{GCQ}}}  = 2000$. Furthermore, $\mathcal{M}_0 \left(  \cdot  \right) = \mathcal{M}_I \left(  \cdot  \right)$ are obtained from \cite[Eq. (2.58)]{SimonBook} with $N_{{\rm{GHQ}}}  = 5$.} \label{Fig_PathLossNum}
\end{figure}

In Figs. \ref{Fig_Ray}--\ref{Fig_RiceLogN}, we have considered dense cellular networks ($\lambda \ge 0.1$) and large path--loss exponents ($\alpha \ge 4$). The reason of this choice is mainly due to the long time needed to obtained Monte Carlo simulations for less dense cellular networks and for smaller path--loss exponents. However, it is important to verify the accuracy of the proposed MGF--based approach for more practical densities of BSs and for a wider range of path--loss exponents. In general, practical densities for macro BSs deployments are of the order of $\lambda \approx 10^{-6}$ \cite{AndrewsNov2011}, \cite{HeathTSP2012}, \cite{AndrewsOct2012}. Thus, to test numerical accuracy and stability of the MGF--based approach, we consider densities of BSs in the set $\lambda  = \left\{ {10^{ - 6} ,10^{ - 4} ,10^{ - 2} ,10^{ - 1} } \right\}$, in order to study sparse, medium, and dense deployments. As far as the path--loss exponent is concerned, we consider values in the set $\alpha = \left\{ {2.05,2.2,2.4,3,4,5} \right\} > 2$, which cover typical propagation environments for cellular applications \cite[Table 2.2]{GoldsmithBook}, \cite[Ch. 2, Sec. 5]{StuberBook}. The results of this study are shown in Figs. \ref{Fig_Density45}--\ref{Fig_PathLossNum}.

In Figs. \ref{Fig_Density45} and \ref{Fig_Density23}, we compare Monte Carlo simulations with the MGF--based approach. In particular, numerical results are obtained by using both the (exact) two--fold integral in \emph{Theorem} \ref{Theo_1} and the (approximated) single--integral in \emph{Corollary} \ref{Cor_2} in order to test complexity and accuracy of both methods. As far as Monte Carlo simulations are concerned, it is worth mentioning that only some SNR points (markers) are shown in the figures. The missing SNR points are not shown because of the long simulation time and the need to consider very large simulation areas to get accurate estimates of the average rate. The simulation time increases, in general, for more sparse cellular networks and for smaller path--loss exponents. The SNR points shown in the figures are those for which accurate estimates can be obtained with the simulation setup described above. The numerical examples confirm the very good accuracy and the numerical stability of the MGF--based approach for all cellular network setups. By comparing the curves obtained using \emph{Theorem} \ref{Theo_1} and \emph{Corollary} \ref{Cor_2}, we notice that the latter is very accurate except in the transition (corner) region from noise-- to interference--limited operating conditions, where the average rate reaches the asymptote calculated in \emph{Corollary} \ref{Cor_3} (further comments are available below where the high--SNR scenario is discussed). The reason of this (in practice negligible) numerical inaccuracy originates from the non--smooth transition introduced by the adoption of the Heaviside function in \emph{Corollary} \ref{Cor_3}.

In Fig. \ref{Fig_PathLossNum}, we compare the average rate as a function of the path--loss exponent. Only numerical results obtained from the MGF--based approach are shown in this figure. The curves are obtained by using the two--fold integral in \emph{Theorem} \ref{Theo_1}. As far as the application of \emph{Corollary} \ref{Cor_2} is concerned, its accuracy for $\alpha \ge 2.4$ is shown in Figs. \ref{Fig_Density45} and \ref{Fig_Density23}. For $\alpha < 2.4$, it is less practical to use the Meijer G--function in \emph{Corollary} \ref{Cor_2} since for such values of $\alpha$ we would have $\alpha _N  \gg 1$ and $\alpha _D  \gg 1$, and, thus, computation time and numerical accuracy would highly depend on the specific implementation of the Meijer G--function. On the other hand, \emph{Theorem} \ref{Theo_1} provides very accurate estimates in a few seconds (for each SNR point), as discussed in Section \ref{Pcov_vs_MGF}. Figure \ref{Fig_PathLossNum} clearly shows that \emph{Theorem} \ref{Theo_1} provides reliable numerical estimates for the considered set of path--loss exponents, including $\alpha  \approx 2$. As far as the performance trend is concerned, Fig. \ref{Fig_PathLossNum} shows a very different behavior for dense ($\lambda = 10^{-2}$) and medium/sparse ($\lambda = 10^{-4}$) cellular networks. In dense cellular networks, the higher the path--loss exponent the better the average rate regardless of the operating SNR. On the other hand, in medium/sparse cellular networks two SNR regions can be identified: i) for low--SNR (noise--limited regime), the lower the path--loss exponent the higher the average rate. This is due to that fact that the useful signal undergoes a lower attenuation and that the aggregate interference is negligible compared to the additive noise; and ii) for high--SNR (interference--limited regime), the higher the path--loss exponent the higher the average rate. This is due to the fact that the additive noise is negligible compared to the aggregate interference and that the interfering BSs undergo a larger attenuation, which has a more pronounced effect on the average rate than the larger attenuation undergone by the useful signal. These results are in agreement with intuition and confirm the usefulness of the proposed MGF--based approach for cellular networks analysis and design.

For ease of comparison with Monte Carlo simulations, in the following only large path--loss exponents and dense cellular networks deployments are considered.
\begin{figure}[!t]
\centering
\includegraphics [width=\columnwidth] {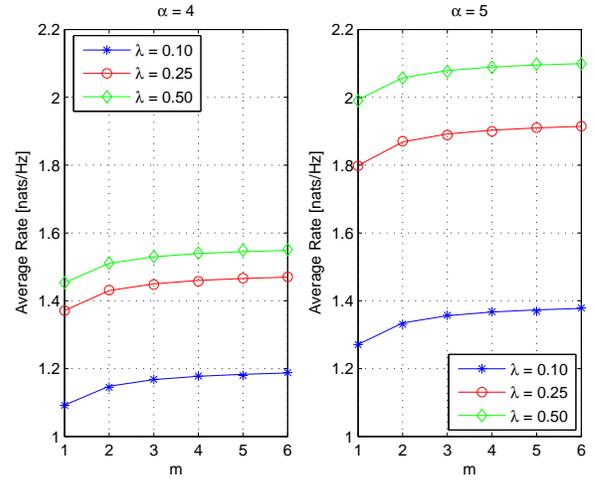}
\vspace{-0.60cm} \caption{Average rate of a single--tier cellular network over Nakagami--\emph{m} fading ($F_B=1$, SNR = 10 dB). Markers show Monte Carlo simulations. Solid lines show the analytical framework, which is computed by using \emph{Theorem} \ref{Theo_1}, \emph{Corollary} \ref{Cor_2} with $\varepsilon  = 0.05$, \emph{Proposition} \ref{Prop_1} with $\Omega = 1$, and \emph{Remark} \ref{Remark_1} with $N_{{\rm{GCQ}}}  = 2000$. Furthermore, $\mathcal{M}_0 \left(  \cdot  \right) = \mathcal{M}_I \left(  \cdot  \right)$ are obtained from \cite[Eq. (2.22)]{SimonBook}.} \label{FigSNR_NakM}
\end{figure}
\begin{figure}[!t]
\centering
\includegraphics [width=\columnwidth] {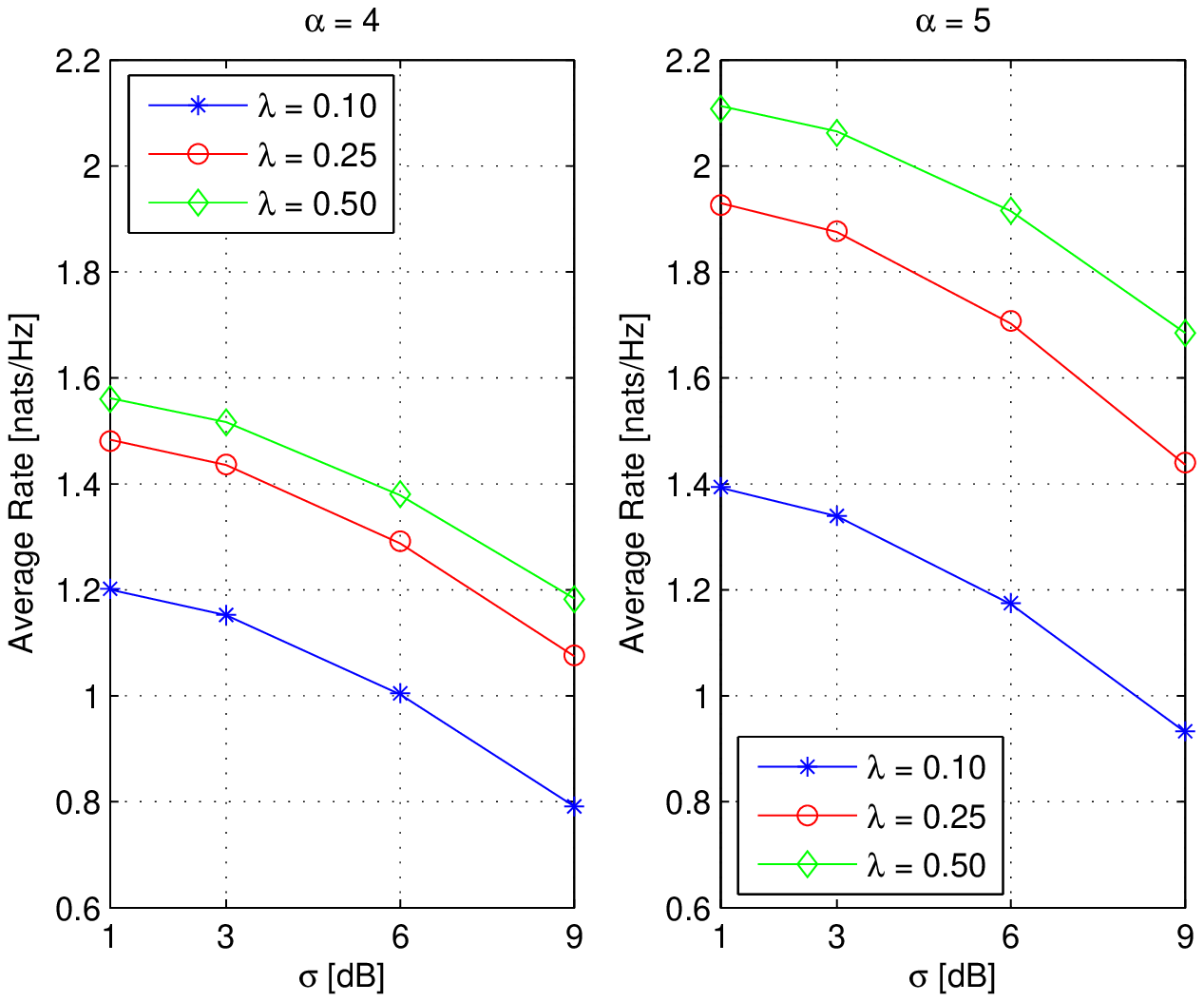}
\vspace{-0.60cm} \caption{Average rate of a single--tier cellular network over Log--Normal fading ($F_B=1$, SNR = 10 dB). Markers show Monte Carlo simulations. Solid lines show the analytical framework, which is computed by using \emph{Theorem} \ref{Theo_1}, \emph{Corollary} \ref{Cor_2} with $\varepsilon  = 0.05$, \emph{Proposition} \ref{Prop_2} with $\mu  =  - \ln \left( {10} \right)\sigma ^2 /20 \, {\rm{dB}}$ and $N_{{\rm{GHQ}}}  = 5$, and \emph{Remark} \ref{Remark_1} with $N_{{\rm{GCQ}}}  = 2000$. Furthermore, $\mathcal{M}_0 \left(  \cdot  \right) = \mathcal{M}_I \left(  \cdot  \right)$ are obtained from \cite[Eq. (2.54)]{SimonBook} with $N_{{\rm{GHQ}}}  = 5$.} \label{FigSNR_LogN}
\end{figure}
\begin{figure}[!t]
\centering
\includegraphics [width=\columnwidth] {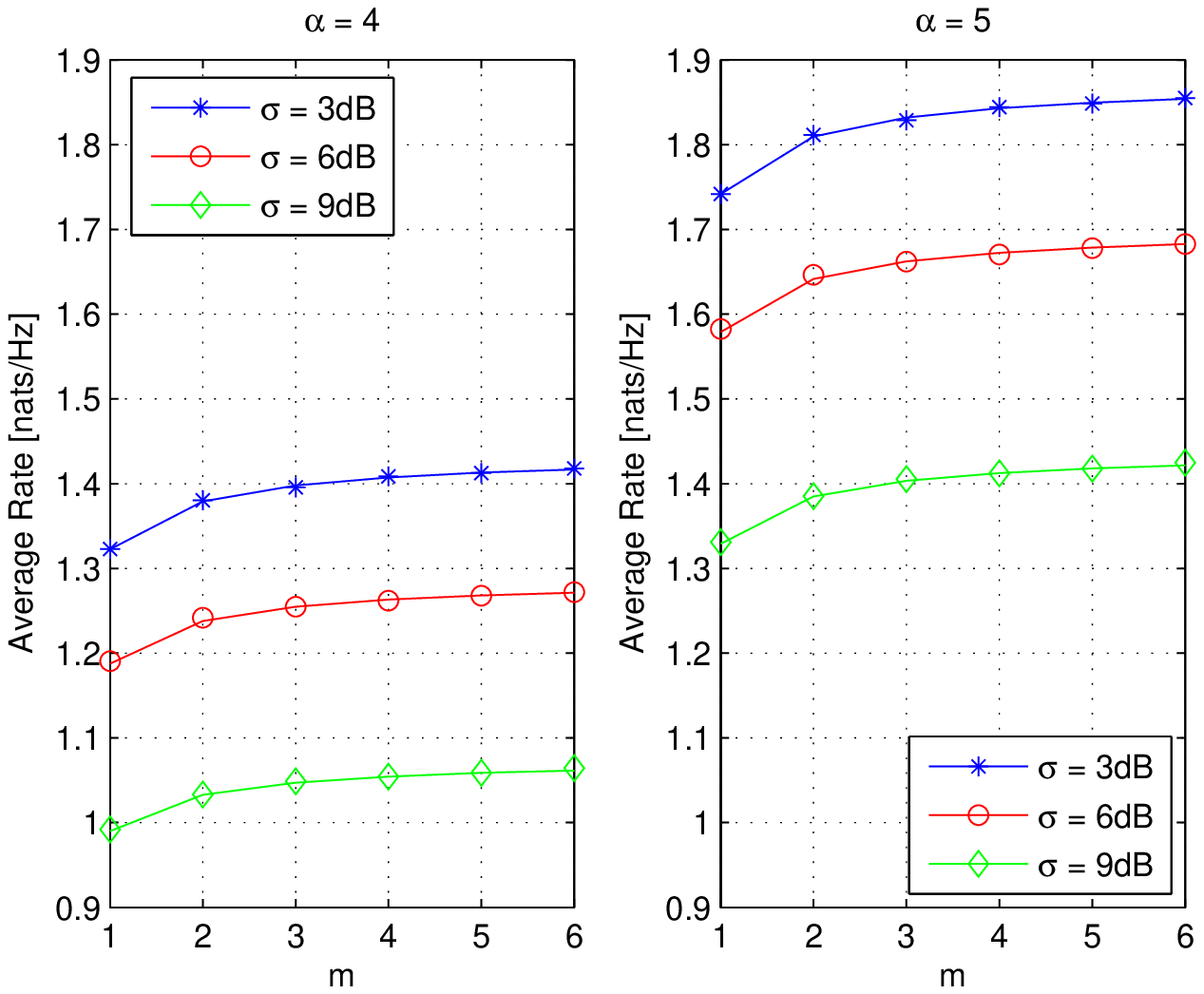}
\vspace{-0.60cm} \caption{Average rate of a single--tier cellular network over composite Nakagami--\emph{m} and Log--Normal fading ($F_B=1$, SNR = 10 dB, and $\lambda = 0.25$). Markers show Monte Carlo simulations. Solid lines show the analytical framework, which is computed by using \emph{Theorem} \ref{Theo_1}, \emph{Corollary} \ref{Cor_2} with $\varepsilon  = 0.05$, \emph{Proposition} \ref{Prop_3} with $\mu  =  - \ln \left( {10} \right)\sigma ^2 /20 \, {\rm{dB}}$ and $N_{{\rm{GHQ}}}  = 5$, and \emph{Remark} \ref{Remark_1} with $N_{{\rm{GCQ}}}  = 2000$. Furthermore, $\mathcal{M}_0 \left(  \cdot  \right) = \mathcal{M}_I \left(  \cdot  \right)$ are obtained from \cite[Eq. (2.58)]{SimonBook} with $N_{{\rm{GHQ}}}  = 5$.} \label{FigSNR_NakMLogN}
\end{figure}
\begin{figure}[!t]
\centering
\includegraphics [width=\columnwidth] {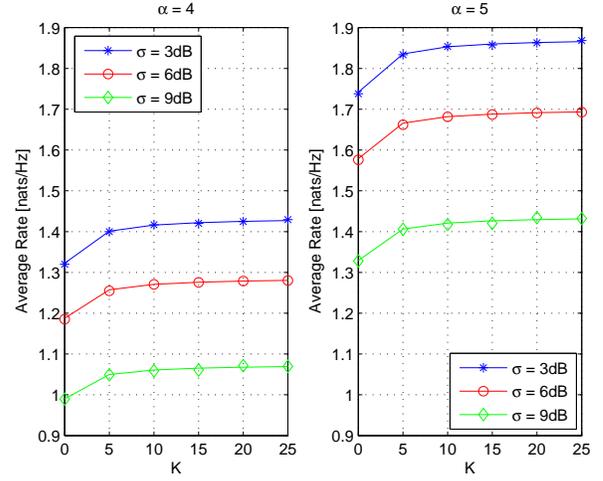}
\vspace{-0.60cm} \caption{Average rate of a single--tier cellular network over composite Rice and Log--Normal fading ($F_B=1$, SNR = 10 dB, and $\lambda = 0.25$). Markers show Monte Carlo simulations. Solid lines show the analytical framework, which is computed by using \emph{Theorem} \ref{Theo_1}, \emph{Corollary} \ref{Cor_2} with $\varepsilon  = 0.05$, \emph{Proposition} \ref{Prop_4} with $\mu  =  - \ln \left( {10} \right)\sigma ^2 /20 \, {\rm{dB}}$ and $N_{{\rm{GHQ}}}  = 5$, and \emph{Remark} \ref{Remark_1} with $N_{{\rm{GCQ}}}  = 2000$. In \emph{Proposition} \ref{Prop_4}, the series in (\ref{Eq_16}) is truncated to the first 100 terms. The application of \emph{Remark} \ref{Remark_2} provides the same result and accuracy, but with less computational complexity. Furthermore, $\mathcal{M}_0 \left(  \cdot  \right) = \mathcal{M}_I \left(  \cdot  \right)$ are obtained from (\ref{APP5__Eq_1}) with $N_{{\rm{GHQ}}}  = 5$.} \label{FigSNR_RiceLogN}
\end{figure}
\paragraph{Impact of Fading Model and Fading Parameters} In Figs. \ref{FigSNR_NakM}--\ref{FigSNR_RiceLogN}, the average rate of Nakagami--\emph{m}, Log--Normal, composite Nakagami--\emph{m} and Log--Normal, and composite Rice and Log--Normal fading is shown, respectively, for a single--tier cellular network and for different choices of the fading parameters. Also in this case, the framework provides very accurate estimates. More specifically, we observe that: i) the average rate is slightly sensitive to \emph{m} and \emph{K} fading parameters of Nakagami--\emph{m}, composite Nakagami--\emph{m} and Log--Normal, and composite Rice and Log--Normal distributions, as well as that it increases for less severe fading ($m$ increases) and in the presence of a stronger line--of--sight component ($K$ increases); and ii) the average rate strongly depends on the shadowing standard deviation $\sigma$ of Log--Normal, composite Nakagami--\emph{m} and Log--Normal, and composite Rice and Log--Normal distributions, as well as that it decreases significantly for more severe shadowing ($\sigma$ increases).
\begin{figure}[!t]
\centering
\includegraphics [width=\columnwidth] {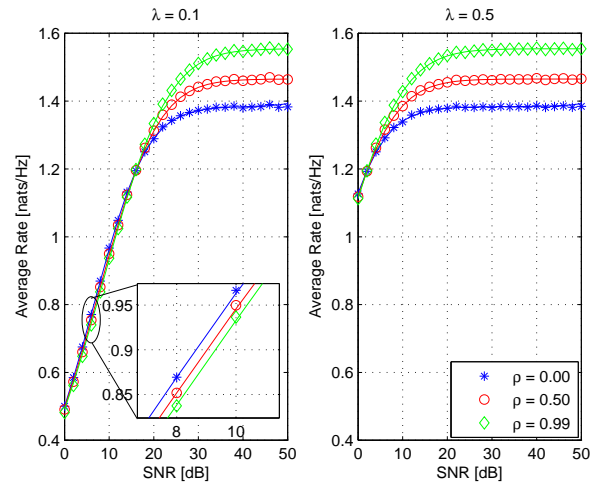}
\vspace{-0.60cm} \caption{Average rate of a single--tier cellular network over correlated composite Nakagami--\emph{m} and Log--Normal fading ($F_B=1$). Markers show Monte Carlo simulations. Solid lines show the analytical framework, which is computed by using (\ref{Eq_20bis}) with $N_{{\rm{GHQ}}}  = 5$, \emph{Theorem} \ref{Theo_1}, \emph{Corollary} \ref{Cor_2} with $\varepsilon  = 0.05$, \emph{Proposition} \ref{Prop_3} with ($m=2.5$, $\sigma=6 \, {\rm{dB}}$, $\mu  =  - \ln \left( {10} \right)\sigma ^2 /20 \, {\rm{dB}}$, $\alpha = 4$) and $N_{{\rm{GHQ}}}  = 5$, and \emph{Remark} \ref{Remark_1} with $N_{{\rm{GCQ}}}  = 1000$. Furthermore, $\mathcal{M}_0 \left(  \cdot  \right) = \mathcal{M}_I \left(  \cdot  \right)$ are obtained from \cite[Eq. (2.58)]{SimonBook} with $N_{{\rm{GHQ}}}  = 5$.} \label{Fig_CorrLogN_1}
\end{figure}
\begin{figure}[!t]
\centering
\includegraphics [width=\columnwidth] {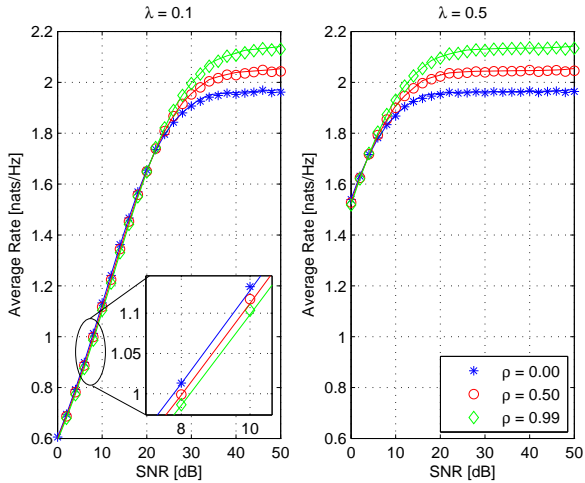}
\vspace{-0.60cm} \caption{Average rate of a single--tier cellular network over correlated composite Nakagami--\emph{m} and Log--Normal fading ($F_B=1$). Markers show Monte Carlo simulations. Solid lines show the analytical framework, which is computed by using (\ref{Eq_20bis}) with $N_{{\rm{GHQ}}}  = 5$, \emph{Theorem} \ref{Theo_1}, \emph{Corollary} \ref{Cor_2} with $\varepsilon  = 0.05$, \emph{Proposition} \ref{Prop_3} with ($m=2.5$, $\sigma=6 \, {\rm{dB}}$, $\mu  =  - \ln \left( {10} \right)\sigma ^2 /20 \, {\rm{dB}}$, $\alpha = 5$) and $N_{{\rm{GHQ}}}  = 5$, and \emph{Remark} \ref{Remark_1} with $N_{{\rm{GCQ}}}  = 1000$. Furthermore, $\mathcal{M}_0 \left(  \cdot  \right) = \mathcal{M}_I \left(  \cdot  \right)$ are obtained from \cite[Eq. (2.58)]{SimonBook} with $N_{{\rm{GHQ}}}  = 5$.} \label{Fig_CorrLogN_2}
\end{figure}
\paragraph{Correlated Log--Normal Shadowing} In Figs. \ref{Fig_CorrLogN_1} and \ref{Fig_CorrLogN_2}, numerical examples in the presence of shadowing correlation over a composite Nakagami--\emph{m} and Log--Normal fading channel are shown. We observe that the proposed approach for correlated Log--Normal shadowing is very accurate for different choices of the correlation coefficient. Furthermore, by comparing Figs. \ref{Fig_CorrLogN_1} and \ref{Fig_CorrLogN_2} with Fig. \ref{Fig_NakMLogN}, we note that the framework in Section \ref{Correlated_LogN} reduces to the independent case for $\rho = 0$. More specifically, the figures show the following performance trends: i) for high--SNR, the average rate is independent of the density of BSs regardless of the shadowing correlation coefficient; ii) for low--SNR, the average rate slightly decreases if the shadowing correlation coefficient increases; and iii) for high--SNR, the average rate increases if the shadowing correlation coefficient increases. Even though it may seem counterintuitive that the average rate increases with shadowing correlation, this result seems to originate from the BSs association policy adopted in the present paper (the probe mobile terminal is associated with the closest BS). Finally, we emphasize that (apparently) counterintuitive trends in the presence of Log--Normal shadowing have been observed in other papers, \emph{e.g.}, \cite{AndrewsNov2011} and \cite{BaszczyszynJune2012} where it is shown that the coverage probability increases and that the blocking probability is not always increasing with the shadowing standard deviation, respectively. This confirms, once again, the importance of taking into account correlated Log--Normal shadowing for accurate performance prediction.
\begin{figure}[!t]
\centering
\includegraphics [width=\columnwidth] {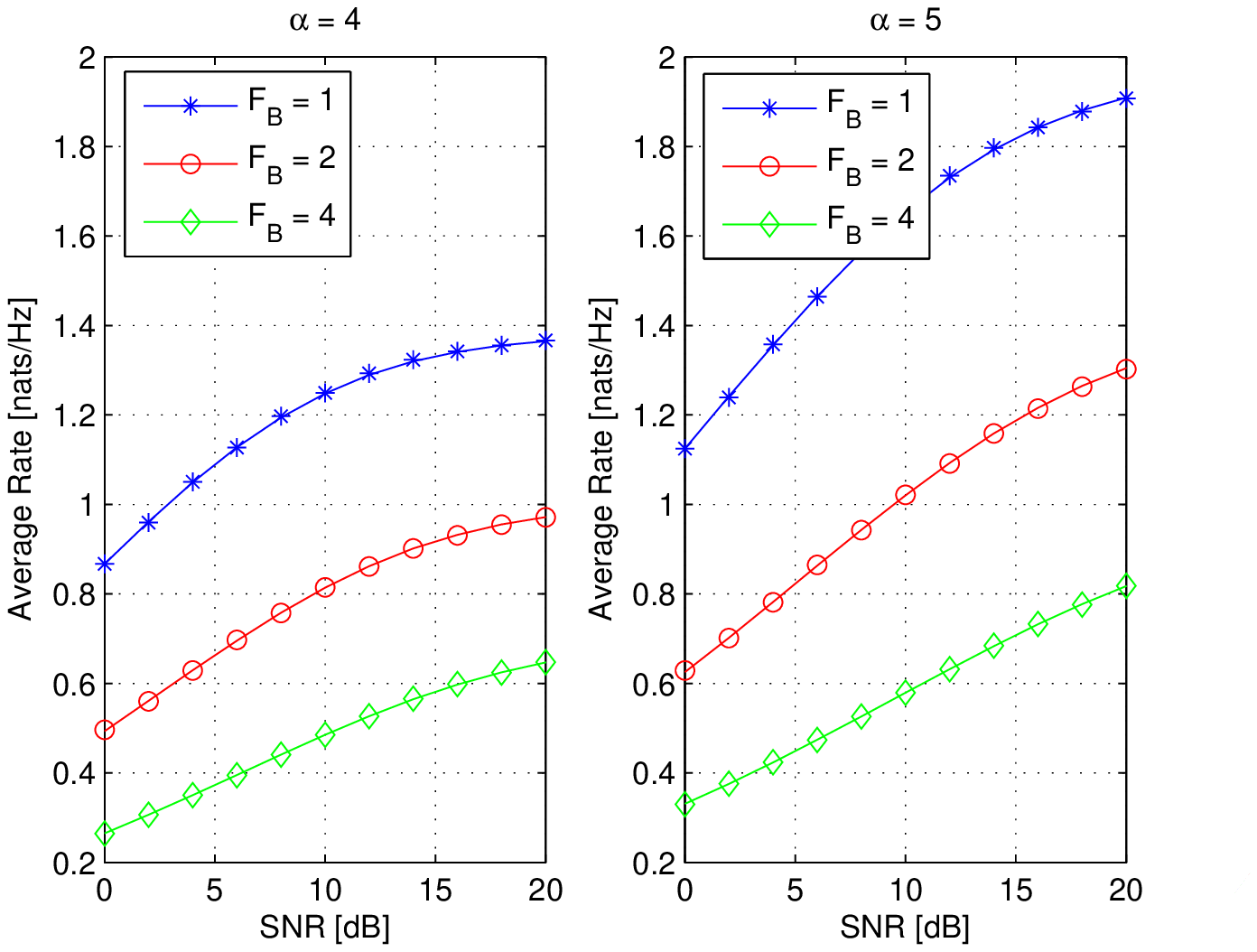}
\vspace{-0.60cm} \caption{Average rate of a single--tier cellular network with frequency reuse over composite Nakagami--\emph{m} and Log--Normal fading ($\lambda = 0.25$). Markers show Monte Carlo simulations. Solid lines show the analytical framework, which is computed by using \emph{Theorem} \ref{Theo_1}, \emph{Corollary} \ref{Cor_2} with $\varepsilon  = 0.05$, \emph{Proposition} \ref{Prop_3} with ($m=2.5$, $\sigma=6 \, {\rm{dB}}$, $\mu  =  - \ln \left( {10} \right)\sigma ^2 /20 \, {\rm{dB}}$) and $N_{{\rm{GHQ}}}  = 5$, and \emph{Remark} \ref{Remark_1} with $N_{{\rm{GCQ}}}  = 2000$. Furthermore, $\mathcal{M}_0 \left(  \cdot  \right) = \mathcal{M}_I \left(  \cdot  \right)$ are obtained from \cite[Eq. (2.58)]{SimonBook} with $N_{{\rm{GHQ}}}  = 5$.} \label{FigFreqReuse_NakMLogN}
\end{figure}
\paragraph{Impact of Frequency Reuse} In Fig. \ref{FigFreqReuse_NakMLogN}, the impact of frequency reuse on the average rate of a single--tier cellular network over composite Nakagami--\emph{m} and Log--Normal fading is studied. Our numerical analysis confirms the findings in Section \ref{FrequencyReuse_1Tier}, and that frequency reuse is detrimental for the average rate. As suggested in \cite{AndrewsNov2011}, frequency reuse significantly improves the coverage probability. The analysis of this trade--off is out of the scope of the present paper but is currently being investigated by the authors.
\begin{figure}[!t]
\centering
\includegraphics [width=\columnwidth] {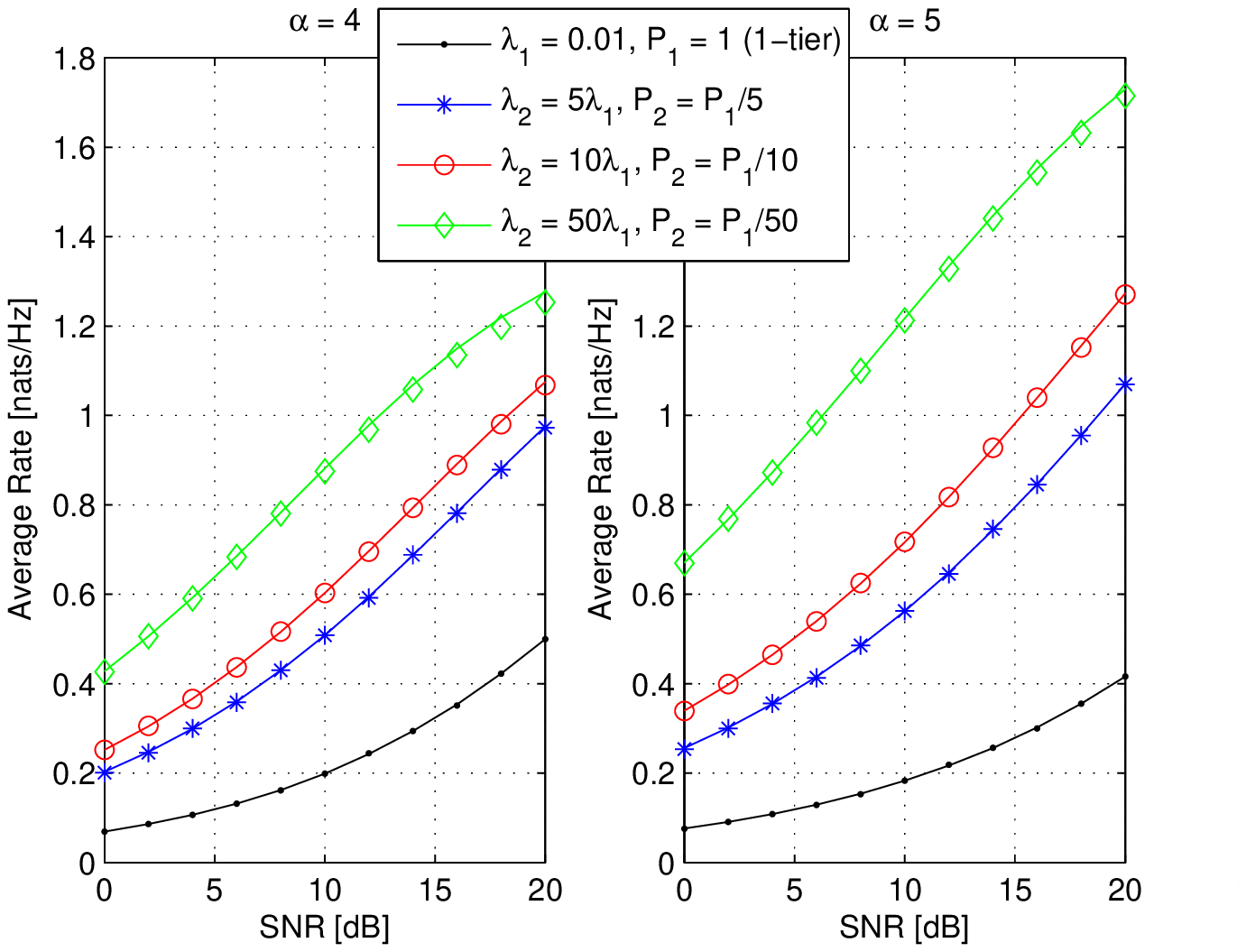}
\vspace{-0.60cm} \caption{Average rate of a two--tier cellular network over composite Nakagami--\emph{m} and Log--Normal fading ($F_B^{(t)}=1$ for $t=1,2$, and ${\rm{SNR}} = {\rm{SNR}}_1$). Markers show Monte Carlo simulations. Solid lines show the analytical framework, which is computed by using \emph{Corollary} \ref{Cor_6}, \emph{Corollary} \ref{Cor_2} with $\varepsilon  = 0.05$, \emph{Proposition} \ref{Prop_3} with ($m=2.5$, $\sigma=6 \, {\rm{dB}}$, $\mu  =  - \ln \left( {10} \right)\sigma ^2 /20 \, {\rm{dB}}$) and $N_{{\rm{GHQ}}}  = 5$, and \emph{Remark} \ref{Remark_1} with $N_{{\rm{GCQ}}}  = 2000$. Furthermore, $\mathcal{M}_{t, 0} \left(  \cdot  \right) = \mathcal{M}_{t,I} \left(  \cdot  \right)$ for $t=1,2$ are obtained from \cite[Eq. (2.58)]{SimonBook} with $N_{{\rm{GHQ}}}  = 5$.} \label{Fig2Tier_NakMLogN}
\end{figure}
\begin{figure}[!t]
\centering
\includegraphics [width=\columnwidth] {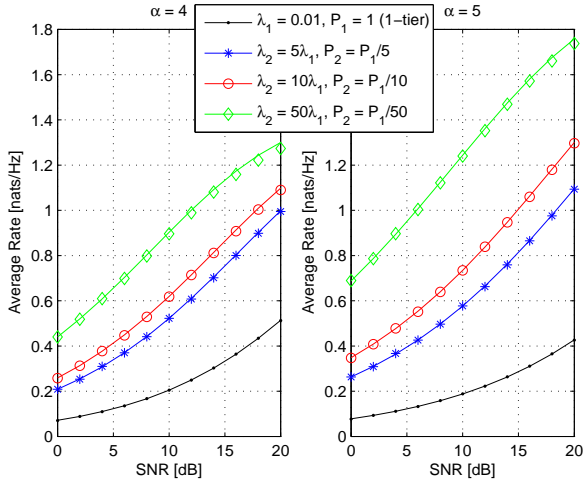}
\vspace{-0.60cm} \caption{Average rate of a two--tier cellular network over composite Rice and Log--Normal fading ($F_B^{(t)}=1$ for $t=1,2$, and ${\rm{SNR}} = {\rm{SNR}}_1$). Markers show Monte Carlo simulations. Solid lines show the analytical framework, which is computed by using \emph{Corollary} \ref{Cor_6}, \emph{Corollary} \ref{Cor_2} with $\varepsilon  = 0.05$, \emph{Proposition} \ref{Prop_4} with ($K=10$, $\sigma=6 \, {\rm{dB}}$, $\mu  =  - \ln \left( {10} \right)\sigma ^2 /20 \, {\rm{dB}}$) and $N_{{\rm{GHQ}}}  = 5$, and \emph{Remark} \ref{Remark_1} with $N_{{\rm{GCQ}}}  = 2000$. In \emph{Proposition} \ref{Prop_4}, the series in (\ref{Eq_16}) is truncated to the first 100 terms. The application of \emph{Remark} \ref{Remark_2} provides the same result and accuracy, but with less computational complexity. Furthermore, $\mathcal{M}_{t,0} \left(  \cdot  \right) = \mathcal{M}_{t,I} \left(  \cdot  \right)$ for $t=1,2$ are obtained from (\ref{APP5__Eq_1}) with $N_{{\rm{GHQ}}}  = 5$.} \label{Fig2Tier_RiceLogN}
\end{figure}
\paragraph{Framework Validation for Multi--Tier Cellular Networks} In Fig. \ref{Fig2Tier_NakMLogN} and Fig. \ref{Fig2Tier_RiceLogN}, the average rate of a two--tier cellular network over composite Nakagami--\emph{m} and Log--Normal and composite Rice and Log--Normal fading is analyzed, respectively. As an illustrative example, we consider the situation where the BSs of every tier transmit with a power that is inversely proportional to their spatial density. This is a reasonable choice if, \emph{e.g.}, the first tier is used to model macro BSs and the second tier is used to model femto BSs. We observe, as expected, that the average rate significantly increases when the BSs density of the lower tier increases. The framework provides a very good accuracy. Furthermore, we note a negligible difference between the two fading models for the chosen set of parameters.
\paragraph{High--SNR Scenario} Finally, we observe that, in all figures, the average rate increases with the SNR tending towards a horizontal asymptote for high--SNR. By direct inspection, the reader can verify that the horizontal asymptote coincides with the average rate that can be computed by using \emph{Corollary} \ref{Cor_3} and \emph{Corollary} \ref{Cor_8} with $\mathcal{\tilde R}_t^{\left( {{\rm{SNR}}_\infty  } \right)}  = \mathcal{\tilde R}_t^{\left( {\lambda _\infty  } \right)}$ in \emph{Corollary} \ref{Cor_9}. In other words, the SNR region where the average rate is flat corresponds to the interference--limited operating regime. As a consequence, we conclude that the simple formulas given in these corollaries are quite accurate for ${\rm{SNR}} \ge {\rm{SNR}}^*$, where ${\rm{SNR}}^*$ is the corner point where the average rate starts approaching the horizontal asymptote. For example, ${\rm{SNR}}^* \approx 20 {\rm{dB}}$ for $\alpha = 4$ in Fig. \ref{Fig_Ray}. In general, ${\rm{SNR}}^*$ depends on the density of BSs, the path--loss exponent, and the fading channel model. The reader can identify ${\rm{SNR}}^*$ for different cellular setups by direct inspection of all the figures shown in the paper. Figure \ref{Fig_PathLossNum} shows many cellular networks deployments of interest.

By observing the asymptotic behavior of the average rate for high--SNR, an interesting problem to investigate is whether ${\rm{SNR}}^*$ is lower or greater than the typical operating SNR of current cellular networks deployments. This question is interesting because many papers assume, for analytical tractability, that the additive noise is always negligible and that, as a consequence, cellular networks are interference--limited. All the figures shown in our paper confirm that the interference--limited assumption is accurate only if the operating SNR is greater than ${\rm{SNR}}^*$. In order to assess whether the operating condition ${\rm{SNR}} \ge {\rm{SNR}}^*$ is usually verified, we investigate, as an example, the same setup as in \cite{AndrewsOct2012}, which holds for typical cellular networks. More specifically, we consider: i) ${\rm{SNR}} = {P \mathord{\left/{\vphantom {P {\sigma _N^2 }}} \right.\kern-\nulldelimiterspace} {\sigma _N^2 }}$; ii) $\sigma _N^2  = {W \mathord{\left/ {\vphantom {W {L_0 }}} \right. \kern-\nulldelimiterspace} {L_0 }}$; iii) $W = k_B T_0 B_0 =  - 104{\rm{ dBm}} = 4 \cdot 10^{ - 14} {\rm{ Watt}}$ is the noise power, where $k_B  = {\rm{1}}.{\rm{38}} \cdot 10^{ - 23} {\rm{ Joule/Kelvin}}$ is the Boltzmann's constant, $T_0  = 290{\rm{ Kelvin}}$ is the noise temperature, $B_0 = 10{\rm{ MHz}}$ is the receiver bandwidth; and iv) $L_0  =  - 38.5{\rm{ dB}} = 1.41 \cdot 10^{ - 4}$ is the path--loss at a reference distance of one meter. Accordingly, we have ${\rm{SNR}} = {P \mathord{\left/{\vphantom {P {\sigma _N^2 }}} \right. \kern-\nulldelimiterspace} {\sigma _N^2 }} = \left( {{{L_0 } \mathord{\left/ {\vphantom {{L_0 } W}} \right. \kern-\nulldelimiterspace} W}} \right)P = 3.52 \cdot 10^9 P$. By assuming typical transmit powers equal to $P_{{\rm{Macro}}}  = 40{\rm{ Watt}}$, $P_{{\rm{Micro}}}  = 6.3{\rm{ Watt}}$, $P_{{\rm{Pico}}}  = 0.13{\rm{ Watt}}$, and $P_{{\rm{Femto}}}  = 0.05{\rm{ Watt}}$ for macro, micro, pico, and femto BSs \cite{HeathTSP2012}, \cite{Auer}, we obtain ${\rm{SNR}}_{{\rm{Macro}}}  = 111.50{\rm{ dB}}$, ${\rm{SNR}}_{{\rm{Micro}}}  = 103.45{\rm{ dB}}$, ${\rm{SNR}}_{{\rm{Pico}}}  = 86.60{\rm{ dB}}$, and ${\rm{SNR}}_{{\rm{Femto}}}  = 82.45{\rm{ dB}}$, respectively. By comparing these typical operating SNRs with the ${\rm{SNR}}^*$ shown in our figures, we conclude that the condition ${\rm{SNR}} \ge {\rm{SNR}}^*$ is well satisfied for many system setups analyzed in the paper. For example, let us consider the setups shown in Fig. \ref{Fig_PathLossNum}. We observe that ${\rm{SNR}} \ge {\rm{SNR}}^*$ is always verified in dense ($\lambda = 10^{-2}$) cellular networks for every path--loss exponents, and in medium/sparse ($\lambda = 10^{-4}$) cellular networks for low path--loss exponents ($\alpha < 4$ in the figure). On the other hand, for larger path--loss exponents the condition may not be verified for some types of BSs. This confirms that typical cellular networks deployments can be approximated, in most cases, to be interference--limited, and that the simple frameworks in \emph{Corollary} \ref{Cor_3} and \emph{Corollary} \ref{Cor_8} can be used for first--order performance analysis, design, and optimization.
\section{Conclusion} \label{Conclusion}
In this paper, we have introduced a comprehensive mathematical framework for the analysis of the average rate of multi--tier cellular networks whose BSs are assumed to be randomly distributed according to a PPP spatial distribution. The framework is applicable to general fading channel models with arbitrary fading parameters. Numerically efficient and stable algorithms to compute some transcendental functions, such as the Meijer G--function, have been proposed. The framework needs the computation of either single-- or two--fold integrals for general fading distributions and arbitrary path--loss exponents. Furthermore, shadowing correlation can be taken into account with another extra numerical integral. The framework can handle random frequency reuse, and it simplifies significantly for interference--limited cellular networks and for high--SNR setups. Extensive Monte Carlo simulations have confirmed the accuracy of the proposed analytical methodology for various fading distributions and cellular deployments.
\appendices
\begin{figure*}[!t]
\setcounter{equation}{37}
\begin{equation}
\label{APP1A__Eq_2}
\begin{split}
 \mathcal{\bar R}\left( z; \lambda \right) & = \frac{1}{{{\rm{SNR}}}}\int\nolimits_0^{ + \infty } {\xi ^{\frac{\alpha }{2} - 1} \exp \left\{ { - \pi \lambda \mathcal{Z}_I \left( z \right)\xi } \right\}\exp \left\{ { - \frac{z}{{{\rm{SNR}}}}\xi ^{\frac{\alpha }{2}} } \right\}d\xi }  \\
 & \mathop  \le \limits^{\left( a \right)} \frac{1}{{{\rm{SNR}}}}\int\nolimits_0^{ + \infty } {\xi ^{\frac{\alpha }{2} - 1} \exp \left\{ { - \pi \lambda \mathcal{Z}_I \left( z \right)\xi } \right\}d\xi } \mathop  = \limits^{\left( b \right)} \frac{1}{{{\rm{SNR}}}} \Gamma \left( {\frac{\alpha }{2}} \right)\left[ {\pi \lambda \mathcal{Z}_I \left( z \right)} \right]^{ - \frac{\alpha }{2}}  \\
 \end{split}
\end{equation}
\normalsize \hrulefill \vspace*{0pt}
\end{figure*}
\section{Proof of \emph{Theorem} \ref{Theo_1}} \label{APP__Theo_1}
By using \cite[Lemma 1]{Hamdi2010} with $N=M=1$, we have:
\setcounter{equation}{31}
\begin{equation}
\label{APP1__Eq_0}
\begin{split} & {\mathbb{E}}\left\{ {\ln \left( {1 + \frac{X}{{Y + 1}}} \right)} \right\}\\
& \hspace{0.75cm} = \int\nolimits_0^{ + \infty } {\frac{{\mathcal{M}_Y \left( z \right) - \mathcal{M}_{X,Y} \left( z \right)}}{z}\exp \left\{ { - z} \right\}dz} \\
& \hspace{0.75cm} \mathop  = \limits^{\left( a \right)} \int\nolimits_0^{ + \infty } {\frac{{\mathcal{M}_Y \left( z \right)\left[ {1 - \mathcal{M}_X \left( z \right)} \right]}}{z}\exp \left\{ { - z} \right\}dz} \end{split}
\end{equation}
\noindent where: i) $X$ and $Y$ are arbitrary non--negative random variables; ii) $\mathcal{M}_{X,Y} \left( z \right) = {\mathbb{E}}\left\{ {e^{ - z\left( {X + Y} \right)} } \right\}$ is the MGF of random variable $X+Y$; and iii) (a) holds if $X$ and $Y$ are independent.

From (\ref{APP1__Eq_0}) with $X = {{\left( {Pg_0 \xi ^{ - \alpha } } \right)} \mathord{\left/ {\vphantom {{\left( {Pg_0 \xi ^{ - \alpha } } \right)} {\sigma _N^2 }}} \right. \kern-\nulldelimiterspace} {\sigma _N^2 }}$, $Y = {{I_{{\rm{agg}}} \left( \xi  \right)} \mathord{\left/ {\vphantom {{I_{{\rm{agg}}} \left( \xi  \right)} {\sigma _N^2 }}} \right. \kern-\nulldelimiterspace} {\sigma _N^2 }}$, and by taking into account that, conditioning upon $\xi$, $X$ and $Y$ are independent, the expectation in (\ref{Eq_7}) can be re--written as follows:
\setcounter{equation}{32}
\begin{equation}
\label{APP1__Eq_1}
\begin{split} &{\mathbb{E}}\left\{ {\ln \left( {1 + \frac{{Pg_0 \xi ^{ - \alpha } }}{{\sigma _N^2  + I_{{\rm{agg}}} \left( \xi  \right)}}} \right)} \right\} \\ & = \int\nolimits_0^{ + \infty } {\frac{{\exp \left\{ { - z} \right\}}}{z} \mathcal{M}_{I_{{\rm{agg}}} } \left( {z;\xi } \right)\left[ {1 - \mathcal{M}_0 \left( {{\rm{SNR}}\xi ^{ - \alpha } z} \right)} \right]dz} \end{split}
\end{equation}

We note that the identity in (\ref{APP1__Eq_1}) avoids the need of computing Pcov and makes our analytical development totally different from current practice \cite{BaccelliSep2009}, \cite{AndrewsNov2011}, and \cite{AndrewsOct2012}.

From (\ref{APP1__Eq_1}), it is apparent that a closed--form expression of $\mathcal{M}_{I_{{\rm{agg}}} } \left( { \cdot ;\xi } \right)$ is needed. This is the MGF of the aggregate interference, which is generated by all the interferers that lie outside a disk of radius $\xi$. In other words, due to the tier and BS association policy, ${\rm{MT}}_0$ has an exclusion zone around it where no interfering BSs are located and are allowed to transmit. This is called exclusion region \cite{Rabbachin2011}. The MGF of ${I_{{\rm{agg}}} \left( \cdot  \right)}$ has been studied in \cite{MGF_ExclusionRegion} for a generic annular region with radii $A$ and $B$. In particular, $\mathcal{M}_{I_{{\rm{agg}}} } \left( { \cdot ;\xi } \right)$ can be obtained from \cite[Eq. (6)]{MGF_ExclusionRegion} by letting $A \to \xi$ and $B \to + \infty$, as follows:
\setcounter{equation}{33}
\begin{equation}
\label{APP1__Eq_2}
\begin{split} \mathcal{M}_{I_{{\rm{agg}}} } \left( {s;\xi } \right) & = \exp \left\{ {\pi \lambda \xi ^2 } \right\}\exp \left\{ { - \pi \lambda \xi ^2 \mathcal{M}_I \left( {s\xi ^{ - \alpha } } \right)} \right\} \\ &\times \exp \left\{ { - \pi \lambda s^{{2 \mathord{\left/
 {\vphantom {2 \alpha }} \right.
 \kern-\nulldelimiterspace} \alpha }} \Gamma \left( {1 - \frac{2}{\alpha }} \right){\mathbb{E}}\left\{ {g_b^{{2 \mathord{\left/
 {\vphantom {2 \alpha }} \right.
 \kern-\nulldelimiterspace} \alpha }} } \right\}} \right\} \\ &\times \exp \left\{ {\pi \lambda s^{{2 \mathord{\left/
 {\vphantom {2 \alpha }} \right.
 \kern-\nulldelimiterspace} \alpha }} {\mathbb{E}}\left\{ {g_b^{{2 \mathord{\left/
 {\vphantom {2 \alpha }} \right.
 \kern-\nulldelimiterspace} \alpha }} \Gamma \left( {1 - \frac{2}{\alpha },sg_b \xi ^{ - \alpha } } \right)} \right\}} \right\} \end{split}
\end{equation}

In \cite{MGF_ExclusionRegion}, the expectation over the fading distribution of the interference channels is not computed in closed--form and only bounds for Rayleigh fading are provided. To the best of our knowledge, there is no analytical framework that provides an exact and closed--form expression of the MGF in (\ref{APP1__Eq_2}) for general fading channels. In what follows, we provide a general methodology to this end. This is a contribution of this paper.

By using \cite[Eq. (6.5.3)]{AbramowitzStegun}, \cite[Eq. (6.5.4)]{AbramowitzStegun}, and \cite[Eq. (6.5.29)]{AbramowitzStegun}, we have:
\setcounter{equation}{34}
\begin{equation}
\label{APP1__Eq_3}
\begin{split} \Gamma \left( {z,x} \right) & = \Gamma \left( z \right) - \gamma \left( {z,x} \right) \\ &= \Gamma \left( z \right) - \Gamma \left( z \right)x^z \exp \left\{ { - x} \right\}\sum\limits_{k = 0}^{ + \infty } {\frac{{x^k }}{{\Gamma \left( {z + k + 1} \right)}}} \end{split}
\end{equation}

By substituting (\ref{APP1__Eq_3}) in (\ref{APP1__Eq_2}), $\mathcal{M}_{I_{{\rm{agg}}} } \left( { \cdot ;\xi } \right)$ simplifies to:
\setcounter{equation}{35}
\begin{equation}
\label{APP1__Eq_4}
\begin{split} \mathcal{M}_{I_{{\rm{agg}}} } \left( {s;\xi } \right) & = \exp \left\{ {\pi \lambda \xi ^2 } \right\}\exp \left\{ { - \pi \lambda \xi ^2 \mathcal{M}_I \left( {s\xi ^{ - \alpha } } \right)} \right\} \\ & \times \exp \left\{ { - \pi \lambda \xi ^2 \mathcal{T}_I \left( {s\xi ^{ - \alpha } } \right)} \right\} \end{split}
\end{equation}
\noindent where $\mathcal{T}_I \left(  \cdot  \right)$ is given in (\ref{Eq_9}). Closed--form expressions of $\mathcal{T}_I \left(  \cdot  \right)$ are available in \emph{Propositions} \ref{Prop_1}--\ref{Prop_4}.

Finally, by substituting (\ref{APP1__Eq_4}) in (\ref{APP1__Eq_1}), the average rate in (\ref{Eq_7}) simplifies to (\ref{Eq_8}) by using the change of variable $y = z\xi ^{ - \alpha }$ and by applying the integration by parts to the integral in $\xi$. This concludes the proof. \hfill $\Box$
\section{Proof of \emph{Remark} \ref{Remark_0}} \label{APP__Remark_0}
From (\ref{Eq_8}), by using the change of variable ${\rm{SNR}}y = z$, we have:
\setcounter{equation}{36}
\begin{equation}
\label{APP1A__Eq_1}
\mathcal{R} = \int\nolimits_0^{ + \infty } {\frac{{1 - \mathcal{M}_0 \left( z \right)}}{{\mathcal{Z}_I \left( z \right)}} \frac{dz}{z}}  - \frac{\alpha }{2}\int\nolimits_0^{ + \infty } {\frac{{1 - \mathcal{M}_0 \left( z \right)}}{{\mathcal{Z}_I \left( z \right)}}\mathcal{\bar R}\left( z; \lambda \right)dz}
\end{equation}
\noindent where $\mathcal{\bar R}\left( \cdot; \cdot \right)$ is given in (\ref{APP1A__Eq_2}) at the top of this page, and (a) follows by taking into account that $0 \le \exp \left\{ { - \left( {{z \mathord{\left/ {\vphantom {z {{\rm{SNR}}}}} \right. \kern-\nulldelimiterspace} {{\rm{SNR}}}}} \right)\xi ^{{\alpha  \mathord{\left/ {\vphantom {\alpha  2}} \right. \kern-\nulldelimiterspace} 2}} } \right\} \le 1$ for every choice of the parameters, while (b) follows from \cite[Eq. (2.1.1.1)]{Prudnikov_Vol4}.

As a consequence, the average rate in (\ref{APP1A__Eq_1}) is upper-- and lower--bound as follows:
\setcounter{equation}{38}
\begin{equation}
\label{APP1A__Eq_3}
\int\nolimits_0^{ + \infty } {\frac{{1 - \mathcal{M}_0 \left( z \right)}}{{\mathcal{Z}_I \left( z \right)}}\frac{dz}{z}}  - \mathcal{\bar R}\left( \lambda  \right) \le \mathcal{R} \le \int\nolimits_0^{ + \infty } {\frac{{1 - \mathcal{M}_0 \left( z \right)}}{{\mathcal{Z}_I \left( z \right)}}\frac{dz}{z}}
\end{equation}
\noindent where:
\setcounter{equation}{39}
\begin{equation}
\label{APP1A__Eq_4}
\begin{split}\mathcal{\bar R}\left( \lambda  \right) & = \frac{1}{{{\rm{SNR}}}}\frac{\alpha }{2}\Gamma \left( {\frac{\alpha }{2}} \right)\left( {\pi \lambda } \right)^{ - \frac{\alpha }{2}} \\ & \times \int\nolimits_0^{ + \infty } {\left[ {1 - \mathcal{M}_0 \left( z \right)} \right]\mathcal{Z}_I^{ - \left( {\frac{\alpha }{2} + 1} \right)} \left( z \right)dz} \end{split}
\end{equation}

Equation (\ref{Eq_9A}) immediately follows from (\ref{APP1A__Eq_4}) by observing that $\mathop {\lim }\nolimits_{\lambda  \to  + \infty } \mathcal{\bar R}\left( \lambda  \right) = 0$ for every $\alpha > 2$. This concludes the proof. \hfill $\Box$
\section{Proof of \emph{Proposition} \ref{Prop_1}} \label{APP__Prop_1}
By using \cite[Eq. (2.21)]{SimonBook} and \cite[Eq. (2.2.1.2)]{Prudnikov_Vol4}, $\mathcal{M}_I^{\left( k \right)} \left( \cdot \right)$ in (\ref{Eq_9}) can be computed as follows:
\setcounter{equation}{40}
\begin{equation}
\label{APP2__Eq_1}
\mathcal{M}_I^{\left( k \right)} \left( s \right) = \frac{1}{{\Gamma \left( m \right)}}\left( {\frac{m}{\Omega }} \right)^m \left( {s + \frac{m}{\Omega }} \right)^{ - \left( {m + k + 1} \right)} \Gamma \left( {m + k + 1} \right)
\end{equation}

By substituting (\ref{APP2__Eq_1}) in $\mathcal{T}_I \left(  \cdot  \right)$ in (\ref{Eq_9}) and by using the identity \cite[Eq. (15.1.1)]{AbramowitzStegun}:
\setcounter{equation}{41}
\begin{equation}
\label{APP2__Eq_2}
\begin{split} &\sum\limits_{k = 0}^{ + \infty } {\frac{{\Gamma \left( {k + A} \right)s^{x + k} }}{{\Gamma \left( {k + B} \right)\left( {s + C} \right)^{z + k} }}}  \\ & \hspace{1.5cm} = \frac{{s^x }}{{\left( {s + C} \right)^z }}\frac{{\Gamma \left( A \right)}}{{\Gamma \left( B \right)}}{}_2F_1 \left( {A,1,B,\frac{s}{{s + C}}} \right) \end{split}
\end{equation}
\noindent with $A$, $B$, $C$, $x$, and $s$ being positive constants, we eventually obtain (\ref{Eq_13}) with some algebraic manipulations and using the identity $\Gamma \left( {z + 1} \right) = z\Gamma \left( z \right)$. This concludes the proof. \hfill $\Box$
\begin{figure*}[!t]
\setcounter{equation}{45}
\begin{equation}
\label{APP5__Eq_1}
\left\{ \begin{split}
 & f_{g_b } \left( x \right) \approx \left( {1 + K} \right)\exp \left\{ { - K} \right\}\frac{1}{{\sqrt \pi  }}\sum\limits_{n = 1}^{N_{{\rm{GHQ}}} } {\tilde w_n \tilde \omega _n \exp \left\{ { - \left( {1 + K} \right)\tilde \omega _n x} \right\}I_0 \left( {2\sqrt {K\left( {1 + K} \right)\tilde \omega _n x} } \right)}  \\
 & \mathcal{M}_{g_b } \left( s \right)\mathop  \approx \limits^{\left( a \right)} \frac{1}{{\sqrt \pi  }}\sum\limits_{n = 1}^{N_{{\rm{GHQ}}} } {\tilde w_n \frac{{1 + K}}{{1 + K + \left( {{s \mathord{\left/
 {\vphantom {s {\tilde \omega _n }}} \right.
 \kern-\nulldelimiterspace} {\tilde \omega _n }}} \right)}}\exp \left\{ { - s\frac{{{K \mathord{\left/
 {\vphantom {K {\tilde \omega _n }}} \right.
 \kern-\nulldelimiterspace} {\tilde \omega _n }}}}{{1 + K + \left( {{s \mathord{\left/
 {\vphantom {s {\tilde \omega _n }}} \right.
 \kern-\nulldelimiterspace} {\tilde \omega _n }}} \right)}}} \right\}}  \\
 \end{split} \right.
\end{equation}
\normalsize \hrulefill \vspace*{0pt}
\end{figure*}
\begin{figure*}[!t]
\setcounter{equation}{46}
\begin{equation}
\label{APP5__Eq_2}
\begin{split}
 \mathcal{M}_I^{\left( k \right)} \left( s \right) & \approx \left( {1 + K} \right)\exp \left\{ { - K} \right\}\frac{1}{{\sqrt \pi  }}\sum\limits_{n = 1}^{N_{{\rm{GHQ}}} } {\left[ {\tilde w_n \tilde \omega _n \int\nolimits_0^{ + \infty } {x^{k + 1} \exp \left\{ { - \left[ {s + \left( {1 + K} \right)\tilde \omega _n } \right]x} \right\}I_0 \left( {2\sqrt {K\left( {1 + K} \right)\tilde \omega _n x} } \right)dx} } \right]}  \\
 & \mathop  = \limits^{\left( a \right)} \left( {1 + K} \right)\exp \left\{ { - K} \right\}\frac{1}{{\sqrt \pi  }}\sum\limits_{n = 1}^{N_{{\rm{GHQ}}} } {\tilde w_n \tilde \omega _n \Gamma \left( {k + 2} \right)\left[ {s + \left( {1 + K} \right)\tilde \omega _n } \right]^{ - \left( {k + 2} \right)} {}_1F_1 \left( {k + 2,1,\frac{{K\left( {1 + K} \right)}}{{1 + K + \left( {{s \mathord{\left/
 {\vphantom {s {\tilde \omega _n }}} \right.
 \kern-\nulldelimiterspace} {\tilde \omega _n }}} \right)}}} \right)}  \\
 \end{split}
\end{equation}
\normalsize \hrulefill \vspace*{0pt}
\end{figure*}
\begin{figure*}[!t]
\setcounter{equation}{47}
\begin{equation}
\label{APP5__Eq_3}
\begin{split}\mathcal{T}_I \left( s \right) & \approx \left( {1 + K} \right)\exp \left\{ { - K} \right\}\Gamma \left( {1 - \frac{2}{\alpha }} \right)\frac{1}{{\sqrt \pi  }} \\ & \times \sum\limits_{n = 1}^{N_{{\rm{GHQ}}} } {\left[ {\tilde w_n \tilde \omega _n \sum\limits_{k = 0}^{ + \infty } {\frac{{\Gamma \left( {2 + k} \right)}}{{\Gamma \left( {2 - \frac{2}{\alpha } + k} \right)}}\frac{{s^{k + 1} }}{{\left[ {s + \left( {1 + K} \right)\tilde \omega _n } \right]^{2 + k} }}{}_1F_1 \left( {k + 2,1,\frac{{K\left( {1 + K} \right)}}{{1 + K + \left( {{s \mathord{\left/
 {\vphantom {s {\tilde \omega _n }}} \right.
 \kern-\nulldelimiterspace} {\tilde \omega _n }}} \right)}}} \right)} } \right]} \end{split}
\end{equation}
\normalsize \hrulefill \vspace*{0pt}
\end{figure*}
\section{Proof of \emph{Proposition} \ref{Prop_2}} \label{APP__Prop_2}
By using the approximate expression of the PDF of Log--Normal random variables in \cite[Table IV]{AlouiniFading_B}, \emph{i.e.}, $f_{g_b } \left( x \right) \approx \left( {{1 \mathord{\left/ {\vphantom {1 {\sqrt \pi  }}} \right. \kern-\nulldelimiterspace} {\sqrt \pi  }}} \right)\sum\nolimits_{n = 1}^{N_{{\rm{GHQ}}} } {\tilde w_n \delta \left( {x - 10^{{{\left( {\sqrt 2 \sigma \tilde s_n  + \mu } \right)} \mathord{\left/ {\vphantom {{\left( {\sqrt 2 \sigma \tilde s_n  + \mu } \right)} {10}}} \right. \kern-\nulldelimiterspace} {10}}} } \right)}$, $\mathcal{M}_I^{\left( k \right)} \left( \cdot \right)$ in (\ref{Eq_9}) can be computed as follows:
\setcounter{equation}{42}
\begin{equation}
\label{APP3__Eq_1}
\begin{split} \mathcal{M}_I^{\left( k \right)} \left( s \right) &\approx \frac{1}{{\sqrt \pi  }}\sum\limits_{n = 1}^{N_{{\rm{GHQ}}} } \tilde w_n 10^{\left( {k + 1} \right){{\left( {\sqrt 2 \sigma \tilde s_n  + \mu } \right)} \mathord{\left/
 {\vphantom {{\left( {\sqrt 2 \sigma \tilde s_n  + \mu } \right)} {10}}} \right.
 \kern-\nulldelimiterspace} {10}}} \\ & \times \exp \left\{ { - 10^{{{\left( {\sqrt 2 \sigma \tilde s_n  + \mu } \right)} \mathord{\left/
 {\vphantom {{\left( {\sqrt 2 \sigma \tilde s_n  + \mu } \right)} {10}}} \right.
 \kern-\nulldelimiterspace} {10}}} s} \right\} \end{split}
\end{equation}

By substituting (\ref{APP3__Eq_1}) in $\mathcal{T}_I \left(  \cdot  \right)$ in (\ref{Eq_9}) and by using the identity \cite[Eq. (13.1.2)]{AbramowitzStegun}:
\setcounter{equation}{43}
\begin{equation}
\label{APP3__Eq_2}
\sum\limits_{k = 0}^{ + \infty } {\frac{{s^{x + k} }}{{\Gamma \left( {k + B} \right)}}}  = \frac{{s^x }}{{\Gamma \left( B \right)}}{}_1F_1 \left( {1,B,s} \right)
\end{equation}
\noindent with $B$ and $z$ being positive constants, we obtain (\ref{Eq_14}) after some algebra. This concludes the proof. \hfill $\Box$
\section{Proof of \emph{Proposition} \ref{Prop_3}} \label{APP__Prop_3}
Similar to Appendix \ref{APP__Prop_2}, in order to have an analytically tractable expression of PDF and MGF of composite Nakagami--\emph{m} and Log--Normal fading, we use the approximation of the Log--Normal distribution that is based on the Gauss--Hermite quadrature in \cite[Eq. (2.58)]{SimonBook} and \cite[Table V]{AlouiniFading_A}. Accordingly, $\mathcal{M}_I^{\left( k \right)} \left( \cdot \right)$ in (\ref{Eq_9}) can be computed as follows:
\setcounter{equation}{44}
\begin{equation}
\label{APP4__Eq_1}
\begin{split} \mathcal{M}_I^{\left( k \right)} \left( s \right) &\approx \frac{1}{{\sqrt \pi  }}\frac{{m^m }}{{\Gamma \left( m \right)}} \\ & \times \sum\limits_{n = 1}^{N_{{\rm{GHQ}}} } {\tilde w_n \tilde \omega _n^m \left( {s + m\tilde \omega _n } \right)^{ - \left( {m + k + 1} \right)} \Gamma \left( {m + k + 1} \right)} \end{split}
\end{equation}

By substituting (\ref{APP4__Eq_1}) in $\mathcal{T}_I \left(  \cdot  \right)$ in (\ref{Eq_9}), from (\ref{APP2__Eq_2}) we eventually get (\ref{Eq_15}). This concludes the proof. \hfill $\Box$
\section{Proof of \emph{Proposition} \ref{Prop_4}} \label{APP__Prop_4}
Similar to Appendix \ref{APP__Prop_2}, in order to have an analytically tractable expression of PDF and MGF of composite Rice and Log--Normal fading, we use the approximation of the Log--Normal distribution that is based on the Gauss--Hermite quadrature in \cite[Table IV]{AlouiniFading_B}. In particular, from \cite[Eq. (6)]{Loo} we obtain (\ref{APP5__Eq_1}) shown at the top of this page, where (a) is obtained by using \cite[Eq. (2.17)]{SimonBook}. Thus, $\mathcal{M}_I^{\left( k \right)} \left( \cdot \right)$ in (\ref{Eq_9}) can be computed as shown in (\ref{APP5__Eq_2}) at the top of this page, where (a) is obtained by using \cite[Eq. (6.631)]{AbramowitzStegun}. By substituting (\ref{APP5__Eq_2}) in $\mathcal{T}_I \left(  \cdot  \right)$ in (\ref{Eq_9}), we obtain (\ref{APP5__Eq_3}) shown at the top of this page.

Since the infinite series in (\ref{APP5__Eq_3}) is not fast converging, we elaborate further $\mathcal{T}_I \left(  \cdot  \right)$ in order to obtain a better expression for simple numerical computation. For $K \ne 0$, this can be obtained by first replacing ${}_1F_1 \left( { \cdot , \cdot , \cdot } \right)$ with its series expansion in \cite[Eq. (13.1.2)]{AbramowitzStegun}, and then computing the infinite series in $k$ by using (\ref{APP2__Eq_2}). Eventually, we obtain (\ref{Eq_16}) with some algebraic manipulations and using the identity $\Gamma \left( {z + 1} \right) = z\Gamma \left( z \right)$. The case $K=0$ can be obtained by noting that ${}_1F_1 \left( {A,1,0} \right) = 1$ for every $A$, and by applying the same procedure as in Appendix \ref{APP__Prop_3} with $m=1$. This concludes the proof. \hfill $\Box$
\section{Proof of \emph{Corollary} \ref{Cor_2}} \label{APP__Cor_2}
Let $\mathcal{U}^{\left( {{\rm{asymptote}}} \right)} \left( z \right) = \lim _{z \to 0^ +  } \mathcal{U}\left( z \right)$ in (\ref{Eq_17}). This limit can be computed as follows:
\setcounter{equation}{48}
\begin{equation}
\label{APP6__Eq_1}
\begin{split} \mathcal{U}^{\left( {{\rm{asymptote}}} \right)} \left( z \right) & = \mathop {\lim }\limits_{z \to 0^ +  } G_{\alpha _N ,\alpha _D }^{\alpha _D ,\alpha _N } \left( {z\left| {\begin{array}{*{20}c}
   {\Delta \left( {\alpha _N , - \nu _\alpha  } \right)}  \\
   {\Delta \left( {\alpha _D ,0} \right)}  \\
\end{array}} \right.} \right) \\ & \mathop  = \limits^{\left( a \right)} \mathop {\lim }\limits_{\zeta  \to +\infty } G_{\alpha _D ,\alpha _N }^{\alpha _N ,\alpha _D } \left( {\zeta \left| {\begin{array}{*{20}c}
   {1 - \Delta \left( {\alpha _D ,0} \right)}  \\
   {1 - \Delta \left( {\alpha _N , - \nu _\alpha  } \right)}  \\
\end{array}} \right.} \right) \end{split}
\end{equation}
\noindent where $\zeta  = {1 \mathord{\left/ {\vphantom {1 z}} \right. \kern-\nulldelimiterspace} z}$ and (a) is obtained by using \cite[Eq. (8.2.2.14)]{Prudnikov_Vol3}.

From (\ref{APP6__Eq_1}), (\ref{Eq_17}) can be obtained by using \cite[Theorem 1.8.3]{MathaiSaxena}, and more explicitly by using \cite[Eq. 1.8.8]{MathaiSaxena} whose parameters are defined in \cite[Eq. 1.4.2]{MathaiSaxena} and \cite[Eq. 1.4.7]{MathaiSaxena}, as well as by taking into account that $\lim _{z \to 0^ +  } {}_pF_q \left( {a,b,z} \right) = 1$, and, thus, the generalized hypergeometric function ${}_pF_q \left( { \cdot , \cdot , \cdot } \right)$ can be neglected. This concludes the proof. \hfill $\Box$
\begin{biography} {Marco Di Renzo} (S'05--AM'07--M'09) was born in L'Aquila, Italy, in 1978. He received the Laurea (cum laude) and the Ph.D. degrees in Electrical and Information Engineering from the Department of Electrical and Information Engineering, University of L'Aquila, Italy, in April 2003 and in January 2007, respectively.

From August 2002 to January 2008, he was with the Center of Excellence for Research DEWS, University of L'Aquila, Italy.
From February 2008 to April 2009, he was a Research Associate with the Telecommunications Technological Center of Catalonia (CTTC), Barcelona, Spain.
From May 2009 to December 2009, he was an EPSRC Research Fellow with the Institute for Digital Communications (IDCOM), The University of Edinburgh, Edinburgh, United Kingdom (UK).

Since January 2010, he has been a Tenured Researcher (``Charg\'e de Recherche Titulaire'') with the French National Center for Scientific Research (CNRS), as well as a faculty member of the Laboratory of Signals and Systems (L2S), a joint research laboratory of the CNRS, the \'Ecole Sup\'erieure d'\'Electricit\'e (SUP\'ELEC), and the University of Paris--Sud XI, Paris, France. His main research interests are in the area of wireless communications theory. He is a Principal Investigator of three European--funded research projects (Marie Curie ITN--GREENET, Marie Curie IAPP--WSN4QoL, and Marie Curie ITN--CROSSFIRE).

Dr. Di Renzo is the recipient of the special mention for the outstanding five--year (1997--2003) academic career, University of L'Aquila, Italy;
the THALES Communications fellowship for doctoral studies (2003--2006), University of L'Aquila, Italy; the Best Spin--Off Company Award (2004), Abruzzo Region, Italy; the Torres Quevedo award for research on ultra wide band systems and cooperative localization for wireless networks (2008--2009), Ministry of Science and Innovation, Spain; the ``D\'erogation pour l'Encadrement de Th\`ese'' (2010), University of Paris--Sud XI, France; the 2012 IEEE CAMAD Best Paper Award from the IEEE Communications Society; and the 2012 Exemplary Reviewer Award from the IEEE WIRELESS COMMUNICATIONS LETTERS of the IEEE Communications Society. He currently serves as an Editor of the IEEE COMMUNICATIONS LETTERS.
\end{biography}
\begin{biography} {Alessandro Guidotti} received the Master degree (magna cum laude) and the Ph.D. degree in Telecommunications Engineering from the University of Bologna in 2008 and 2012, respectively. He is now research grantee (Post--Doctoral Fellow) at the University of Bologna.

In 2011 and 2012, he was a Visiting Researcher at SUPELEC (Paris, France) working on stochastic geometry for interference modeling in distributed wireless networks. His research interests are spectrum management, cognitive radio, interference characterization, and stochastic geometry.

He has experience in ESA and FP7 projects. He represented the Italian Administration in CEPT SE43 meetings. He received the Lions Club Prize for the Best Master Thesis in 2008.
\end{biography}
\begin{biography} {Giovanni E. Corazza} is a Full Professor at the Alma Mater Studiorum--University of Bologna, Member of the Alma Mater Executive Board, former Head of the Department of Electronics, Computer Science, and Systems (DEIS), and founder of the Marconi Institute for Creativity (2011).

He was Chairman of the School for Telecommunications in the years 2000--2003, Chairman of the Advanced Satellite Mobile Systems Task Force (ASMS TF), Founder and Chairman of the Integral Satcom Initiative (ISI), a European Technology Platform devoted to Satellite Communications. In the years 1997--2012, he served as Editor for Communications Theory and Spread Spectrum for the IEEE TRANSACTIONS ON COMMUNICATIONS.

He is author of more than 260 papers, and received the Marconi International Fellowship Young Scientist Award in 1995, the 2009 IEEE Satellite Communications Distinguished Service Award, the 2002 IEEE VTS Best System Paper Award, and the Best Paper Award at the 1998 IEEE ISSSTA, at the 2011 IEEE ICT, and at the 2005 ISWCS. He has been the General Chairman of the IEEE ISSSTA 2008, ASMS 2004, ASMS 2006, ASMS 2008, ASMS 2010 conferences.

His research interests are wireless and satellite communications, estimation and synchronization, spread spectrum and multi--carrier transmission, upper layer coding, navigation and positioning, scientific creative thinking.
\end{biography}
\end{document}